\documentclass[reqno,12pt]{book}
\usepackage{amsmath} 
\usepackage{amsfonts} 
\usepackage{amssymb}
\usepackage[bf]{caption2}
\usepackage{bm}
\usepackage{graphicx,epsfig}
\usepackage[dvips]{color}
\usepackage{amssymb,amsmath,amsthm}

\newtheorem{prop}{Proposition}

\newcommand{\gth}[1]{\textsf{T}_\textsf{#1}}
\newtheorem{theorem}{Theorem}
\newtheorem{lemma}{Lemma}
\newtheorem{corollary}{Corollary}
\newtheorem{definition}{Definition}
\def\d{{\mathrm{d}}}
\def\implies{\Rightarrow}
\def\arctanh{{\mathrm{arctanh}}}
\def\SIM{\triangleq}
\def\txt{\textstyle}
\def\lint{\hbox{\Large $\displaystyle\int$}} 

\begin{document}
\bibliographystyle{plain}
\begin{titlepage}

\center


{\Huge \bf Some Exact Solutions}\\[0.3cm]
{\Huge \bf in}\\[0.3cm]
{\Huge \bf General Relativity}\\[1.3cm]
{\sf \large
Master's thesis\\[1mm]

by\\[1mm]

\Large Petarpa Boonserm\\[1cm]


\normalsize
A thesis submitted to the \\
Victoria University of Wellington \\
in fulfilment of the requirements \\
for the degree of \\
Master of Science in Mathematics\\[0.5cm]


}
\begin{minipage}[t][4cm][c]{10cm}
\begin{center}
\includegraphics{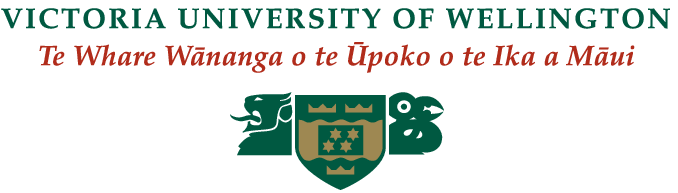}\\
2005
\end{center}
\end{minipage}

Supervisor:
Dr. Matt Visser\\[1cm]

\end{titlepage}

\thispagestyle{empty}
\frontmatter
\centerline{\bf \large Abstract}

\bigskip

In this thesis four separate problems in general relativity are considered, divided into two separate themes: coordinate conditions and perfect fluid spheres. Regarding coordinate conditions we present a pedagogical discussion of how the appropriate use of coordinate conditions can lead to simplifications in the form of the spacetime curvature --- such tricks are often helpful when seeking specific exact solutions of the Einstein equations. Regarding perfect fluid spheres we present several methods of transforming any given perfect fluid sphere into a possibly new perfect fluid sphere.  This is done in three qualitatively distinct manners: The first set of solution generating theorems apply in Schwarzschild curvature coordinates, and are phrased in terms of the metric components: they show how to transform one static spherical perfect fluid spacetime geometry into another. A second set of solution generating theorems extends these ideas to other coordinate systems (such as isotropic, Gaussian polar, Buchdahl, Synge, and exponential coordinates), again working directly in terms of the metric components. Finally, the solution generating theorems are rephrased in terms of the TOV equation and density and pressure profiles. Most of the relevant calculations are carried out analytically, though some numerical explorations are also carried out.
\chapter{Acknowledgements}
I would like to thank my supervisor, Dr.~Matt Visser for his guidance and 
support. I am lucky to have such an active and hard working supervisor. 
I really appreciate the effort he made for my thesis, making sure that 
things got done properly and on time. I would also like to thank Silke 
Weinfurtner, who I collaborated with on several projects, for her help, 
support, and encouragement.

I am grateful to the School of Mathematics, Statistics, and Computer Science 
for providing me with an office and all the facilities, and the Thai 
Government Scholarship that provided me with funding.

I also would like to say thanks to my family who were also very supportive 
and listened to me.

Finally, I would like to extend my gratitude to Brijesh Dongol for always 
being there to listen to me. I am not sure that this thesis would have been 
finished without the support he showed.

\chapter{Preface}


This thesis looks at four problems in general relativity:
\begin{itemize}
\item
Coordinate conditions in general relativity:
A method of looking for nice solutions of the Einstein equations.
\item
Develop several new transformation theorems that map perfect fluid spheres into perfect fluid spheres.
\item
Develop a systematic way of classifying the set of all perfect fluid spheres.
\item
Rephrase all these theorems which apply in Schwarzschild coordinates,
directly in terms of the pressure profile and density profile.
\end{itemize}

General relativity (GR) or general relativity theory (GRT) is a theory of gravitation discovered by Albert Einstein. The fundamental physical postulate of GR is that the presence of matter causes curvature in the spacetime in which it exists. This curvature is taken to be the gravitational field produced by the matter. Einstein's field equation gives the mathematical description of how the matter and curvature are related. Moreover, once this curvature is given, GR describes how other objects (such as planets and light beams) move in this gravitational field via the geodesic equation.

In addition, general relativity states that clocks run slower in strong gravitational fields (or highly accelerated frames), predicting a gravitational redshift. It also predicts the existence of gravitational lensing, gravitational waves, gravitomagnetism, the Lense-Thirring effect, and relativistic precession of orbiting bodies.

The first problem is looking for nice solutions of the Einstein equations. However,
Before stating Einstein's equation, we need to briefly describe the concept of special relativity, otherwise general relativity will be hard to understand.

\section*{Structure of the thesis}
This thesis has been written with the goal of being accessible to people with a basic background in general relativity, especially in coordinate conditions in general relativity and perfect fluid spheres.

This thesis is made up of six chapters and three appendices. Two of appendices are papers published on work relating to this thesis. One was produced as a collaboration with Dr.~Matt Visser, Silke Weinfurtner, Celine Cattoen and Tristan Faber. The second was produced in collaboration with Dr.~Matt Visser and Silke Weinfurtner. At the time of writing all papers have been published.

\tableofcontents
\listoffigures
\listoftables
\mainmatter
\chapter{General introduction}
\section{Introduction}
General relativity (GR) is the theory of space, time and gravitation formulated by Einstein in 1915. General relativity is a beautiful theory, however it is often regarded as a very deep and difficult theory. It has been considerably developed since the late 1950s. Furthermore, in the mid 1960s, the modern theory of gravitational collapse, singularities, and black holes has been developed.
In this thesis, we present the coordinate condition in general relativity, we will discuss the recent advances and developments in methods of looking for a nice solutions of Einstein equation. In addition, we also present the perfect fluid spheres in general relativity. Understanding the concept of perfect fluid spheres are important. They are first approximations to attempt at building a realistic model for a general relativistic star. We introduce the ``algorithmic" techniques that permit one to generate large classess of perfect fluid spheres in a purely mechanical way. Furthermore, we will extend these algorithmic ideas, by proving several solution-generating theorems of varying levels of complexity. We shall then explore the formal properties of these solution-generating theorems and then will use these theorems to classify some of the previously known exact solutions. Indeed, we develop several new transformation theorems that map perfect fluid spheres into perfect fluid spheres. Additionally, we develop a systematic way of classifying the set of all perfect fluid spheres.
The remaining part of this thesis will be devoted to rephrasing all these theorems, which originally apply in Schwarzschild coordinates, directly in terms of the pressure profile and density profile.

Now we would like to give a mathematically precise formulation of the ideas in Einstein's equation. Before stating Einstein's equation, we need to describe in brief the concept of special relativity, otherwise general relativity will be hard to understand.

\section{Einstein's equation}
To start we give an exposition of elementary topics in the geometry of space, and the spatial tensorial character of physical laws in prerelativity physics.
We formulate general relativity and provide a motivation for Einstein's equation, which relates the geometry of spacetime to the distribution of matter in the universe.
\subsection{General and Special Covariance}
Assume that space has the manifold  structure of $\mathbb{R}^3$ and the association of points of space with elements $(x^1, x^2, x^3)$ of $\mathbb{R}^3$ can be accomplished by construction of a \emph{rigid rectilinear grid} of metersticks. We call the coordinates of space derived in this manner as \emph{Cartesian coordinates} ~\cite{Wald}.
The distance, $S$, between two points, $x$ and $\bar{x}$, defined in terms of Cartesian coordinates by
\begin{equation}
\label{distance1}
S^2 = (x^1 - \bar{x}^1)^2 + (x^2 - \bar{x}^2)^2 + (x^3 - \bar{x}^3)^2
\end{equation}
This formula is the distance between two points.
\\
Referred to equation (\ref{distance1}), the distance between two \emph{nearby} points is
\begin{equation}
(\delta S)^2 = (\delta x^1)^2 + (\delta x^2)^2 + (\delta x^3)^2
\end{equation}
Therefore, the metric of space is given by
\begin{equation}
\d s^2 = (\d x^1)^2 + (\d x^2)^2 + (\d x^3)^2
\end{equation}
In the Cartesian coordinate basis, we derive
\begin{equation}
\d s^2 = \sum_{a, b} \; h_{a b}(\d x^{a}) (\d x^{b})
\end{equation}
with $h_{a b}$ = diag (1, 1, 1). This definition of $h_{ab}$ is independent of choice of Cartesian coordinate system.

When the components of the metric in the Cartesian coordinate basis are constants, we get
\begin{equation}
\partial_a h_{bc} = 0 .
\end{equation}
The space is the manifold $\mathbb{R}^3$ which possesses a flat Riemann metric. We are able to use the geodesics of the flat metric to construct a Cartesian coordinate system. We use the fact that initially parallel geodesics remain parallel because the cuvature vanishes.
\section{Special Relativity}
\subsection{Minkowski space-time}
We introduce a $4$-\emph{dimensional} continuum called space-time in which an event has coordinates $(t, x, y, z)$ \cite{D'Inverno}.

Minkowski space-time is defined as a 4-\emph{dimensional} manifold provided with a \emph{flat} metric of signature $+2$. By the definiton, since the metric is flat, there exists a special coordinate system covering the whole manifold in which the metric is diagonal, with diagonal elements equal to $\pm 1$. For convenience, we prefer to use the convention that lower case latin indices run from $0$ to $3$. The special coordinate system is called a Minkowski coordinate system and is written
\begin{equation}
(x^a) = (x^0, x^1, x^2, x^3) = (c \, t, x, y, z).
\end{equation}
In special relativity \cite{Wald}, it is assumed that spacetime has the manifold structure of $\mathbb{R}^4$.

The spacetime interval, $\mathcal{S}$, between two events $x$ and $\bar{x}$ defined by
\begin{equation}
\label{interval1}
\mathcal{S} = -(x^0 - \bar{x}^0)^2 + (x^1 - \bar{x}^1)^2 + (x^2 - \bar{x}^2)^2 + (x^3 - \bar{x}^3)^2
\end{equation}
in units where $c =1$.

From the equation (\ref{interval1}), we are able to defind the \emph{metric of spacetime} $\eta_{ab}$ by
\begin{equation}
\d s^2 = \sum_{a, b = 0}^3 \; \eta_{a b} \, \d x^{a} \, \d x^{b}
\end{equation}
with $\eta_{a b}$ = diag (-1, 1, 1, 1), where ${x^{a}}$ is any global inertial coordinate system.

Therefore, the ordinary derivative operator, $\partial_a$, of the global inertial coordinates satisfies
\begin{equation}
\partial_a \eta_{bc} = 0
\end{equation}
The curvature of $\eta_{ab}$ vanishes.
In addition, we can parameterize timelike curves by \emph{proper time}, $\tau$, defined by
\begin{equation}
\label{proper_time}
\tau = \int (-\eta_{ab} \; T^a \, T^b)^{1/2} \, \d t,
\end{equation}
where $t$ = arbitrary parameterizaton of the curve, and $T^a$ = the tangent to the curve in this parameterization.

The tangent vector $u^a$ to a timelike curve parameterized by $\tau$ is defined by the 4-\emph{velocity} of the curve.
The square of any 4-vector is an invariant and so 
\begin{equation}
u^a u_a = - c^2,
\end{equation}
where $c =1$ so we get,
\begin{equation}
u^a u_a = -1.
\end{equation}
In the absence of external forces, its 4-\emph{velocity} will satisfy the equation of motion,
\begin{equation}
\label{u_a}
u^a \partial_a u^b = 0,
\end{equation}
where $\partial_a$ is the derivative operator associated with $\eta_{ab}$. In addition, when forces are present, the equation (\ref{u_a}), $u^a \partial_a u^b$ is nonzero. Furthermore, all material particles have a parameter known as ``rest mass", $m$, which appears as a parameter in the equations  of motion when forces are present. We can define the energy momentum 4-vector, $p^{a}$, of a particle of mass $m$ by
\begin{equation}
\label{momentum1}
p^a = m \, u^a .
\end{equation}
Finally, we can define the energy of particle which is measured by an observer -- present at the site of the particle -- as
\begin{equation}
\label{Energy}
E = -p_{a} \, v^{a},
\end{equation}
where $v^{a}$ is the 4-velocity of the observer.
\\
In special relativity, the energy is the ``time component" of the 4-vector, $p^{a}$. At the rest frame, a particle with respect to the observer, equation (\ref{Energy}) reduces to the familiar formula $E = m \, c^2$. When the spacetime metric, $\eta_{ab}$ is flat, and the parallel transport is path independent, we are able to define the energy of a particle as measured by an observer who is not present at the site of the particle and has 4-velocity parallel to that of the distant observer.
\subsection{The Stress-Energy Tensor}
In special relavity, we define the \emph{energy-momentum} 4-vector of a particle of mass $m$ 
as in equation (\ref{momentum1}).
\\
From \cite{D'Inverno}, the Minkowski line element takes the form 
\begin{equation}
\d s^2 = - \d t^2 + \d x^2 + \d y^2 + \d z^2
\end{equation}
We can write this in tensorial form as
\begin{equation}
\d s^2 = \eta_{ab} \, \d x^a \d x^b,
\end{equation}
We take $\eta_{ab}$ to denote the Minkowski metric
\begin{equation}
\eta_{ab} \equiv
\left[
\begin{array}{cccc}
  -1&0 &0 &0\\
  0   &+1    &0&0\\
  0     &0     &+1 &0\\
  0   &0  &0  &+1\\
\end{array}
\right] 
= \text{diag} \, (-1, +1, +1, +1)
\end{equation}
In relativistic units the equation for the proper time satisfies
\begin{equation}
\d \tau^2 = - \d s^2.
\end{equation}
Now we present proper time $\tau$ relates to coordinate time $t$ for any observer whose velocity at time $t$ is ${v}$, where
\begin{equation}
v = \left({\d x \over \d t}, {\d y \over \d t}, {\d z \over \d t}\right).
\end{equation}
So we have,
\begin{eqnarray}
\d \tau^2 &=& - \d s^2 = - (- c^2 \, \d t^2 + \d x^2 + \d y^2 + \d z^2) 
\nonumber
\\
&=& - \d t^2 \left[- c^2 + \left({\d x \over \d t}\right)^2 + \left({\d y \over \d t}\right)^2 + \left({\d z \over \d t}\right)^2 \right]
\nonumber
\\
&=& - c^2 \, \d t^2 \left\{- 1 + {1 \over c^2} \left[\left({\d x \over \d t}\right)^2 + \left({\d y \over \d t}\right)^2 + \left({\d z \over \d t}\right)^2 \right]\right\}
\nonumber
\\
&=& c^2 \, \d t^2 \left(1 - {v^2 \over c^2} \right)
\nonumber
\\
\d \tau &=& \left(1 - {v^2 \over c^2} \right)^{1/2} \, c \, \d t
\end{eqnarray}
The time-component of the energy-momentum vector does represent the energy of the particle
\begin{equation}
p^0 = {E \over c}, \qquad \gamma = \left(1 - {v^2 \over c^2 }\right)^{-1/2},  \qquad \text{and} \qquad E = mc^2 \gamma.
\end{equation}
The space-components are the components of the three-dimensional momentum
\begin{equation}
\text{p} = m \gamma \text{v}.
\end{equation}





A perfect fluid is defined to be a continuous distribution of matter with stress energy tensor $T_{ab}$ of the form
\begin{equation}
T_{ab} = \rho u_{a} u_{b} + p (\eta_{ab} + u_{a} u_{b}),
\end{equation}
where $u^{a}$ is the 4-velocity of  the fluid, $\rho$ is the mass-energy density in the rest-frame of fluid, and $p$ is the the pressure in the rest-frame of the fluid.

When there is no external forces, the equation of motion of a perfect fluid is simply
\begin{equation}
\partial^{a} \, T_{ab} = 0.
\end{equation}
Consider $\partial^{a} T_{ab}$, we can write this in terms of $\rho$, $p$, and $u^a$ as
\begin{eqnarray}
\label{partial_T_ab}
\partial^{a} T_{ab} &=& \partial^{a}\left[\rho u_a u_b + p (\eta_{ab} + u_a u_b)\right] 
\nonumber
\\
&=&(\partial^{a} \rho) u_a u_b + (\partial^{a} p) (\eta_{ab} + u_a u_b)
\nonumber
\\
&&+ (\rho + p) (\partial^{a} u_a) u_b + (\rho + p) u_a (\partial^{a} u_b) 
\nonumber
\\
&=& \left[(u^a \partial_a \rho)  + (\rho + p) \partial^a u_a \right] u_b
\nonumber
\\
&& + \left[(\rho + p) u^a \partial_a u_b + \partial^a p (\eta_{ab} + u_a u_b)\right]
\end{eqnarray}
For equation (\ref{partial_T_ab}) we can project the resulting equation parallel and perpendicular to $u^b$, we find:
\begin{equation}
\label{real_1}
\left[(u^a \partial_a \rho)  + (\rho + p) \partial^a u_a \right] = 0,
\end{equation}
\begin{equation}
\label{real_2}
\left[(\rho + p) u^a \partial_a u_b + \partial^a p (\eta_{ab} + u_a u_b)\right] = 0.
\end{equation}
In the non-relativistic limit, when $p \ll \rho$ , $u^{\mu} = (1, \overrightarrow{v})$, and $v \, {\txt \d p \over \txt \d t} \ll |\overrightarrow{\nabla} p|$, equation (\ref{real_1}) becomes,
\begin{eqnarray}
u^a \, \partial_{a} \rho + \rho \, \partial^a \, u_a &=& 0
\nonumber
\\
\Rightarrow \;\;\;\, \partial_t \rho + \overrightarrow{v} \cdot \overrightarrow{\nabla} \, \rho + \rho \overrightarrow{\nabla} \cdot \overrightarrow{v} &=& 0
\nonumber
\\
\Rightarrow\quad \qquad \qquad\partial_t \rho  + \overrightarrow{\nabla} \cdot (\rho \overrightarrow{v}) &=& 0
\end{eqnarray}
and equation (\ref{real_2}) becomes
\begin{equation}
\rho \left\{{\partial \overrightarrow{v} \over \partial t} + (\overrightarrow{v} \cdot \overrightarrow{\nabla}) \overrightarrow{v} \right\} = -\overrightarrow{\nabla} p
\end{equation}
\subsection{Relativistic hydrodynamics}
Ordinary hydrodynamics is dealing with two basic equations \cite{Matt1}:

The equation of continuity
\begin{equation}
\partial_{t} \rho + \nabla \cdot (\rho \overrightarrow{v}) = 0
\end{equation}
This relates to the density and velocity of the fluid. It is equivalent to the conservation of mass.

The Euler equation is the fluid mechanics equivalent to Newton's second law, it relates the acceleration of a particle following the flow
\begin{equation}
\overrightarrow{a} = {\d v \over \d t} = {\partial \overrightarrow{v} \over  \partial t} + (\overrightarrow{v} \cdot \overrightarrow{\nabla}) \overrightarrow{v}
\end{equation}
We can write it in term of force density $\overrightarrow{f}$ and mass density $\rho$:
\begin{equation}
\overrightarrow{a} =  {\partial \overrightarrow{v} \over  \partial t} + (\overrightarrow{v} \cdot \overrightarrow{\nabla}) \overrightarrow{v} = {\overrightarrow{f} \over \rho}
\end{equation}

These laws have relativistic generalizations which are:
\begin{eqnarray}
\nabla_{a} (\rho \, v^a) &=& 0
\\
A^a &=& V^b \, \nabla_{b} \, V^a
\end{eqnarray}
where $A^a$ is now a 4-vector field of 4-accelerations.

The relativistic continuity equation yields
\begin{equation}
{\partial (\rho \, \gamma) \over \partial t} = \nabla \cdot (\rho \, \gamma \, \overrightarrow{v})
\end{equation}
We interpret $\rho$ as proportional to the number density of particles as measured by an observer moving with the fluid. Indeed, the $\gamma$ factor is corresponding to the fact that Lorentz contraction ``squashes" in the direction of motion, therefore, as seen by an observer moving with respect to the fluid the number density of particles is $\rho \, \gamma$.

Now for the 4-acceleration
\begin{equation}
A^{i} = (\gamma \, \partial_{t} + \gamma \, [\overrightarrow{v} \cdot \overrightarrow{\nabla}]) \, [\gamma \, v^{i}]
\end{equation}
and
\begin{equation}
A^{0} = (\gamma \, \partial_{t} + \gamma \, [\overrightarrow{v} \cdot \overrightarrow{\nabla}]) \, [\gamma]
\end{equation}
The standard Newtonian results is reproduced at the low velocity, where $\gamma \to 1$.

The special theory of relativity only deals with flat spacetime and the motion of objects is usually treated in terms of Lorentz transformations and translations from inertial frame to another. The general theory of relativity extends it to deal with non-inertial frames, and via Einstein's equations with curved spacetimes as well.
We need to understand the basic concepts of special relativity before  turning to general relativity---as otherwise the mathematical constructions used in general relativity would appear rather unmotivated.
\section{Conclusion}
\subsection{Postulational formulation of special relativity}
There are two sets of postulates which are useful to generalize the general
theory \cite{D'Inverno}.
\begin{itemize}
\item[\bf {Postulate I.}]
Space and time are represented by a 4-dimensional manifold provided
with a symmetric affine connection, $\Gamma^{a}{}_{bc}$, and a metric tensor,
$g_{ab}$, which is satisfied
as follows:
\item[(i)]
$g_{ab}$ is non-singular with signature $- + + +$;
\item[(ii)]
$\nabla_{c} \, g_{ab} = 0;$
\item[(iii)]
$R^a{}_{bcd} = 0.$
\end{itemize}
The Postulate states that $\Gamma^{a}{}_{bc}$ is the metric connection and
that
the metric is flat.
\begin{itemize}
\item[\bf {Postulate II.}]
There exist privileged classes of curves in the manifold singled out as
follows:
\item[(i)]
ideal clocks travel along timelike curves and measure the parameter  $\tau$
defined by $\d \tau^2 = -g_{ab} \, \d x^a \, \d x^b$;
\item[(ii)]
free particles travel along timelike geodesics;
\item[(iii)]
light rays travel along null geodesics.
\end{itemize}
The first part of the second postulate makes physical the distinction
between
space and time in the manifold. In Minkowski coordinates, it distinguishes
the coordinate $x^0$ from the other three as the ``time'' coordinate.
Furthermore, it states that the proper time $\tau$ which a clock
measures is in accordance with the clock hypothesis. The rest of Postulate 
II
singles out the privileged curves that free particles and light rays travel
along.
\newpage
\subsection{The correspondence principle}
\begin{figure}[!ht]
\centering
\includegraphics[scale = 0.5]{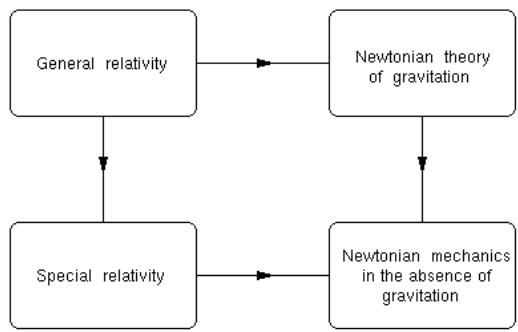}
\caption{\label{correspondence}This structure shows the correspondence
principle for general relativity.}
\end{figure}

Any new theory is consistent with any acceptable earlier theories within
their range
of validity. General relativity must agree on the one hand with special
relativity in the absence of gravitation and on the other hand with
Newtonian
gravitational theory in the limit of weak gravitational fields and low
velocities by comparing with the speed of light. This gives rise to a
correspondence principle, as in figure (\ref{correspondence}), where arrows
indicate directions of increased specialization.
\subsection{Discussion}
Special relativity is a physical theory published in 1905 by Albert
Einstein \cite{Wikipedia}. The theory is a special case of Einstein's theory
of relativity where the effects of acceleration and of gravity can be
ignored. Special relativity can be cast into an abstract geometrical form 
based on the use of a flat 4-manifold called Minkowski space. Indeed, most of this
can easily be handled using high school algebra, however this is not
necessarily the most useful way of doing things --- especially when you then
want to generalize to the curved spacetime of the general relativity. Most of
the special relativity, especially the foundations, can be adequately 
treated using matrix algebra \cite{Matt1}.

The theory of special relativity asserts that spacetime is the manifold
$\mathbb{R}^4$ with a flat metric of Lorentz signature defined on it. In
special relativity, the principle of general covariance states that the
spacetime metric, $\eta_{ab}$, is the only quantity relating to spacetime
structure can appear in any physical laws. In addition, the laws of physics
in special relativity satisfy the principle of special covariance with
respect to the proper Poincar\'{e} transformation.

In special relativity, energy is described in terms of the ``time component" 
of 4-vector $p^{a}$. The continuous matter distributions in special relativity
are described by a symmetric tensor $T_{ab}$ called the stress energy
momentum tensor.

In this chapter we introduce the theory of special relativity, which deals 
with the motion of objects as seen from inertial frames. General relativity 
extends this to cover non-inertial frames as well. As the theory of general 
relativity relies on the theory of special relativity, we present the 
concept of special relativity first, so general relativity is easier to 
understand.


\chapter{Coordinate conditions in general relativity}
\label{part_1}
\section{Introduction}
   General relativity is a beautiful physical theory, invented for describing the gravitational field and the
equations it obeys. General relativity is phrased 
in the language of differential geometry. It is by construction
coordinate independent.

General relativity does not have a preferred coordinate system,
and the coordinates must be constructed along with the spacetime.
The mathematics and physics cannot depend on the particular choice
of coordinate system you choose to place on the manifold. 
However, the choice of the coordinate system can be useful in other
ways. It may simplify the mathematical calculation and the physics interpretation of what is going on.

For example, if the physics is time independent and spherically 
symmetric, it is extremely useful to choose a coordinate system 
that respects this property.

Reference ~\cite{D'Inverno} describes several coordinate conditions that we can consider to solve the vacuum field equations

\begin{equation} 
G_{ab}=0
\end{equation}
for $g_{ab}$. At first sight, the problem seems well posed. Indeed,
there are 10 equations for the 10 unknowns $g_{ab}$.
However, the equations are not independent, they are
connected by 4 differential constraints through the contracted 
Bianchi identities

\begin{equation}
\nabla_{b}G^{ab} \equiv 0.
\end{equation}

We therefore seem to have a problem of under-determinacy, when there are fewer equations than unknowns. We cannot expect complete determinancy for any set $g_{ab}$,
since they can be transformed with fourfold freedom by a coordinate transformation:

\begin{equation}
X^{a} \to X^{\prime a} = X^{\prime a}(x) \quad\quad (a = 0,1,2,3).
\end{equation}

In fact, we can use this coordinate freedom to impose four conditions
on the $g_{ab}$. These are known as coordinate conditions or gauge
conditions. We can give an example by introducing
Gaussian normal coordinates in which

\begin{equation}
g_{00} \to +1,\qquad\qquad g_{0i} \to  0 \quad\quad (i = 1, 2, 3).
\end{equation}

The remaining six unknowns $g_{\alpha\beta}$ can be determined
by six independent equations in (1).

The assumption of spherical symmetry of a spacetime means that
each event belongs to a two dimensional submanifold with 
spherical symmetry.
This two dimensional submanifold can be given the metric

\begin{equation}
\d s^2 = R^2(\d\theta^2 + \sin^2\d\phi^2)
\end{equation}
where $\theta$ and $\phi$ are the standard angular coordinates
on the sphere. There must exist (at least locally) two additional 
coordinates $t$ and $r$ labelling the different spheres.

They can be chosen to be orthogonal to the sphere and also mutually 
orthogonal, $t$ to be time-like and $r$ space-like. 
The metric thus reads
\begin{equation}
g_{ik} = \mathrm{diag}\left(-c^2 e^{2\phi},e^{2\Lambda},R^2,R^2\sin^2\theta\right)
\end{equation}
where $\Lambda = \Lambda(r,t)$, $\phi = \phi(r,t)$ and $R = R(r,t)$.
If ${\txt \partial R \over \txt \partial r}$ $\neq$ 0, it is possible to rescale $R$
to be identical with $r$. However, this special choice of the coordinate 
system need not always be useful in particular cases.

In general relativity, we express physics in terms of tensors. General relativity is a geometrical theory. Specifically, we will deal with a Riemannian manifold which defines a metric. We used the metric to define a distance and length (norm) of vectors. We have infinitesimal distance  or interval which is defined as
\begin{equation}
\label{Riemann_1}
(\d s)^2 = \d s^2 = g_{ab}\; \d x^a \d x^b
\end{equation}
where, $\d s$ in ({\ref{Riemann_1}}) is called a ``line element", $g_{ab}$ is called a metric form. In addition, $g_{ab}$ is called ``covariant" metric and it has an inverse of the form $g^{ab}$ which is a ``contravariant" metric.
\section{Independent components: Riemann's argument}
The quick way to see how much we can get away with is to simply count
the number of independent components of the metric tensor. In \emph{n} 
dimensions, since the metric is a symmetric \emph{T}(0,2) tensor, it has 
\emph{n(n+1)/2} independent components. Since there are \emph{n} coordinates, 
we can eliminate up to \emph{n} of the metric components by suitably choosing
our coordinates. This leaves
\[\emph{n(n+1)/2 - n = n(n-1)/2}\]
true ``independent" functions in the metric. This argument is 
due to Riemann.

The number of free components of the Riemann tensor is strictly limited by all symmetries when we deal with a [torsion-free] metric connexion.

\subsection{For $n=1$}
If \emph{n} = 1, then this implies there are zero truly independent components
in the metric (see the proof below): In any 1-dimensional Riemann 
manifold we can always choose coordinates so that the metric is flat.
(A space-time is said to be flat if $R^i{}_{jkl} = 0$ at every point in it.)
\newpage
\begin{proof}
\begin{equation*}
\d s^2 = f(x) \; \d x^2 = g_{xx}(x) \; \d x^2.
\end{equation*}
Let
\begin{equation*}
y = \int\sqrt{f(x)} \; \d x,
\end{equation*}
then
\begin{equation*}
\frac{\d y}{\d x} = \sqrt{f(x)},
\end{equation*}
which implies
\begin{equation*}
\d y = \sqrt{f} \; \d x.
\end{equation*}
So finally
\begin{equation*}
\d y^2 = f \; \d x^2 = \d s^2.
\end{equation*}
\end{proof}
\noindent
i.e. we always can choose a coordinate \emph{y} such that $\d s^2 = \d y^2$ in any 1-dimension 
manifold

\subsection{For $n = 2$}
If \emph{n} = 2, then this Riemann's argument implies there is exactly one truly independent
component in the metric: In any 2-dimensional Riemann manifold it is a 
standard result that \emph{locally} we can always choose coordinates to make
the metric conformally flat.

Indeed, when $n=2$ the Riemann tensor has only one independent component, essentially the Ricci Scalar, $R$.
We have (in any coordinate system)
\begin{eqnarray}
R_{abcd} &=& \frac{1}{2} R \;(g_{ac} \; g_{bd} - g_{ad} \; g_{bc})
\nonumber
\\
R_{ab} &=& \frac{1}{2} R \;g_{ab}
\nonumber
\end{eqnarray}
\begin{description} 
\item[Theorem:] Any 2-dimensional Riemann manifold is locally conformally flat
\end{description}
\[g_{ij} = \exp[2\phi(x)]\;\delta_{ij}\]

\subsection{For $n = 3$}
If \emph{n} = 3, then Riemann's argument implies there are exactly three truly independent 
components in the metric: In any 3-dimensional Riemann manifold it is a 
standard result (typically attributed to Darboux) that \emph{locally} we can always choose coordinates to make 
the metric diagonal:
\[
g_{ij} =
\left[
\begin{array}{ccc}
  g_{11}&0     &0\\
  0     &g_{22}&0\\
  0     &0     &g_{33}
\end{array}
\right]
\]
That is, Riemann 3-manifolds have metric that are always locally diagonalizable.
If $n = 3$, then the Riemann tensor has only six independent components, essentially $R_{ab}$.
Indeed
\begin{equation}
R_{abcd} = -2\{g_{a[d}R_{c]b}+g_{b[c}R_{d]a}\}-R\; g_{a[c}g_{d]b}
\end{equation}
That is
\begin{eqnarray}
R_{abcd} &=& \{g_{ac} \; R_{bd}+g_{bd} \; R_{ac}-g_{ad} \; R_{bc}-g_{bc} \; R_{ad}\}
\nonumber
\\
&&-\frac{1}{2} R\; \{g_{ac} \; g_{bd}-g_{ad} \; g_{bc}\} 
\end{eqnarray}
\subsection{For $n = 4$}
If $n = 4$ Riemann's argument implies there are exactly six truly independent components
in the metric. This means that for a generic manifold in $n = 4$ you \emph{cannot} assert that
\[
g_{ij} =
\left[
\begin{array}{cccc}
  g_{11}&0     &0     &0\\
  0     &g_{22}&0     &0\\
  0     &0     &g_{33}&0\\
  0     &0     &0     &g_{44}
\end{array}
\right]
\]
because this metric has too few free functions (four rather than six). So in general we cannot
in 4 dimensions choose patches to make the metric patchwise diagonal. In local coordinate patches, there are often non-zero off-diagonal elements. For a truly general metric there \emph{must} be at least 2 off diagonal elements. For instance, it might be true that we
can always set:
\[
g_{ij} =
\left[
\begin{array}{cccc}
  g_{11}&g_{12}     &0     &0\\
  g_{12}&g_{22}     &0     &0\\
  0     &0          &g_{33}&g_{34}\\
  0     &0          &g_{34}&g_{44}
\end{array}
\right]
\]
This at least has the correct counting properties, and it would be nice to know if this 
sort of thing is always possible. Another thing that can always be done \emph{locally}
in 4 dimensions is to set:
\[
g_{ij} =
\left[
\begin{array}{cccc}
  1     &0     &0     &0\\
  0     &g_{22}&g_{23}&g_{24}\\
  0     &g_{23}&g_{33}&g_{34}\\
  0     &g_{24}&g_{34}&g_{44}
\end{array}
\right]
\]
This is called a Gaussian normal coordinate patch.

If $n = 4$ the Riemann tensor has only twenty independent components. Ten of them are the 
Ricci tensor $R_{ab}$ and the other ten are hidden in the ``Weyl tensor''.
\begin{equation}
C_{abcd} = R_{abcd}+\{g_{a[d}R_{c]b}+g_{b[c}R_{d]a}\}+\frac{1}{3}Rg_{a[c}g_{d]b}  
\end{equation}
That is
\begin{equation}
R_{abcd} = C_{abcd} + \frac{1}{2}\{g_{ac}R_{bd}+g_{bd}R_{ac}-g_{ad}R_{bc}-g_{bc}R_{ad}\}-\frac{1}{6}R\{g_{ac}g_{bd}-g_{ad}g_{bc}\}  
\end{equation}

\section{Lorentzian manifolds}
The Lorentz metric is used to formulate the free-space Maxwell equations in a 4-dimensional pseudo Riemann space and also used in describing mechanical phenomena in a mass-free Riemann space. The line element and metric are 
then assumed to be
\begin{eqnarray}
\d s^2 &=& - (\d x^0)^2 + (\d x^1)^2 + (\d x^2)^2 + (\d x^3)^2
\nonumber
\\
&=& -c^2 \d t^2+(\d x^2+\d y^2+\d z^2)
\end{eqnarray}
This particular metric tensor will be called the Lorentz metric tensor, and the corresponding 
metric, the Lorentz metric.
\[
g_{ab} =
\left[
\begin{array}{cccc}
  -1& 0& 0& 0\\
   0& 1& 0& 0 \\
   0& 0& 1& 0\\
   0& 0& 0& 1
\end{array}
\right]
\]
The signature of a metric tensor is the number of positive and negative eigenvalues of the metric.
That is, the corresponding real symmetric matrix is diagonalized, and the diagonal entries 
of each sign counted. If the matrix is $n\,\times\,n$, the possible number of positive signs may take any value $p$ from $0$ to $n$. Let $q=n-p$, then the signature may be denoted either by a pair of integers such as $(p,q)$, or as the bracketed sum $(p+q)$, or as a explicit list such as $- + + +$, or even as the number $s = |p -q|$. (Unfortunately all of these different notations are in common use).
A Lorentzian metric is one with signature $(p,1) = (n-1,1)$ or sometimes $(1,q) = (1, n-1)$. For a Lorentzian signature:

\subsection*{(0+1) dimension}
For a Riemannian geometry in (0+1) dimensions,
this is trivial to get Lorentzian geometry
\[g_{11} = -\exp[2\phi(t)]
\]
\subsection*{(1+1) dimensions}
For a Riemannian geometry in (2+0) dimensions we can always locally choose coordinates such that
\[g_{ij} = \exp[2\phi(t,x)] \; \delta_{ij}\]
For Lorentzian geometry in (1+1) dimensions this becomes
\[g_{ij} = \exp[2\phi(t,x)] \; \eta_{ij}\]
where
\[
\eta_{ij}=
\left[
\begin{array}{cc}
  -1&0\\
   0&+1
\end{array}
\right]
\]
\[g_{ij} = \exp[2\phi(t,x)]
\left[
\begin{array}{cc}
  -1&0\\
   0&+1
\end{array}
\right]
\]
so
\[
g_{ij} =
\left[
\begin{array}{cc}
  -\exp[2\phi(t,x)]&0\\
   0             &+\exp[2\phi(t,x)]
\end{array}
\right]
\]
We can find the Ricci scalar $n$ in term of $\phi$, its derivative, $\eta$, the Ricci tensor and Riemann tensor.
The Ricci scalar is

\begin{equation}
R = -  \,\frac{2\left\{-\left({\txt \d ^2 \over \txt \d x^2} \, \phi(t,x)\right) +\left({\txt \d ^2 \over \txt \d t^2} \, \phi(t,x)\right)\right\}}{\txt \exp(2 \phi(t,x))}
\end{equation}
And we can calculate the Ricci tensor, its non-zero components are:


\begin{eqnarray}
R_{11} &=& -\left({\d ^2 \over \d x^2} \, \phi(t,x)\right) + \left({\d ^2 \over \d t^2} \, \phi(t,x)\right)
\\
R_{22} &=& \left({\d ^2 \over \d x^2} \, \phi(t,x)\right) - \left({\d ^2 \over \d t^2} \, \phi(t,x)\right)
\end{eqnarray}
Similarly, we can calculate the Riemann tensor, its non-zero components are:
\begin{equation}
R_{1212} = - \exp(2 \phi(t,x)) \, \left(-\left({\d ^2 \over \d x^2} \, \phi(t,x) \right) + \left({\d ^2 \over \d t^2} \, \phi(t,x) \right)\right)
\end{equation}
In Lorentzian geometry, it is possible to make another coordinate choice by setting
\[
g_{ij} =
\left[
\begin{array}{cc}
  -[1-v^2]&-v\\
  - v     &+1
\end{array}
\right]
\]
where \emph{v = v(t,x)}, so
\[
g_{ij} =
\left[
\begin{array}{cc}
  -[1-v(t,x)^2]&-v(t,x)\\
  -v(t,x)       &+1
\end{array}
\right]
\]
In terms of this coordinate system we have:
The Ricci scalar is
\begin{equation}
R = -2 \left({\partial^2 \over \partial x \, \partial t} \, v(t,x) \right) - 2 \left({\partial \over \partial x} \, v(t,x) \right)^2 - 2 v(t,x) \left({\partial^2 \over \partial x^2} \, v(t,x) \right)
\end{equation}
And we can calculate the Ricci tensor, its non-zero components are:
\begin{equation}
R_{11} = - \left(-1 + v(t,x)\right)^2 \left\{\left({\partial^2 \over \partial x \, \partial t} \, v(t,x)\right) +  \left({\partial \over \partial x} \, v(t,x)\right)^2 + v(t,x) \left({\partial^2 \over \partial x^2} \, v(t,x)\right)\right\}
\end{equation}
\begin{equation}
R_{12} = v(t,x) \left\{\left({\partial^2 \over \partial x \, \partial t} \, v(t,x)\right) + \left({\partial \over \partial x} \, v(t,x)\right)^2 + v(t,x) \left({\partial^2 \over \partial x^2} \, v(t,x)\right)\right\}
\end{equation}
\begin{equation}
R_{22} = - \left({\partial^2 \over \partial x \, \partial t} \, v(t,x)\right) - \left({\partial \over \partial x} \, v(t,x)\right)^2 - v(t,x) \left({\partial^2 \over \partial x^2} \, v(t,x)\right)
\end{equation}
Similary, we can calculate the Riemann tensor, its non-zero components are:
\begin{equation}
R_{1212} = - \left({\partial^2 \over \partial x \, \partial t} \, v(t,x)\right) - \left({\partial \over \partial x} \,v(t,x) \right)^2 - v(t,x) \left({\partial^2 \over \partial x^2} \, v(t,x)\right)
\end{equation}
\subsection*{(2+1) dimensions}
The simplest special coordinate patch that it is always possible to set up yields
\[
g_{ij} =
\left[
\begin{array}{ccc}
  -{h_{0}}^2&0                &0\\
   0                &{h_{1}}^2&0\\
   0                &0                &{h_{2}}^2
\end{array}
\right],
\]
where $h_{i} = h_{i}(t,x,y)$, so
\[
g_{ij} =
\left[
\begin{array}{ccc}
  -h_{0}(t,x,y)^2&0                &0\\
   0                &h_{1}(t,x,y)^2&0\\
   0                &0                &h_{2}(t,x,y)^2
\end{array}
\right].
\]
Alternatively we can always locally set up patches of the form
\[
g_{ij} =
\left[
\begin{array}{ccc}
  -[N^2-v^2]&v_{1}&v_{2}\\
  v_{1}     &1    &0\\
  v_{2}     &0    &1
\end{array}
\right]
\]
where 
$v_{i} = v_{i}(t,x,y),N = N(t,x,y),v^2 = v_{1}^2+v_{2}^2.$
\\
So the metric is
\[
g_{ij} =
\left[
\begin{array}{ccc}
  -[N(t,x,y)^2-(v_{1}^2(t,x,y)+v_{2}^2(t,x,y))]&v_{1}(t,x,y)&v_{2}(t,x,y)\\
  v_{1}(t,x,y)                           &1           &0\\
  v_{2}(t,x,y)                           &0           &1
\end{array}
\right]
\]
Coordinate patches of this type are called Painleve--Gullstrand coordinates.
Another alternative, in (2+1) dimensions, is that we can always locally set up patches of the form
\[
g_{ij} =
\left[
\begin{array}{ccc}
  -[1-v^2]&v_{1}           &v_{2}\\
  v_{1}     &\exp[2\phi]    &0\\
  v_{2}     &0             &\exp[2\phi]
\end{array}
\right]
\]
where $\phi = \phi(t,x,y), v^2 = \exp[-2\phi] \; (v_{1}^2+v_{2}^2)$.
\\
The inverse metric is then:
\[
g^{ij} =
\left[
\begin{array}{ccc}
  -1&v_{1} \exp[-2\phi]   &v_{2} \exp[-2\phi]\\
  v_{1} \exp[-2\phi]     &(\exp[2 \phi] - v_{1}^2)\;\exp[-4\phi]  & -v_{1}v_{2}\;\exp[-4\phi]\\
  v_{2} \exp[-2\phi]     &-v_{1} v_{2} \exp[-4\phi] & (\exp[2\phi]-v_{2}^2) \exp[-4\phi]
\end{array}
\right]
\]
In (2+1) dimensions we can calculate the Ricci scalar, the Ricci tensor and
the Riemann tensor. However, all results are too complex to analyze in detail.

\subsection*{(3+1) dimensions}
This is the physically interesting case, and this is the case that will be most difficult to analyze in detail, and the remainder of the Chapter is devoted to this case.

\section{The block diagonal ansatz}
\begin{itemize}
\item
It is certainly true that we can always locally set:
\[
g_{\mu\nu} = 
\left[\begin{array}{cc|cc}
g_{11}&g_{12}&0&0\\
g_{12}&g_{22}&0&0\\
\hline
0&0&g_{33}&g_{34}\\
0&0&g_{34}&g_{44}
\end{array}
\right]
\]
\item 
This at least has the correct counting properties.
\end{itemize} 
Now we can take the above as an \emph{ansatz} and calculate the Ricci tensor, and then the Einstein tensor, respectively.
\\
Specifically, We write the four coordinates as $x^\mu=(y^a,z^i)$,
with
\[
x^1=y^1; \quad x^2=y^2; \qquad\qquad x^3 = z^1; \quad x^4 = z^2.
\]
Then
\[
g_{\mu\nu}(x) = g_{\mu\nu}(y,z) =
\left[
\begin{array}{cc|cc}
g_{11}(y,z)&g_{12}(y,z)&0&0\\
g_{12}(y,z)&g_{22}(y,z)&0&0\\
\hline
0&0&g_{33}(y,z)&g_{34}(y,z)\\
0&0&g_{34}(y,z)&g_{44}(y,z)
\end{array}
\right]
\]
That is
\[
 g_{\mu\nu}(y,z) =
\left[
\begin{array}{c|c}
h_{ab}(y,z)&0\\
\hline
0&f_{ij}(y,z)
\end{array}
\right]
\]
This means the 4 dimensional metric has split up into four $2\times2$
blocks, two of them trivial (zero). This will make the Ricci tensor relatively easy to calculate.

The inverse tensor is very simple
\[
g^{\mu\nu} = 
\left[
\begin{array}{c|c}
h^{ab}(y,z)&0\\
\hline
0&f^{ij}(y,z)
\end{array}
\right]
\]
where $h^{ab}$ and $f^{ij}$ are two dimensional inverses of the two
$2\times2$ matrices $h_{ab}$ and $f_{ij}$. 

For definiteness, we assume the $h$ subspace has signature (1+1) [one
time direction plus one space direction] while the $f$ subspace has
signature (2+0) [two space directions]. But this subtle point does not
really affect the subsequent analysis.

The Christoffel symbols can be computed from
\begin{equation}
\label{Christoffel}
\Gamma^\mu{}_{\sigma\rho}(y,z) =
g^{\mu\nu} \left\{
g_{\nu(\sigma,\rho)} - {1\over2} g_{\sigma\rho,\nu}
\right\}
\end{equation}
and now we can spilt them up into little $2\times2$ blocks
\[
\Gamma^\mu{}_{\sigma\rho}(y,z) = 
\left[
\begin{array}{cc|cc}
\Gamma^\mu{}_{11}(y,z)&\Gamma^\mu{}_{12}(y,z)&\Gamma^\mu{}_{13}(y,z)&\Gamma^\mu{}_{14}(y,z)\\
\Gamma^\mu{}_{12}(y,z)&\Gamma^\mu{}_{22}(y,z)&\Gamma^\mu{}_{23}(y,z)&\Gamma^\mu{}_{24}(y,z)\\
\hline
\Gamma^\mu{}_{13}(y,z)&\Gamma^\mu{}_{23}(y,z)&\Gamma^\mu{}_{33}(y,z)&\Gamma^\mu{}_{34}(y,z)\\
\Gamma^\mu{}_{14}(y,z)&\Gamma^\mu{}_{24}(y,z)&\Gamma^\mu{}_{34}(y,z)&\Gamma^\mu{}_{44}(y,z)
\end{array}
\right]
\]
Some of these blocks should have a simple interpretation in terms of the
Christoffel symbols of the 2 dimensional metrics $h_{ab}$ and $f_{ij}$.

For example
\begin{eqnarray}
\Gamma^a{}_{bc}(y,z) 
&=&
g^{a\nu} \left\{ g_{\nu(b,c)}   -{1\over2} g_{bc,\nu} \right\}
= 
g^{ad} \left\{ g_{d(b,c)}   -{1\over2} g_{bc,d} \right\}
\nonumber
\\
&=& h^{ad} \left\{ h_{d(b,c)}   -{1\over2} h_{bc,d} \right\}
= [\Gamma(h)]^a{}_{bc}(y,z).
\end{eqnarray}
That is, this portion of the 4 dimensional Christoffel symbol can be
calculated simply by looking at the 2 dimensional Christoffel symbol
corresponding to the two dimensional metric $h_{ab}$. Similarly, we can calculate this:
\begin{eqnarray}
\Gamma^i{}_{jk}(y,z) 
&=& g^{i\nu} \left\{ g_{\nu(j,k)}   -{1\over2} g_{jk,\nu} \right\}
= 
g^{il} \left\{ g_{l(j,k)}   -{1\over2} g_{jk,l} \right\}
\nonumber
\\
&=& h^{il} \left\{ h_{l(j,k)}   -{1\over2} h_{jk,l} \right\}
=[\Gamma(f)]^i{}_{jk}(y,z).
\end{eqnarray}
The complications of course come with the other pieces, such as
\begin{eqnarray}
\Gamma^i{}_{bc}(y,z) &=&
g^{i\nu} \left\{ g_{\nu(b,c)} -{1\over2} g_{bc,\nu} \right\}
= 
f^{ij} \left\{ g_{j(b,c)}   -{1\over2} g_{bc,j} \right\}
\nonumber
\\
&=& f^{ij} \left\{ 0   -{1\over2} h_{bc,j} \right\}
= 
-{1\over2}  f^{ij}  h_{bc,j}.
\end{eqnarray}
Similarly, and we can calculate this:
\begin{eqnarray}
\Gamma^a{}_{jk}(y,z) &=&
g^{a\nu} \left\{ g_{\nu(j,k)} -{1\over2} g_{jk,\nu} \right\}
= h^{ab} \left\{ g_{b(j,k)}   -{1\over2} g_{jk,b} \right\}
\nonumber
\\
&=& h^{ab} \left\{ 0   -{1\over2} f_{jk,b} \right\}
= -{1\over2}  h^{ab}  f_{jk,b}.
\end{eqnarray}
This part of the 4-dimensional Christoffel symbol is rather simple but
does not have a simple interpretation in terms of the 2 dimensional
sub-geometries.

There are four other sub-blocks to calculate:
\begin{eqnarray}
\Gamma^a{}_{bk}(y,z) &=&
g^{a\nu} \left( g_{\nu(b,k)}-{1\over2} g_{bk,\nu}\right)
=
g^{ac} \left( g_{c(b,k)}-{1\over2} g_{bk,c}\right)
\\
&=& g^{ac} \left({1\over2} g_{cb,k} + {1\over2} g_{ck,b} 
-{1\over2} g_{bk,c}\right)
= h^{ac} \left({1\over2} h_{cb,k} \right) = {1\over2}  h^{ac} h_{cb,k}
\nonumber
\end{eqnarray}
Whence
\begin{equation}
\Gamma^a{}_{jc}(y,z)  ={1\over2}  h^{ab} h_{bc,j}
\end{equation}

Similarly
\begin{eqnarray}
\Gamma^i{}_{jc}(y,z) &=&
g^{i\nu} \left( g_{\nu(j,c)}-{1\over2} g_{jc,\nu}\right)
=
g^{ik} \left( g_{k(j,c)}-{1\over2} g_{jc,k}\right)
\\
&=& g^{ik} \left({1\over2} g_{kj,c} + {1\over2} g_{kc,j} 
-{1\over2} g_{jc,k}\right) = f^{ik} \left({1\over2} f_{kj,c} \right) = {1\over2}  f^{ik} \; f_{jk,c}
\nonumber
\end{eqnarray}
Whence
\begin{equation}
\Gamma^i{}_{bk}(y,z)  = {1\over2}  f^{ij} \; f_{jk,b}
\end{equation}

Various symmetries should make all four of these results obvious once
any \emph{one} of them has been calculated.
   \subsection{Collected results for the connexion}
\label{sub-block}
There are a total of 8 sub-blocks to deal with:
\begin{eqnarray}
\Gamma^a{}_{bc}(y,z) &=&  [\Gamma(h)]^a{}_{bc}(y,z).
\nonumber
\\
\Gamma^i{}_{jk}(y,z) &=& [\Gamma(f)]^i{}_{jk}(y,z).
\nonumber
\\
\Gamma^i{}_{bc}(y,z) &=&  -{1\over2}  f^{ij} \; h_{bc,j}.
\nonumber
\\
\Gamma^a{}_{jk}(y,z) &=& -{1\over2}  h^{ab}  \;f_{jk,b}.
\nonumber
\\
\Gamma^a{}_{bk}(y,z) &=& {1\over2}  h^{ac} \;h_{cb,k}.
\nonumber
\\
\Gamma^a{}_{jc}(y,z) &=& {1\over2}  h^{ab} \;h_{bc,j}.
\nonumber
\\
\Gamma^i{}_{jc}(y,z) &=&  {1\over2}  f^{ik} \;f_{jk,c}.
\nonumber
\\
\Gamma^i{}_{bk}(y,z) &=& {1\over2}  f^{ij} \;f_{jk,b}.
\nonumber
\end{eqnarray}

   \subsection{Back to the Ricci tensor: qualitative results}
   

Once we have all eight sub-blocks, we have the entire 4 dimensional
Christoffel symbol.

The Ricci tensor of $g_{\mu\nu}$ can then be calculated from \cite{Misner}, page 224
\begin{equation}
\label{Ricci_Misner}
R_{\mu\nu} =
\Gamma^\sigma{}_{\mu\nu,\sigma}
 -
\Gamma^\sigma{}_{\mu\sigma,\nu} 
+
\Gamma^\sigma{}_{\rho\sigma} \;\Gamma^\rho{}_{\mu\nu} 
-
\Gamma^\rho{}_{\sigma\mu}\;\Gamma^\sigma{}_{\rho\nu}.
\end{equation}
\begin{itemize}
\item
It is convenient to first calculate
$\Gamma^\sigma{}_{\rho\sigma}$
\end{itemize}
In fact, it is a standard result that
\begin{equation}
\Gamma^\sigma{}_{\rho\sigma} = \left(\ln \sqrt{-g}\right)_{,\rho} 
= {1\over2} {g_{,a}\over g}
\end{equation}
(see, for instance, equation (8.51a) on page 222 of \cite{Misner}).
Then we can simplify:
\begin{equation}
R_{\mu\nu} =  {1\over\sqrt{-g}} \;
\left( \sqrt{-g} \;\Gamma^\sigma{}_{\mu\nu}\right)_{,\sigma}
- \left(\ln \sqrt{-g} \right)_{,\mu\nu}
- \Gamma^\rho{}_{\sigma\mu}\;\Gamma^\sigma{}_{\rho\nu},
\end{equation}
(see, for instance, equation (8.51b) on page 222 of \cite{Misner}).
This is the explicit formula for the Ricci tensor.

To interpret the Ricci tensor, write it in $2\times2$ sub-blocks
\[
R_{\mu\nu}(y,z) = 
\left[
\begin{array}{cc|cc}
R_{11}(y,z)&R_{12}(y,z)&R_{13}(y,z)&R_{14}(y,z)\\
R_{12}(y,z)&R_{22}(y,z)&R_{23}(y,z)&R_{24}(y,z)\\
\hline
R_{13}(y,z)&R_{23}(y,z)&R_{33}(y,z)&R_{34}(y,z)\\
R_{14}(y,z)&R_{24}(y,z)&R_{34}(y,z)&R_{44}(y,z)
\end{array}
\right]
\]
That is
\[
 R_{\mu\nu}(y,z) =
\left[
\begin{array}{c|c}
R_{ab}(y,z)&R_{ai}(y,z)\\
\hline
R_{jb}(y,z)&R_{ij}(y,z)
\end{array}
\right]
\]

Some of these blocks should have a simple interpretation in terms of the
Ricci tensor of the 2 dimensional metrics $h_{ab}$ and $f_{ij}$.
For instance we expect
\[
[R(g)]_{ab} = [R(h)]_{ab} + ???
\]
\[
[R(g)]_{ij} = [R(f)]_{ij} + ???
\]
where the extra terms (the ???) involve $y$ derivatives of $f$ and $z$
derivatives of $h$. That is, we expect the full 4 dimensional Ricci
tensor of these particular $2\times 2$ sub-blocks to be related to the
Ricci tensors of the 2 dimensional metrics $f$ and $h$, but with some
additional terms.

For the off-diagonal block
\[
[R(g)]_{ia} = ???
\]
It will probably involve $y$ derivatives of $f$ and $z$ derivatives of $h$.

Once you have the Ricci tensor, the Ricci scalar is easy
\[
R(g) = g^{\mu\nu} R_{\mu\nu} = h^{ab} R_{ab} + f^{ij} R_{ij} 
= R(h) + ??? + R(f) + ???
\]
That is
\[
R(g) = R(h) + R(f) + ???
\]
That is, we expect the full 4 dimensional Ricci scalar to be related to
the sum of the Ricci tensors of the 2 dimensional metrics $f$ and $h$,
but with some additional terms.

Then for the four dimensional Einstein tensor
\begin{equation}
\label{Einstein_form}
G_{\mu\nu} = R_{\mu\nu} - {1\over2} \; R \; g_{\mu\nu}
\end{equation}
we can sub-divide the $2\times2$ sub blocks as
\[
[G(g)]_{ab} = [G(h)]_{ab}  - {1\over2} \; R(f) \; h_{ab} + ???
\]
\[
[G(g)]_{ij} = [G(f)]_{ij}  - {1\over2} \; R(g) \; f_{ij} + ???
\]
But the Einstein tensor of any two dimensional manifold is zero; this
is one of the special features of two dimensions, so
\[
[G(h)]_{ab} = 0 = [G(f)]_{ij} 
\]
and we have
\[
[G(g)]_{ab} =  - {1\over2} \; R(f) \; h_{ab} + ???
\]
\[
[G(g)]_{ij} = - {1\over2} \; R(g) \; f_{ij} + ???
\]
Finally, for the off-diagonal
\[
[G(g)]_{aj} = [R(g)]_{aj} = ???
\]
Of course it is all these ??? pieces that give the important
mathematics and physics. Again, we are not really sure what to expect,
they will probably involve $y$ derivatives of $f$ and $z$ derivatives
of $h$.

   \subsection{Calculating the Ricci tensor - diagonal pieces}
\def\tr{\mathrm{tr}}
Start from the general expression for the Ricci tensor of the metric
$g_{\mu\nu}$:
\begin{equation}
R_{\mu\nu} =  {1\over\sqrt{-g}} \;
\left( \sqrt{-g} \;\Gamma^\sigma{}_{\mu\nu}\right)_{,\sigma}
- \left(\ln \sqrt{-g} \right)_{,\mu\nu}
- \Gamma^\rho{}_{\sigma\mu}\;\Gamma^\sigma{}_{\rho\nu}.
\end{equation}
Note that, since the metric is block diagonal,
\begin{equation}
g = \det[g_{\mu\nu}] = \det[h_{ab}]\;\det[f_{ij}] = h\; f.
\end{equation}
Now consider
\begin{equation}
[R(g)]_{ab} =  {1\over\sqrt{-g}} \;
\left( \sqrt{-g} \;\Gamma^\sigma{}_{ab}\right)_{,\sigma}
- \left(\ln \sqrt{-g} \right)_{,ab}
- \Gamma^\rho{}_{\sigma a}\;\Gamma^\sigma{}_{\rho b}
\end{equation}
which we expand as
\begin{eqnarray}
[R(g)]_{ab} &=& 
{1\over\sqrt{-h\;f}} \;
\left( \sqrt{-h\;f} \;\Gamma^c{}_{ab}\right)_{,c}
+
{1\over\sqrt{-h\;f}} \;
\left( \sqrt{-h\;f} \;\Gamma^k{}_{ab}\right)_{,k}
- \left(\ln \sqrt{-h\;f} \right)_{,ab}
\nonumber
\\
&&- \Gamma^c{}_{d a}\;\Gamma^d{}_{c b}
- \Gamma^c{}_{k a}\;\Gamma^k{}_{c b}
- \Gamma^j{}_{k a}\;\Gamma^k{}_{j b}
- \Gamma^j{}_{d a}\;\Gamma^d{}_{j b}
\nonumber
\end{eqnarray}
Then
\begin{eqnarray}
[R(g)]_{ab} &=&
{1\over\sqrt{-h}} \;
\left( \sqrt{-h} \;\Gamma^c{}_{ab}\right)_{,c} +
{1\over2} {f_{,c}\over f} \;\Gamma^c{}_{ab}
+
{1\over\sqrt{-h\;f}} \;
\left( \sqrt{-h\;f} \;\Gamma^k{}_{ab}\right)_{,k}
\nonumber
\\
&&- \left(\ln \sqrt{-h} \right)_{,ab}
- \left(\ln \sqrt{f} \right)_{,ab}
\nonumber
\\
&&- \Gamma^c{}_{d a}\;\Gamma^d{}_{c b}
- \Gamma^c{}_{k a}\;\Gamma^k{}_{c b}
- \Gamma^j{}_{k a}\;\Gamma^k{}_{j b}
- \Gamma^j{}_{d a}\;\Gamma^d{}_{j b}
\nonumber
\end{eqnarray}
which we regroup as
\begin{eqnarray}
[R(g)]_{ab} &=&
{1\over\sqrt{-h}} \;
\left( \sqrt{-h} \;\Gamma^c{}_{ab}\right)_{,c} 
- \left(\ln \sqrt{-h} \right)_{,ab}
- \Gamma^c{}_{d a}\;\Gamma^d{}_{c b}
\nonumber
\\
&&+
{1\over2} {f_{,c}\over f} \;\Gamma^c{}_{ab}
+
{1\over\sqrt{-h\;f}} \;
\left( \sqrt{-h\;f} \;\Gamma^k{}_{ab}\right)_{,k}
- \left(\ln \sqrt{f} \right)_{,ab}
\nonumber
\\
&&- \Gamma^c{}_{k a}\;\Gamma^k{}_{c b}
- \Gamma^j{}_{k a}\;\Gamma^k{}_{j b}
- \Gamma^j{}_{d a}\;\Gamma^d{}_{j b}
\end{eqnarray}
But the first line is just $[R(h)]_{ab}$, the 2-dimensional Ricci tensor
calculated using the 2-metric $h_{ab}$. The other terms can be
slightly rearranged as
\begin{eqnarray}
[R(g)]_{ab} &=& [R(h)]_{ab} 
- {1\over2} \left\{  
\left(\ln {f} \right)_{,ab} -  {f_{,c}\over f} \;\Gamma^c{}_{ab}
\right\}
+
{1\over\sqrt{-h\;f}} \;
\left( \sqrt{-h\;f} \;\Gamma^k{}_{ab}\right)_{,k}
\nonumber
\\
&&- \Gamma^c{}_{k a}\;\Gamma^k{}_{c b}
- \Gamma^j{}_{k a}\;\Gamma^k{}_{j b}
- \Gamma^j{}_{d a}\;\Gamma^d{}_{j b}
\end{eqnarray}
Now use colons $:$ to denote 2-dimensional covariant derivatives using
the 2-metric $h$. Then
\begin{eqnarray}
[R(g)]_{ab} &=& [R(h)]_{ab} 
- {1\over2} \left(\ln {f} \right)_{:ab} 
+
{1\over2} {h_{,k}\over h} \; \Gamma^k{}_{ab}
+
{1\over\sqrt{f}} \;\left( \sqrt{f} \;\Gamma^k{}_{ab}\right)_{,k}
\nonumber
\\
&&- \Gamma^c{}_{k a}\;\Gamma^k{}_{c b}
- \Gamma^j{}_{k a}\;\Gamma^k{}_{j b}
- \Gamma^j{}_{d a}\;\Gamma^d{}_{j b}
\end{eqnarray}
Now we start to insert the various pieces.
\\
Note that
\[
{1\over\sqrt{f}} \;\left( \sqrt{f} \;\Gamma^k{}_{ab}\right)_{,k}
= - {1\over2}  
{1\over\sqrt{f}} \;\left( \sqrt{f} \; f^{kj} \; \partial_j h_{ab}\right)_{,k}
=
 - {1\over2}  
{1\over\sqrt{f}} \;\left( \sqrt{f} \; f^{ij} \; \partial_i h_{ab}\right)_{,j}
\]
We can write this more compactly as
\begin{equation}
{1\over\sqrt{f}} \;\left( \sqrt{f} \;\Gamma^k{}_{ab}\right)_{,k}
=
 - {1\over2}  \; \Delta_f\, h_{ab}
\end{equation}
where by $\Delta_f$ we mean the two-dimensional Laplacian in the $f$
subspace, treating the $h$ indices as though they were scalars.
\\
Also note that
\begin{equation}
\Gamma^j{}_{k a}\;\Gamma^k{}_{j b}
= {1\over 4} f^{jm} \;f_{mk,a} \;f^{kn} \;f_{nj,b}
\end{equation}
which we can re-write as
\begin{eqnarray}
\Gamma^j{}_{k a}\;\Gamma^k{}_{j b} &=& {1\over 4} f^{ij} \;f_{jk,a} \;f^{kl} \; f_{li,b}
\nonumber
\\
&=& {1\over 4} \tr \left[ 
{\bf f}^{-1} \; {\bf f}_{,a}  \; {\bf f}^{-1} \; {\bf f}_{,b} 
\right]
\end{eqnarray}
Next, we start looking at
\begin{eqnarray}
\Gamma^c{}_{k a}\;\Gamma^k{}_{c b} &=&
-{1\over 4} ( h^{cd} \; h_{da,k}) \; ( f^{kj} \; h_{bc,j})
\nonumber
\\
&=&
-{1\over 4} h^{cd} \; (  h_{ad,i} \; f^{ij} \; h_{bc,j})
\end{eqnarray}
Similarly,
\begin{equation}
\Gamma^j{}_{d a}\;\Gamma^d{}_{j b}
=
-{1\over 4} h^{cd} \; (  h_{ad,i} \;  f^{ij} \; h_{bc,j})
\end{equation}
Finally
\begin{equation}
{h_{,k}\over h} \; \Gamma^k{}_{ab} = 
-{1\over2} {h_{,i}\over h}\; f^{ij} h_{ab,j}
\end{equation}
Now pull all these pieces together. We get
\begin{eqnarray}
[R(g)]_{ab} &=& [R(h)]_{ab} 
- {1\over2} \left(\ln {f} \right)_{:ab}   - {1\over2}  \; \Delta_f\, h_{ab}
-{1\over 4} \tr \left[ 
{\bf f}^{-1} \; {\bf f}_{,a}  \; {\bf f}^{-1} \; {\bf f}_{,b} 
\right]
\nonumber
\\
&&-{1\over4}  {h_{,i}\over h}\; f^{ij} h_{ab,j}
+ {1\over 2} h^{cd} \; (  h_{ad,i} \;  f^{ij} \; h_{bc,j})
\end{eqnarray}
There are still simplifications we could make. For instance
\begin{equation}
\left(\ln {f} \right)_{:ab} = \left( {f_{:a}\over f}\right)_{:b} 
= {f_{:ab}\over f} - {f_{:a} \over f} \; {f_{:b}\over f}
\end{equation}
Using this we get rid of the logarithm. Then finally
\begin{eqnarray}
[R(g)]_{ab} &=& [R(h)]_{ab} 
- {1\over2} {f_{:ab}\over f} 
+{1\over2}  {f_{:a} \over f} \; {f_{:b}\over f}
- {1\over2}  \; \Delta_f\, h_{ab}
-{1\over 4} \tr \left[
{\bf f}^{-1} \; {\bf f}_{,a}  \; {\bf f}^{-1} \; {\bf f}_{,b} 
\right]
\nonumber
\\
&&-{1\over4}  {h_{,i}\over h}\; f^{ij} h_{ab,j}
+ {1\over 2} h^{cd} \; (  h_{ad,i} \;  f^{ij} \; h_{bc,j})
\end{eqnarray}
For $[R(g)]_{ij}$ we can appeal to symmetry to simply write:
\begin{eqnarray}
[R(g)]_{ij} &=& [R(f)]_{ij} 
- {1\over2} {h_{|ij}\over h} 
+{1\over2}  {h_{|i} \over h} \; {h_{|j}\over h}
- {1\over2}  \; \Delta_h\, f_{ij}
-{1\over 4} \tr \left[ 
{\bf h}^{-1} \; {\bf h}_{,i}  \; {\bf h}^{-1} \; {\bf h}_{,j} 
\right]
\nonumber
\\
&&-{1\over4}  {f_{,a}\over f}\; h^{ab} f_{ij,a}
+ {1\over 2} f^{kl} \; (  f_{ik,a} \;  h^{ab} \; f_{jl,b})
\end{eqnarray}
The vertical bar $|$ now denotes 2-dimensional covariant derivatives in
the $f$ direction, treating the $h$ indices as scalars.
\\
To obtain the Ricci scalar we note
\begin{eqnarray}
h^{ab} \; [R(g)]_{ab} &=& h^{ab} \; [R(h)]_{ab} 
- {1\over2} {\Delta_h\,f\over f} 
+{1\over2}  {f_{:a} \over f} \; {f^{:a}\over f}
- {1\over2}   h^{ab} \; \Delta_f\, h_{ab}
-{1\over 4}  h^{ab} \; \tr \left[ 
{\bf f}^{-1} \; {\bf f}_{,a}  \; {\bf f}^{-1} \; {\bf f}_{,b} 
\right]
\nonumber
\\
&&-{1\over4}  {h_{|i}\over h}\;  {h^{|i}\over h}
+ {1\over 2}  f^{ij} \;
\tr \left[ 
{\bf h}^{-1} \; {\bf h}_{,i}  \; {\bf h}^{-1} \; {\bf h}_{,j} 
\right]
\end{eqnarray}
with a similar formula holding for $h^{ij}\;[R(g)]_{ij}$:
\begin{eqnarray}
f^{ij}\;[R(g)]_{ij} &=& f^{ij}\; [R(f)]_{ij} 
- {1\over2} {\Delta_f\,h\over h} 
+{1\over2}  {h_{|i} \over h} \; {h^{|i}\over h}
- {1\over2}  f^{ij} \; \Delta_h\, f_{ij}
-{1\over 4}  f^{ij} \;   \tr \left[ 
{\bf h}^{-1} \; {\bf h}_{,i}  \; {\bf h}^{-1} \; {\bf h}_{,j} 
\right]
\nonumber
\\
&&-{1\over4}  {f_{|a}\over f}\;  {f^{|a}\over f}
+ {1\over 2}  h^{ab} \; \tr \left[ 
{\bf f}^{-1} \; {\bf f}_{,a}  \; {\bf f}^{-1} \; {\bf f}_{,b} 
\right]
\end{eqnarray}
Then in the Ricci scalar there are a lot of partial cancellations
\begin{eqnarray}
R(g) &=&  h^{ab} \; [R(g)]_{ab}  + f^{ij}\;[R(g)]_{ij} 
\nonumber
\\
&=& R(h) + R(g) 
- {1\over2} {\Delta_h\,f\over f} - {1\over2} {\Delta_f\,h\over h} 
+{1\over4}  {f_{:a} \over f} \; {f^{:a}\over f}
+{1\over4}  {h_{|i} \over h} \; {h^{|i}\over h}
\nonumber
\\
&&- {1\over2}   h^{ab} \; \Delta_f\, h_{ab}
- {1\over2}  f^{ij} \; \Delta_h\, f_{ij}
\nonumber
\\
&&+{1\over 4}  h^{ab} \; \tr \left[ 
{\bf f}^{-1} \; {\bf f}_{,a}  \; {\bf f}^{-1} \; {\bf f}_{,b} 
\right]
+ {1\over 4}  f^{ij} \;
\tr \left[ 
{\bf h}^{-1} \; {\bf h}_{,i}  \; {\bf h}^{-1} \; {\bf h}_{,j} 
\right]
\end{eqnarray}
   \subsection{Off-diagonal part of the Ricci tensor}
   Now we consider what we can say about the off-diagonal piece 
$[R(g)]_{ai}$. Again we will want to start from the general formula
for the Ricci tensor of the metric $g_{\mu\nu}$:
\[
R_{\mu\nu} =  {1\over\sqrt{-g}} \;
\left( \sqrt{-g} \;\Gamma^\sigma{}_{\mu\nu}\right)_{,\sigma}
- \left(\ln \sqrt{-g} \right)_{,\mu\nu}
- \Gamma^\rho{}_{\sigma\mu}\;\Gamma^\sigma{}_{\rho\nu}.
\]
Now consider
\begin{equation}
[R(g)]_{ai} =  {1\over\sqrt{-g}} \;
\left( \sqrt{-g} \;\Gamma^\sigma{}_{ai}\right)_{,\sigma}
- \left(\ln \sqrt{-g} \right)_{,ai}
- \Gamma^\rho{}_{\sigma a}\;\Gamma^\sigma{}_{\rho i}
\end{equation}
which we expand as
\begin{eqnarray}
[R(g)]_{ai} &=& {1\over\sqrt{-h\;f}} \;
\left( \sqrt{-h\;f} \;\Gamma^c{}_{ai}\right)_{,c}
+
{1\over\sqrt{-h\;f}} \;
\left( \sqrt{-h\;f} \;\Gamma^k{}_{ab}\right)_{,k}
- \left(\ln \sqrt{-h\;f} \right)_{,ai}
\nonumber
\\
&&- \Gamma^c{}_{d a}\;\Gamma^d{}_{c i}
- \Gamma^c{}_{k a}\;\Gamma^k{}_{c i}
- \Gamma^j{}_{k a}\;\Gamma^k{}_{j i}
- \Gamma^j{}_{d a}\;\Gamma^d{}_{j i}
\nonumber
\end{eqnarray}
That is
\begin{eqnarray}
[R(g)]_{ai} &=&
{1\over\sqrt{-h\;f}} \;
\left( \sqrt{-h\;f} \;\Gamma^c{}_{ai}\right)_{,c}
+
{1\over\sqrt{-h\;f}} \;
\left( \sqrt{-h\;f} \;\Gamma^k{}_{ab}\right)_{,k}
\nonumber
\\
&&-{1\over2}\left(\ln({-h}) \right)_{,ai}
 -{1\over2}\left(\ln(f) \right)_{,ai} 
\nonumber
\\
&&- \Gamma^c{}_{d a}\;\Gamma^d{}_{c i}
- \Gamma^c{}_{k a}\;\Gamma^k{}_{c i}
- \Gamma^j{}_{k a}\;\Gamma^k{}_{j i}
- \Gamma^j{}_{d a}\;\Gamma^d{}_{j i}
\end{eqnarray}
So that
\begin{eqnarray}
[R(g)]_{ai} &=&
- {1\over2}\left(\ln(f) \right)_{,ai}
+ {1\over2}\left(\ln(f) \right)_{,b} \Gamma^b{}_{ai}
+
{1\over\sqrt{-h}} \;
\left( \sqrt{-h} \;\Gamma^b{}_{ai}\right)_{,b}
\nonumber
\\
&&-{1\over2}\left(\ln({-h}) \right)_{,ai}
+ {1\over2}\left(\ln({-h}) \right)_{,j} \Gamma^j{}_{ai}
+ {1\over\sqrt{f}} \;
\left( \sqrt{f} \;\Gamma^j{}_{ay}\right)_{,j}
\nonumber
\\
&&- \Gamma^c{}_{d a}\;\Gamma^d{}_{c i}
- \Gamma^c{}_{k a}\;\Gamma^k{}_{c i}
- \Gamma^j{}_{k a}\;\Gamma^k{}_{j i}
- \Gamma^j{}_{d a}\;\Gamma^d{}_{j i}
\end{eqnarray}
Now consider
\begin{equation}
\Gamma^b{}_{ai:b} =
{1\over\sqrt{-h}} \; \partial_b\left( \sqrt{-h}\; \Gamma^b{}_{ai} \right)
- \Gamma^c{}_{ab} \;\left( \Gamma^b{}_{ci} \right)
\end{equation}
Here $ \Gamma^b{}_{ai}$ is treated as though it were a $T^1_1$ tensor
with respect to the $h$ subspace, the $i$ index [which lies in the $f$
subspace] is treated as though this quantity were a scalar with
respect to coordinate transformations in the $f$ subspace. This
notation, together with the analogous
\begin{equation}
\Gamma^j{}_{ia|j} = 
{1\over\sqrt{f}} \; \partial_j\left( \sqrt{f}\; \Gamma^j{}_{ia} \right)
- \Gamma^{k}_{ij} \;\left( \Gamma^j{}_{ka} \right)
\end{equation}
allows us to simplify the off-diagonal Ricci tensor as:
\begin{eqnarray}
[R(g)]_{ai} &=&
\Gamma^b{}_{ai:b} + \Gamma^j{}_{ia|j} 
- {1\over2}\left(f_{,a}/f\right)_{,i}
- {1\over2}\left(h_{,i}/h\right)_{,a}
\nonumber
\\
&&+ {1\over2} \left(f_{,b}/f\right) \; \Gamma^b{}_{ai} 
+ {1\over2} \left(h_{,j}/h\right) \; \Gamma^j{}_{ai} 
- {1\over2} h_{ac,l} h^{cd} f^{lj} f_{ji,d}
\end{eqnarray}
To simplify things, we can define two tensors by
\begin{eqnarray}
K_{abi} &=& -{1\over2} h_{ab,i} =  -{1\over2} h_{ab|i} 
\\
L_{ija} &=&  -{1\over2} f_{ij,a} =  -{1\over2} h_{ij:a} 
\end{eqnarray}
with the convention that the $abc...$ indices are raised and lowered using
$h$, and the $ijk...$ indices are raised and lowered using $f$.
\\
(The reason for this notation is that these quantities are some sort of
generalization of the notion of extrinsic curvature.)
\\
With these conventions we have:
\begin{eqnarray}
\Gamma^a{}_{bc}(y,z) &=&  [\Gamma(h)]^a{}_{bc}(y,z).
\nonumber
\\
\Gamma^i{}_{jk}(y,z) &=& [\Gamma(f)]^i{}_{jk}(y,z).
\nonumber
\\
\Gamma^i{}_{bc}(y,z) &=&  K_{bc}{}^i.
\nonumber
\\
\Gamma^a{}_{jk}(y,z) &=& L_{jk}{}^a.
\nonumber
\\
\Gamma^a{}_{bk}(y,z) &=& - K^a{}_{bk}.
\nonumber
\\
\Gamma^a{}_{jc}(y,z) &=& - K^a{}_{cj}.
\nonumber
\\
\Gamma^i{}_{jc}(y,z) &=&  - L^i{}_{jc}.
\nonumber
\\
\Gamma^i{}_{bk}(y,z) &=& - L^i{}_{kb}.
\nonumber
\end{eqnarray}
Which then implies
\begin{eqnarray}
R_{ai} &=& - K^b{}_{ai:b} + K^b{}_{bi:a} - L^j{}_{ia|j} + L^j{}_{ja|i}
\nonumber
\\
&&- {1\over2} {f_{,b}\over f} K^b{}_{ai} 
- {1\over2} {h_{,j}\over f}LK^j{}_{ia} 
- 2 K_{ac}{}^j \; L_{ji}{}^c 
\end{eqnarray}



   \subsection{Discussion}

This form of the metric is physically interesting because there is a good
chance that it really is general for 4 dimensional spacetime, and general
enough to encode the Kerr metric for a rotating black hole.

For this chapter we simply counted the number of independent components of
the metric tensor. In 1-dimension, there are zero truly independent
components in the metric. Furthermore, in any 1-dimensional Riemann manifold
we are able to choose the coordinates to make the metric flat.
For 2-dimensions, Riemann's argument implies that there is only one
truly
independent component in the metric. Indeed, any 2-dimensional Riemann
manifold has a standard result that locally, we can choose coordinates to
make
the metric conformally flat. The Riemann tensor also has one independent
component, essentially the Ricci scalar, $R$.
In addition, for $n = 3$, then Riemann's argument implies that there are
exactly
three truly independent components in the metric. The Riemann 3-manifolds
have metrics that are always locally diagonalizable.
For $n = 4$, the Riemann tensor has only twenty independent components. Ten of
them are the Ricci tensor $R_{ab}$, while the other ten are hidden in the
``Weyl tensor".

We use the Lorentz metric to formulate the free-space Maxwell equations in a
4-dimensional Riemann space, and to describe mechanical phenomena in
a vacuum Riemann space. It is trivial to get a Lorentzian geometry from a
Riemannian geometry in (0+1) dimensions. For a Riemannian geometry in (1+1)
dimensions we are able to find the Ricci scalar, the Ricci tensor and the
Riemann tensor. While, for a Riemannian geometry in (2+1) dimensions it is
more complicated to calculate the Ricci scalar, Ricci tensor and Riemann
tensor.

The case for (3+1) dimensions is physically interesting, but it is the most
difficult to analyze in detail. However, we are able to calculate the Ricci tensor and Einstein tensor in the 4
dimensional metric. We split the metric into four $2 \times 2$ blocks, two
of them trivial (zero), which allows us to calculate the Ricci tensor easily.
Furthermore, the Christoffel symbols can be computed from equation
(\ref{Christoffel}). For the 4-dimensional Christoffel symbol, there is a
total of 8 sub-blocks to deal with (see section (\ref{sub-block})). Now we
are able to calculate the Ricci tensor of $g_{\mu\nu}$
from equation (\ref{Ricci_Misner}). We expect the full 4 dimensional Ricci
tensor to be related to the sum of the Ricci tensors of the 2 dimensional
metrics $f$ and $h$, with some additional terms. In addition, we can
calculate the 4 dimensional Einstein tensor from equation
(\ref{Einstein_form}). This is useful as a way of looking for nice new
solutions of the Einstein equations.
\section{The ADM (Arnowitt-Deser-Misner) ansatz}
Both ADM and reverse-ADM (Kaluza--Klein) decompositions seek to 
split ``spacetime'' into ``space'' plus ``time'' in some 
sort of natural way, and then use this split to describe 
spacetime curvature in terms of the curvature of space and
some additional structure.

In addition, this decomposes the four-metric of spacetime as a time-varying three-metric on a three-dimensional domain together with the lapse and shift, which are scalar-valued and three-vector-valued functions of time and space, respectively.
   \subsection{ADM Decomposition}
   \begin{definition}
In the case of the ADM decomposition one takes
\[
g_{ab} =
\left[\begin{array}{c|c}
-[c^2-v^2]&-v_{j}\\
\hline
-v_{i}    &  h_{ij}         
\end{array}
\right]
\]
{\bf Notation:} Latin indices from the beginning of the alphabet $(a, b, c, ...)$ run from 0-3 and refer to space-time; Latin indices from the middle of the alphabet $(i, j, k ...)$ run from 1-3 and refer to space. We use $X_{;a}$ to denote a space-time covariant derivative and $X_{:i}$ to denote a three-space covariant derivative.
\end{definition}
\noindent
{\bf Interpretation:} The light cones, defined by $\d s^2 = g_{ab} \; \d x^{a} \; \d x^{b} = 0$ take the form 
\[h_{ij} \; (\d x^i - v^i \d t) \; (\d x^j - v^j \; \d t) = c^2 \; \d t^2\]
Here $h_{ij}$ is the 3-metric of space. We take
\[h^{ij} = [h_{ij}]^{-1}\]
to be the inverse 3-metric, and
\[v^2 = h^{ij} \; v_{i} \; v_{j} = h_{ij} \;  v^i \; v^j\]
Then $v^i$ is the 3-velocity of a coordinate system with respect to which $c$ is the ``coordinate speed of light''. (Note that in comparison to usual conventions $v_{i}$ is $minus$ the ``shift vector'', while $c$ is equal to the ``lapse function''.) 

We are able to calculate the inverse metric:
\[
g^{ab} =
\left[\begin{array}{c|c}
-{1 / c^2}   &{v^{j} / c^2}\\
\hline
-{v^{i} / c^2} & h^{ij}-{v^{i} v^{j} / c^2}
\end{array}
\right]
\]
Note that 3 contravariant vectors that point in the ``space'' directions are

$X^a = (0;1,0,0),\quad Y^a = (0;0,1,0),\quad$  and $\quad Z^a = (0;0,0,1).$
\\
So if we want {\bf n} to be a vector that is ``orthogonal'' to space, then we need to pick its covariant components to satisfy

\[n_{a} \propto (1;0,0,0)\]
 If we want {\bf n} to be a $unit$ vector that is ``orthogonal'' to space, then we need to pick its covariant components to satisfy

\[n_{a} = \pm \; (c;0,0,0)\]
Its contravariant components are then

\[n^a = \mp \; \left(\frac{1}{c};\frac{v^i}{c}\right) = \mp \; \frac{(1;v^i)}{c}\]
Choosing $n^a$ to be a $future$ $directed$ unit vector orthogonal to the space slices then fixes  
\[n^a = \left(\frac{1}{c};\frac{v^i}{c}\right) =\frac{(1;v^i)}{c}\] 
and
\[n_{a} = (-c;0,0,0).\]




   \subsection{Reverse-ADM (Kaluza-Klein) Decomposition}
\begin{definition}
The ``reverse-ADM decomposition'' can be found in Landau-Lifshitz and consists of setting
\[
g_{ab} =
\left[\begin{array}{c|c}
-c^2  &-v_{j}\\
\hline
-v_{i}&h_{ij} - {v_{i} \; v_{j} / c^2}
\end{array}
\right]
\]
{\bf Notation:} Again, Latin indices from the begining of the alphabet (such as $a, b, c$,...) run from 0-3 and refer to space-time; Latin indices from the middle of the alphabet $(i, j, k,...)$ run from 1-3 and refer to space. We use $X_{;a}$ to denote a space-time covariant derivative and $X_{;i}$ to denote a three-space covariant derivative.
\end{definition}
\noindent
{\bf Interpretation:} The light cones, defined by $\d s^2 = g_{ab} \; \d x^a \; \d x^b = 0$ now take the form
\[h_{ij} \; \d x^i \;  \d x^j = c^2(\d t + v_{i} \; \d x^{i})^2\]
We again say that $h_{ij}$ is the 3-metric of space, though it's a different 3-metric than in the ADM case. We take
\[h^{ij} = [h_{ij}]^{-1}\]
to be the inverse 3-metric, and
\[v^2 = h^{ij} \; v_{i} v_{j} = h_{ij} \; v^i v^j\]
Then $c$ is the ``two-way coordinate-averaged speed of light'', which is by construction isotropic. See \S 84 on pages 251-255 of \cite{Landau} for details. The physical interpretation of the 3-velocity $v^i$ is not particularly clear, and there seems to be no clean interpretation of the current formalism in terms of ``analogue models'' \cite{Carlos}. Nevertheless it is one of the standard metric decompositions, so let us adopt it and see what happens.
\newpage

\section{Static spacetimes: ADM and reverse-ADM (Kaluza-Klein) Decomposition}
Our strategy is straightforward:

$\bullet$\quad Take any static spacetime, and use the natural time coordinate to slice it into space plus time.

$\bullet$\quad Use the Gauss-Codazzi and Gauss-Weingarten equations to decompose the (3+1)-dimensional spacetime curvature tensor in terms of the 3-dimensional spatial curvature tensor, the extrinsic curvature of the time slice [zero], and the gravitational potential.
\\
In any static spacetime one can decompose the spacetime metric into block diagonal form [1, 2, 3]:
\begin{eqnarray}
\d s^2 &=& g_{\mu\nu} \; \d x^\mu \d x^\nu
\nonumber
\\
&=& -c^2 \d t^2 + h_{ij} \; \d x^i \d x^j.
\end{eqnarray}
Here $c$ and $h_{ij}$ are assumed to be time-independent, they depend only on position in space. This is equivalent to taking
\[
g_{ab} =
\left[\begin{array}{c|c}
-c^2&0\\
\hline
0   &h_{ij}
\end{array}
\right]
\]
{\bf Theorem:}
In any static spacetime coordinates can be found in which the ADM and reverse-ADM (Kaluza-Klein) decompositions are identical.
The proof is by inspection. Simply set $\vec{v}$ = 0 in the general formalism.
\\
Being static tightly constrains the space-time geometry in terms of the three-geometry of space on a constant time slice, and the manner in which this three-geometry is embedded into the spacetime. For example, from \cite{Misner}, page 515, and applying page 518, we have the standard results (which we will re-derive in a slightly more general context below):
\begin{eqnarray}
\label{static1}
^{(3+1)} R_{ijkl} &=& ^{(3)} R_{ijkl}.
\\
\label{static2}
^{(3+1)} R_{tijk} &=& 0.
\\
\label{static3}
^{(3+1)} R_{titj} &=& c\, c_{:i:j}.
\end{eqnarray}
This will generalize when we allow time dependence for $c$ and $h_{ij}$.

Now taking suitable contractions,
\begin{eqnarray}
^{(3+1)}R_{ij} &=& ^{(3)}R_{ij} - {\txt c_{:i:j} \over \txt c}.
\\
^{(3+1)}R_{ti} &=& 0.
\\
^{(3+1)}R_{tt} &=& c\, g^{ij} c_{:i:j} = c \;\; ^{(3)}\, \Delta{c}.
\end{eqnarray}
So that
\begin{equation}
^{(3+1)}R = \; ^{(3)} R - 2 \; \frac{^{(3)}\, \Delta{c}}{c}.
\end{equation}

Finally, for the spacetime Einstein tensor (for bits of this see \cite{Misner}, page 552, noting that in our present case $K_{ij}$ = 0, while in the corresponding formulae of \cite{Misner} they chose c = 1)
\begin{eqnarray}
\label{Einstein1}
^{(3+1)}G_{ij} &=& ^{(3)} G_{ij} - \frac{c_{:i:j}}{c} + g_{ij}\left\{\frac{^{(3)}\, \Delta{c}}{c}\right\}.
\\
\label{Einstein2}
^{(3+1)}G_{ti} &=& 0.
\\
\label{Einstein3}
^{(3+1)}G_{tt} &=& +\frac{c^2}{2} \; ^{(3)}R.
\end{eqnarray}
This decomposition is generic to $any$ static spacetime. (You can check this decomposition against various standard textbooks to make sure the coefficients are correct. For instance see \cite{Adler, Fock, Synge}).
\section{Zero--shift spacetimes: ADM and reverse-ADM (Kaluza-Klein) Decomposition}
If the spacetime is not static but the ``shift vector'' is zero we still have
\begin{equation}
g_{ab} =
\left[\begin{array}{c|c}
-c^2&0\\
\hline
0   &h_{ij} 
\end{array}
\right]
.
\end{equation}
(Such a decomposition is always possible in a globally hyperbolic spacetime.)
\\
But now we will have to permit time dependence for both $c$ and $h_{ij}$. Note that as long as the shift vector is zero the ADM and reversed-ADM decompositions remain identical.

In this case it is useful to define a quantity called the extrinsic curvature \cite{Misner}, the equation (21.67)
\begin{equation}
\label{extrinsic}
K_{ij} = -\frac{1}{2c}\, \frac{\partial{h_{ij}}}{\partial{t}}
\end{equation}
and then
\begin{eqnarray}
^{(3+1)}R_{ijkl} &=& ^{(3)}R_{ijkl} + K_{ik} K_{jl} - K_{il} K_{jk}.
\\
^{(3+1)}R_{tijk} &=& c \; \{K_{ik:j} - K_{ij:k}\}.
\\
^{(3+1)} R_{titj}&=& c\, c_{:ij} + ``mess".
\end{eqnarray}
See \cite{Misner}, equations (21.75) and (21.76). Note $(n \cdot n) = -1$, and the ``$n$-coordinate'' of equation (21.76) is really  ``$Nt$''  [that is, ``$ct$'']. The last equation for $ ^{(3+1)}R_{titj}$ is not given by \cite{Misner}, except implicitly on page 518, and this really is a bit of a mess which we shall explicitly calculate below.

But some pieces of the Riemann, Ricci, and Einstein \emph{can} be immediately read off from \cite{Misner}. For instance from \cite{Misner}, equation (21.8)
\[R_{ti} = G_{ti} = N \left\{K_{ij}\, ^{:j} - tr(K)_{:i}\right\}\]
while from \cite{Misner}, equation (21.77)
\[G_{tt} = \frac{1}{2}N^{2}\left\{^{(3)}R + (trK)^2 - tr(K^2)\right\}\]
   \subsection{Systematic derivation}
\label{systematic}
Christoffel symbols of the first kind:
\begin{eqnarray}
\Gamma_{ttt} &=& g_{t(t,t)} - \frac{1}{2} \; g_{tt,t} = \frac{1}{2} \; g_{tt,t} = -\frac{1}{2} \; (c^2)_{,t} = -c\, c_{,t}
\\
\Gamma_{itt} &=& g_{i(t,t)} - \frac{1}{2} \; g_{tt,i} = -\frac{1}{2} \; g_{tt,t} = +\frac{1}{2} \; (c^2)_{,t} = +c\, c_{,t}
\\
\Gamma_{tti} &=&\Gamma_{tit} =  g_{t(t,i)} - \frac{1}{2} \; g_{ti,t} = \frac{1}{2} \; g_{tt,i} = -\frac{1}{2} \; (c^2)_{,i} = -c\, c_{,i}
\\
\Gamma_{tij} &=& g_{t(i,j)} - \frac{1}{2} \; g_{ij,t} = -\frac{1}{2} \; g_{ij,t} = -\frac{1}{2} \; h_{ij,t} = c\,K_{ij}
\\
\Gamma_{itj} &=&\Gamma_{ijt} = g_{i(t,j)} - \frac{1}{2} \; g_{tj,i} = \frac{1}{2} \; g_{ij,t} = \frac{1}{2} \; h_{ij,t} = - c\,K_{ij}
\\
\Gamma_{ijk} &=& g_{i(j,k)} - \frac{1}{2} \; g_{jk,i} = h_{i(j,k)} - \frac{1}{2} \; h_{jk,i} = \; ^{(3)}\Gamma_{ijk}
\end{eqnarray}
Raising the first index:
\\
Christoffel symbols of the second kind:
\begin{eqnarray}
\Gamma^t{}_{tt} &=& \frac{c_{,t}}{c}
\\
\Gamma^i{}_{tt} &=& +c\,h^{ij}\,c_{,j} = +c\,c^{,j}
\\
\Gamma^t{}_{ti} &=& \Gamma^t{}_{it} = \frac{c_{,i}}{c}
\\
\Gamma^t{}_{ij} &=& -\frac{K_{ij}}{c}
\\
\Gamma^i{}_{tj} &=& \Gamma^i{}_{jr} = - c\,K^i{}_{j}
\\
\Gamma^i{}_{jk} &=& {}^{(3)}\Gamma^i{}_{jk}
\end{eqnarray}

\subsection{Riemann tensor}
Now insert this into the formula for the Riemann tensor.
\begin{equation}
\label{Riemann1}
R^a_{bcd} = -2\left\{\Gamma^a{}_{b[c,d]} - \Gamma^a{}_{e[c}\Gamma^e{}_{|b|d]}\right\}
\end{equation}
where the $|b|$ indicates we do not anti-symmetrize over the $b$. Lowering the $a$ index
\begin{equation}
\label{Riemann2}
R_{abcd} = -2\left\{\Gamma_{ab[c,d]} + g_{ae,[d}\Gamma^e{}_{|b|c]} - \Gamma^a{}_{e[c}\Gamma^e{}_{|b|d]}\right\}
\end{equation}
and rearranging
\begin{equation}
\label{Riemann3}
R_{abcd} = -2\left\{\Gamma_{ab[c,d]} + \Gamma_{ea[c}\,g^{ef}\,\Gamma_{|fb|d]}\right\}
\end{equation}
This is equivalent, for instance, to the unnumbered equation on page 21 of \cite{Dirac}, which can further be simplified to
\begin{equation}
R_{abcd} =\frac{1}{2}\left\{g_{ad,bc} - g_{bd,ac} - g_{ac,bd} + g_{bc,ad}\right\} + \Gamma_{ead}\,g^{ef}\,\Gamma_{fbc} - \Gamma_{eac}\,g^{ef}\,\Gamma_{fbd}
\end{equation}
This is equation (11.6) of \cite{Dirac}. So far this is completely general. If $g_{ab}$ block diagonalizes in the manner $g_{ab} = - c^2 \bigoplus h_{ij}$ then we have
\begin{eqnarray}
R_{abcd} &=& \frac{1}{2} \left\{{g_{ad,bc} - g_{bd,ac} - g_{ac,bd} + g_{bc,ad}}\right\}
- \frac{1}{c^2} \; \Gamma_{tad} \; \Gamma_{tbc}
\nonumber
\\
&& + \frac{1}{c^2} \; \Gamma_{tac} \; \Gamma_{tbd}  + h^{mn} \; \Gamma_{mad} \; \Gamma_{nbc}
 - h^{mn} \; \Gamma_{mac} \; \Gamma_{nbd}
\end{eqnarray}
${\bf R_{ijkl}:}$ If we now look at the space components of the 4-dimensional Riemann tensor we have
\begin{eqnarray}
R_{ijkl} &=& \frac{1}{2}\left\{g_{il,jk} - g_{jl,ik} - g_{ik,jl} + g_{jk,il}\right\}
-\frac{1}{c^2}\,\Gamma_{til}\,\Gamma_{tjk} 
\nonumber
\\
&& + \frac{1}{c^2}\,\Gamma_{tik}\,\Gamma_{tjl} + h^{mn}\,\Gamma_{mil}\,\Gamma_{njk} - h^{mn}\,\Gamma_{mik}\,\Gamma_{njl}
\end{eqnarray}
But most of this is simply the 3-dimensional Riemann tensor
\begin{equation}
R_{ijkl} ={}^{(3)}R_{ijkl} - \frac{1}{c^2}\,\Gamma_{til}\Gamma_{tjk} + \frac{1}{c^2}\,\Gamma_{tik}\,\Gamma_{tjl}
\end{equation}
and the rest we recognize in terms of the $K_{ij}$ we defined earlier, so that
\begin{equation}
R_{ijkl} ={}^{(3)}R_{ijkl} - K_{il}\,K_{jk} + K_{ik}\,K_{jl}
\end{equation}
which is the explicit way of doing things.
\\
${\bf {R_{tijk}:}}$ In a similar manner we can evaluate $R_{tijk}$ as
\begin{eqnarray}
R_{tijk} &=& \frac{1}{2}\left\{g_{tk,ij}- g_{ik,tj} - g_{tj,ik} + g_{ij,tk}\right\}
-\frac{1}{c^2}\,\Gamma_{ttk}\,\Gamma_{tij}
\nonumber
\\
&& + \frac{1}{c^2}\,\Gamma_{ttj}\,\Gamma_{tik} + h^{mn}\,\Gamma_{mtk}\,\Gamma_{nij} - h^{mn}\,\Gamma_{mtj}\,\Gamma_{nik}
\end{eqnarray}
which we can re-write as
\[R_{tijk} = (cK_{ik})_{,j} - (cK_{ij})_{,k} - c_{,j}K_{ik} + C_{,k}K_{ij} - cK^n{}_{k}\,\Gamma_{nij} + c\,K^n{}_{j}\,\Gamma_{nik}\]
so that 
\[R_{tijk} = cK_{ik,j} - cK_{ij,k} - c\,K^n{}_{k}\,\Gamma_{nij} + c\,K^n{}_{j}\,\Gamma_{nik}\]
that is, in terms of 3-space covariant derivatives
\begin{equation}
R_{tijk} = c\left\{K_{ik:j} - K_{ij:k}\right\}
\end{equation}
This is the second part of what we were trying to prove.
\\
${\bf R_{titj}:}$ It only remains to calculate $R_{titj}$. (After all $R_{ttij}, R_{ttti}, R_{tttt}$ and so on are all zero by symmetries of the Riemann tensor.)

So let us calculate
\begin{eqnarray}
R_{titj} &=& \frac{1}{2}\left\{g_{tj,it} - g_{ij,tt} - g_{tt,ij} + g_{it,tj}\right\}
-\frac{1}{c^2}\,\Gamma_{ttj}\,\Gamma_{tit}
\nonumber
\\
&& + \frac{1}{c^2}\,\Gamma_{ttt}\,\Gamma_{tij} + h^{mn}\,\Gamma_{mtj}\,\Gamma_{nit} - h^{mn}\,\Gamma_{mtt}\,\Gamma_{nij}
\nonumber
\end{eqnarray}
So that
\begin{eqnarray}
R_{titj} &=& \frac{1}{2}\left\{-h_{ij,tt} + (c^2)_{,ij}\right\}
\nonumber
\\
&&-\frac{1}{c^2}\,(-c\,c_{,j})\,(-c\,c_{,i}) + \frac{1}{c^2}\,(-c\,c_{,t})\,(c\,K_{ij})
\nonumber
\\
&&+h^{mn}\,(-cK_{mj})\,(-cK_{ni}) - h^{mn}\,(c\,c_{,m})\,\Gamma_{nij}
\nonumber
\end{eqnarray}
That is:
\begin{equation}
R_{titj} = (cK_{ij})_{,t} + c\,c_{,ij} + c_{,i}\,c_{,j} -c_{,j}\,c_{,i} - c_{,t} K_{ij} + c^2(K_{im}\,h^{mn}\,K_{nj}) -(c\,c_{,m})\,\Gamma^m{}_{ij}
\nonumber
\end{equation}
So finally
\begin{equation}
R_{titj} = cK_{ij,t} + c\,c_{:i:j} + (K^2)_{ij}
\end{equation}

\paragraph{Summary:}
Whenever the metric $g_{ab}$ block diagonalizes in the manner $g_{ab} = -c^2\bigoplus h_{ij}$, then even if we allow arbitrary time dependence we have the relativity simple results
\begin{eqnarray}
\label{result1}
R_{ijkl} &=& ^{(3)}R_{ijkl} - K_{il}\,K_{jk} + K_{ik}\,K_{jl}
\\
\label{result2}
R_{tijk} &=& c\,\left\{ K_{ik:j} - K_{ij:k}\right\}
\\
\label{result3}
R_{titj} &=& cK_{ij,t} + c\,c_{:i:j} + (K^2)_{ij}
\end{eqnarray}
where
\[K_{ij} = -\frac{1}{2c} \frac{\partial h_{ij}}{\partial t}\]
Of course once we have deduced these results for the Riemann tensor, results for the Ricci tensor and Einstein tensor can be obtained by simple contraction.
\section{General stationary spacetime: ADM Decomposition}
For a general stationary spacetime we need to use the full ADM decomposition
\[
g_{ab} =
\left[\begin{array}{c|c}
-[c^2 - v^2]&-v_{j}\\
\hline
-v_{i}&h_{ij}
\end{array}
\right]
\]
\[
g^{ab} =
\left[\begin{array}{c|c}
-{1 /c^2}     &-{v^j / c^2}\\
\hline
-{v^{i} / c^2}&h^{ij} - {v^i v^j / c^2}
\end{array}
\right]
\]
We do however have the simplification that all quantities are time-independent. (All time derivatives vanish). Note that the ADM decomposition is in this case quite distinct from the reverse-ADM decomposition.

We still have, thanks to Dirac
\begin{equation}
R_{abcd} = \frac{1}{2}\left\{g_{ad,bc} - g_{bd,ac} - g_{ac,bd} + g_{bc,ad}\right\} + \Gamma_{ead}\,g^{ef}\,\Gamma_{fbc} - \Gamma_{eac}\,g^{ef}\,\Gamma_{fbd}
\end{equation}
But now when we insert the inverse metric
\begin{eqnarray}
R_{abcd} &=& \frac{1}{2}\{g_{ad,bc} - g_{bd,ac} - g_{ac,bd} + g_{bc,ad}\}
\nonumber
\\
&&-\frac{1}{c^2}\,\Gamma_{tad}\,\Gamma_{tbc} + \frac{1}{c^2}\,\Gamma_{tac}\,\Gamma_{tbd}
\nonumber
\\
&&-(v^m/c^2)\{\Gamma_{tad}\,\Gamma_{mbc} + \Gamma_{mad}\,\Gamma_{tbc}\} + (v^m/c^2)\{\Gamma_{mac}\,\Gamma_{tbd} + \Gamma_{tac}\,\Gamma_{mbd}\}
\nonumber
\\
&&+(h^{mn} - v^m v^n/c^2)\,\Gamma_{mad}\,\Gamma_{nbc} - (h^{mn} - v^mv^n/c^2)\,\Gamma_{mac}\,\Gamma_{nbd}
\end{eqnarray}
${\bf R_{ijkl}:}$ Start by looking at the space components $R_{ijkl}$. We have
\begin{eqnarray}
R_{ijkl} &=& \frac{1}{2}\{g_{il,jk} - g_{jl,ik} - g_{ik,lj} + g_{jk,il}\}
\nonumber
\\
&&- \frac{1}{c^2}\,\Gamma_{til}\,\Gamma_{tjk} + \frac{1}{c^2}\,\Gamma_{tik}\,\Gamma_{tjl}
\nonumber
\\
&&-(v^m/c^2)\{\Gamma_{til}\,\Gamma_{mjk} + \Gamma_{mil}\,\Gamma_{tjk}\} + (v^m/c^2)\{\Gamma_{mik}\,\Gamma_{tjl} + \Gamma_{tik}\,\Gamma_{mjl}\}
\nonumber
\\
&&-(v^m v^n/c^2)\,\Gamma_{mil}\,\Gamma_{njk} - (v^m v^n/c^2)\,\Gamma_{mik}\,\Gamma_{njl}
\nonumber
\\
&&+h^{mn}\,\Gamma_{mil}\,\Gamma_{njk} - h^{mn}\,\Gamma_{mik}\,\Gamma_{njl}
\end{eqnarray}
But most of this is simply the 3-dimensional Riemann tensor, and the rest can easily be regrouped as
\begin{equation}
R_{ijkl} = \; ^{(3)}R_{ijkl} - \frac{1}{c^2} (\Gamma_{til}+v^m\Gamma_{mil})\,(\Gamma_{tjk}+v^n\Gamma_{njk})+\frac{1}{c^2}(\Gamma_{tik}+v^m\Gamma_{mik})\,(\Gamma_{tjl}+v^n\Gamma_{njl})
\end{equation}
Now suppose we define
\begin{equation}
K_{ij} = \frac{1}{c}\left\{\Gamma_{tij} + v^m\Gamma_{mij}\right\}
\end{equation}
This is a natural generalization of our previous definition, and would allow us to again write
\begin{equation}
R_{ijkl} =\, ^{(3)}R_{ijkl} - K_{il}\,K_{jk} + K_{ik}\,K_{jl}
\end{equation}
but now for this ``improved'' definition of $K_{ij}$. Let's check what happens. Having delayed things as long as possible, we'll have to start by calculating Christoffel symbols.
   \subsection{Christoffel symbols of first kind}
\label{part2_christ}
(Remember, stationary, no $t$ dependence):
\begin{eqnarray}
\Gamma_{ttt} &=& g_{t(t,t)} - \frac{1}{2} \; g_{tt,t} = \frac{1}{2} \; g_{tt,t} = 0
\\
\Gamma_{itt} &=& g_{i(t,t)} - \frac{1}{2} \; g_{tt,t} = - \frac{1}{2} \; g_{tt,i} = + \frac{1}{2} \; (c^2 - v^2)_{,i}
\\
\Gamma_{tti} &=& \Gamma_{tit} = g_{t(t,i)} - \frac{1}{2} \; g_{ti,t} = \frac{1}{2} \; g_{tt,i} = - \frac{1}{2} \; (c^2 - v^2)_{,i}
\\
\Gamma_{tij} &=& g_{t(i,j)} - \frac{1}{2} \; g_{ij,t} = + \frac{1}{2} \; (g_{ti,j} + g_{tj,i})
\\
\nonumber
&=& -\frac{1}{2} \; (v_{i,j} + v_{j,i}) = -v_{(i,j)}
\\
\Gamma_{itj} &=& \Gamma_{ijt} = g_{i(t,j)} - \frac{1}{2} \; g_{tj,i} = \frac{1}{2} \; (g_{it,j} - g_{tj,i})
\nonumber
\\
&=& - \frac{1}{2} \; (v_{i,j} - v_{j,i}) = -v_{[i,j]}
\\
\Gamma_{ijk} &=& g_{i(j,k)} - \frac{1}{2} \; g_{jk,i} = h_{i(j,k)} - \frac{1}{2} \; h_{jk,i} =\,^{(3)}\Gamma_{ijk}
\end{eqnarray}
Then we have
\begin{equation}
\frac{1}{c}\{\Gamma_{tij} + v^m \; \Gamma_{mij}\} = \frac{1}{c}\left\{-v_{(i,j)} + v_{m}\Gamma^{m}{}_{ij}\right\} = - \frac{1}{c} \; v_{(i:j)}
\end{equation}
so that in this situation
\begin{equation}
K_{ij} = - \frac{1}{c} \; v_{(i:j)}
\end{equation}
This is compatible with the general definition of extrinstic curvature, see \cite{Misner} equation (21.67), note that our $v^{i}$ is $minus$ the shift vector.
\\
\\
${\bf R_{tijk}}$: In a similar manner we can evaluate $R_{tijk}$ as
\begin{eqnarray}
R_{tijk} &=& \frac{1}{2}\left\{g_{tk,ij} - g_{ik,tj} - g_{tj,ik} + g_{ij,tk}\right\}
\nonumber
\\
&&- \frac{1}{c^2}\,\Gamma_{ttk}\,\Gamma_{tij} + \frac{1}{c^2}\,\Gamma_{ttj}\,\Gamma_{tik}
\nonumber
\\
&&-(v^m/c^2) \; \{\Gamma_{ttk}\,\Gamma_{mij} + \Gamma_{mtk}\,\Gamma_{tij}\} + (v^m/c^2) \; \{\Gamma_{mtj}\,\Gamma_{tik} + \Gamma_{ttj}\,\Gamma_{mik}\}
\nonumber
\\
&&+(h^{mn} - v^m v^n /c^2)\,\Gamma_{mtk}\,\Gamma_{nij} - (h^{mn} - v^m v^n/c^2)\,\Gamma_{mtj}\,\Gamma_{nik}
\end{eqnarray}
which we can re-write as
\begin{eqnarray}
R_{tijk} &=& \frac{1}{2}\{-v_{k,ij} + v_{j,ik}\}
\nonumber
\\
&&- \frac{1}{c^2} \; (\Gamma_{ttk} + v^m\Gamma_{mtk})\,(\Gamma_{tij} + v^m\Gamma_{mij}) + \frac{1}{c^2}\,(\Gamma_{ttj} + v^m\Gamma_{mtj})\,(\Gamma_{tik} + v^m\Gamma_{mik})
\nonumber
\\
&&+h^{mn}\,\Gamma_{mtk}\,\Gamma_{nij} - h^{mn}\,\Gamma_{mtj}\,\Gamma_{nik}
\end{eqnarray}
Now
\begin{equation}
\Gamma_{ttk} + v^m\Gamma_{mtk} = - \frac{1}{2}(c^2-v^2)_{,k} - \frac{1}{2}v^m(v_{m,k}-v_{k,m}) = - \frac{1}{2}(c^2-v^2)_{:k} - \frac{1}{2}v^m(v_{m:k}-v_{k:m})
\end{equation}
so that
\begin{equation}
\Gamma_{ttk} + v^m\Gamma_{mtk} = - \frac{1}{2}(c^2)_{,k} + \frac{1}{2}v^m(v_{m:k} + v_{k:m}) = -c \; \{c_{,k} + v^mK_{mk}\}
\end{equation}
So now
\begin{equation}
R_{tijk} = \frac{1}{2} \{-v_{k,ij} + v_{j,ik}\} - (c_{,k} + v^mK_{mk})K_{ij} + (c_{,j} + v^mk_{mj})K_{ik} - v_{[m,k]}\Gamma^m{}_{ij} + v_{[m,j]}\Gamma^m{}_{ik}
\end{equation}
This can be rewritten in several ways: If we remain in a coordinate basis
\begin{equation}
R_{tijk} = v_{[j,k],i} - v_{[m,k]}\Gamma^m{}_{ji} - v_{[j,m]}\Gamma^m{}_{ki} - (c_{,k} + v^mK_{mk})K_{ij} + (c_{,j} + v^mK_{mj})K_{ik}
\end{equation}
whence
\begin{equation}
R_{tijk} = v_{[j:k]:i} - (c_{,k} + v^mK_{mk})K_{ij} + (c_{,j} + v^mK_{mj})K_{ik}
\end{equation}
This is relatively compact, but not so easy to compare to our previous result.

But we can also write
\begin{eqnarray}
2 v_{[j:k]:i} &=& v_{j:k:i} - v_{k:j:i} = v_{j:i:k} + v_{m}\,^{(3)}R^m{}_{jki} - v_{k:i:j} - v_{m}\,^{(3)}R^m{}_{kji}
\nonumber
\\
&=& v_{j:i:k} + v_{i:j:k} - v_{k:i:j} - v_{i:k:j} + v_{m}\left(^{(3)}R^m{}_{jki} -\,^{(3)}R^m{}_{kji} - \,^{(3)}R^m{}_{ijk}\right)
\nonumber
\\
&=& 2\left(v_{(i:j):k} - v_{(i:k):j} - v_{m}\,^{(3)}R^m{}_{ijk}\right)
\nonumber
\\
&=& 2\left([-c K_{ij}]_{:k} - [-c K_{ik}]_{:j}- v_m \; ^{(3)} R^m{}_{ijk}\right)
\end{eqnarray}
which lets us write
\begin{equation}
R_{tijk} = c \; \{[K_{ik}]_{:k} - [K_{ij}]_{:k}\} - v^m\left[^{(3)}R_{mijk} - K_{mk}K_{ij} + K_{mj}K_{ik}\right]
\end{equation}
that is, in view of our previous result,
\begin{equation}
R_{tijk} = c[K_{ik:j} - K_{ij:k}] - v^mR_{mijk}
\end{equation}
This is the tidiest we can make things in a coordinate basis. This is the analogue of \cite{Misner}, equation (21.76).

For future reference define
\begin{equation}
\Omega_{ij} = v_{[i,j]} = v_{[i:j]}
\end{equation}
and note the identity
\begin{equation}
\Omega_{jk:i} = \left[cK_{ik}\right]_{:j} - \left[cK_{ij}\right]_{:k} - v_{m}\,^{(3)}R^m{}_{ijk}
\end{equation}
Note that it sometimes easier to work in what is referred to as a ``non-coordinate basis'' by defining:
\begin{equation}
R_{nijk} \equiv n^a\,R_{aijk}
\end{equation}
in which case
\begin{equation}
R_{nijk} = \frac{1}{c}(1;v^m) \; (R_{tijk};R_{mijk}) = \frac{1}{c}\left[R_{tijk} + v^mR_{mijk}\right] = K_{ik:j} - K_{ij:k}.
\end{equation}
This means you are not really looking in the ``$t$'' direction, but are instead looking in the ``$n$'' direction normal to ``space''.
\\
\\
${\bf R_{titj}}$: Now for the $R_{titj}$ pieces. This is where things might get a little messy. From Dirac's general formula, applied to the present situation
\begin{eqnarray}
R_{titj} &=& \frac{1}{2}\left\{g_{tj,it} - g_{ij,tt} - g_{tt,ij} + g_{it,tj}\right\}
\nonumber
\\
&&- \frac{1}{c^2}\,\Gamma_{ttj}\,\Gamma_{tit} + \frac{1}{c^2}\,\Gamma_{ttt}\,\Gamma_{tij}
\nonumber
\\
&&-(v^m/c^2)\left\{\Gamma_{ttj}\,\Gamma_{mit} + \Gamma_{mtj}\,\Gamma_{tit}\right\} + (v^m/c^2)\left\{\Gamma_{mtt}\,\Gamma_{tij} + \Gamma_{ttt}\,\Gamma_{mij}\right\}
\nonumber
\\
&&+(h^{mn} - v^m v^n/c^2) \,\Gamma_{mtj}\,\Gamma_{nit} - (h^{mn} - v^m v^n/c^2)\,\Gamma_{mtt}\,\Gamma_{nij}
\end{eqnarray}
Because of time independence this greatly simplifies
\begin{eqnarray}
R_{titj} &=& \frac{1}{2}\left\{-g_{tt,ij}\right\} 
- \frac{1}{c^2}\,\left\{ \Gamma_{ttj}\,\Gamma_{tit}\right\}
\nonumber
\\
&&-(v^m/c^2)\left\{\Gamma_{ttj}\,\Gamma_{mit} + \Gamma_{mtj}\,\Gamma_{tit} - \Gamma_{mtt}\,\Gamma_{tij}\right\}
\nonumber
\\
&&+(h^{mn} - v^m v^n/c^2)\Gamma_{mtj}\,\Gamma_{nit} - (h^{mn} - v^m v^n/c^2)\Gamma_{mtt}\,\Gamma_{nij}
\end{eqnarray}
so that
\begin{eqnarray}
R_{titj} &=& \frac{1}{2}\,\{-g_{tt,ij}\} - \frac{1}{c^2}\,\left(\Gamma_{ttj} + v^m\Gamma_{mtj}\right) \left(\Gamma_{tit} + v^m\Gamma_{mit}\right)
\nonumber
\\
&&+(v^m/c^2)\Gamma_{mtt}\,\Gamma_{tij} + h^{mn}\Gamma_{mtj}\,\Gamma_{nit} - h^{mn}\Gamma_{mtt}\,\Gamma_{nij}
\end{eqnarray}
Inserting the known values of the Christoffel symbols (calculated above)
\begin{eqnarray}
R_{titj} &=& \frac{1}{2}\,(c^2-v^2)_{,ij} - (c_{,j} + v^m\,K_{mj})(c_{,i}+v^m\,K_{mi})-\frac{1}{2}\,(v^m/c^2)\,(c^2-v^2)_{,m}\,v_{(i,j)}
\nonumber
\\
&&+h^{mn}\,v_{[m,j]}\,v_{[n,i]} - \frac{1}{2}\,h^{mn}\,(c^2-v^2)_{,m}\,\Gamma_{nij}
\end{eqnarray}
That is
\begin{eqnarray}
R_{titj} &=& \frac{1}{2}\,(c^2-v^2)_{:ij}-(c_{,j}+v^m\,K_{mj})(c_{,i}+v^m\,K_{mi})+\frac{1}{2}\,(v^m/c)\,(c^2-v^2)_{,m}\,K_{ij}
\nonumber
\\
&&+h^{mn}\,v_{[m,j]}\,v_{[n,i]}
\end{eqnarray}
But then
\begin{eqnarray}
R_{titj} &=& c\,c_{:ij} - \frac{1}{2}\,(v^2)_{:ij} - v^m\,[c_{,j}\,K_{mi} + c_{,i}\,K_{mj}]-v^m\,K_{mj}\,v^n\,K_{ni}
\nonumber
\\
&&+v^m\,c_{,m}\,K_{ij}-\frac{1}{2}\,(v^m/c)\,(v^2)_{,m}\,K_{ij}+h^{mn}\,v_{[m,j]}\,v_{[n,i]}
\end{eqnarray}
Which still needs to be simplified further.

For instance
\begin{equation}
\frac{1}{2} (v^2)_{:ij} = \frac{1}{2} (v^k \; v_k)_{:ij} = (v^k \; v_{k:i})_{:j} = (v^k{}_{:j} \; v_{k:i}) + (v^k \; v_{k:i:j})
\end{equation}
But
\begin{eqnarray}
v^k_{:j} \; v_{k:i} &=& h^{kl} \; v_{k:j} \; v_{k:i} = h^{kl} \; (-c K_{kj} + v_{[k:j]}) \; (-cK_{li} + v_{[l:i]})
\nonumber
\\
&=& c^2 (K^2)_{ij} - c (K \Omega - \Omega K)_{ij} + h^{mn} \;  v_{[m,j]} \; v_{[n,i]}
\end{eqnarray}
where with obvious notation $\Omega_{ij} = v_{[i,j]}$ and
\begin{equation}
(K\Omega - \Omega K)_{ij} = K_{ik}\,g^{kl}\,\Omega_{lj} - \Omega_{ik}\,g^{kl}\,K_{lj}
\end{equation}
{\underline{Aside:}} Write $(K\,\Omega - \Omega\, K)_{ij}$ as
\[K g^{-1} \Omega - \Omega \; g^{-1} K\]
then
\begin{eqnarray}
(Kg^{-1}\,\Omega - \Omega\, g^{-1}\, K)^T &=& \Omega^T\,[g^{-1}]^T\,K^T - K^T\,[g^{-1}]^T\,\Omega^T
\nonumber
\\
&=& -\Omega\, g^{-1}\,K + K\,g^{-1}\,\Omega
\end{eqnarray}
so this object is \emph{symmetric}, as long as the minus sign is included.

Then
\begin{eqnarray}
R_{titj} &=& c \; c_{:ij} - (v^k \; v_{k:i:j}) - c^2 (K^2)_{ij} + c (K \Omega - \Omega K)_{ij}
\nonumber
\\
&& -v^m [c_{,j} K_{mi} + c_{,i} K_{mj}] - v^m K_{mj} \; v^n \; K_{ni}
\nonumber
\\
&& + v^m \; c_{,m} K_{ij} - (v^m/c) \; (v^k \; v_{k:m}) \; K_{ij}
\end{eqnarray}
so that
\begin{eqnarray}
R_{titj} &=& c \; c_{:ij} - (v^k \; v_{k:i:j}) - c^2 (K^2)_{ij} + c (K \Omega - \Omega K)_{ij}
\nonumber
\\
&& - v^m[c_{,j} K_{mi} + c_{,i} K_{mj}] -v^m \; K_{mj} \; v^n \; K_{ni}
\nonumber
\\
&& + v^m \; c_{,m} \; K_{ij} + (v^k v^l \; K_{kl}) \; K_{ij}
\end{eqnarray}
Now by the Ricci identities the combination $v^k \; v_{k:i:j}$ is symmetric in $ij$, so that we can write:
\begin{equation}
v^k \; v_{k:i:j} = v^k \; v_{k:(i:j)}
\end{equation}
But now consider
\begin{eqnarray}
v_{k:(i:j)} &=& \frac{1}{2} \left\{ v_{k:i:j} + v_{k:j:i}\right\} = \frac{1}{2} \left\{ ( -c K_{ki} + \Omega_{ki})_{:j} + (-c K_{kj} + \Omega_{kj})_{:i}\right\}
\nonumber
\\
&=& - \frac{1}{2} \left\{(c K_{ki})_{:j} + (c K_{kj})_{:i}\right\} + \frac{1}{2} \left\{ \Omega_{ki:j} + \Omega_{kj:i}\right\}
\end{eqnarray}
Then recall the identity
\begin{equation}
\Omega_{jk:i} = [c K_{ik}]_{:j} - [c K_{ij}]_{:k} - v_m \; ^{(3)} R^m{}_{ijk}
\end{equation}
to deduce
\begin{equation}
v_{k:(i:j)} = [c K_{ij}]_{:k} + \frac{1}{2} v_m \left( \; ^{(3)} R^m{}_{ijk} + \;  ^{(3)} R^m{}_{jik}\right)
\end{equation}
whence
\begin{equation}
v^k \; v_{k:(i:j)} = v^k [c K_{ij}]_{:k} - v^k \; v^l \; ^{(3)} R_{kilj}
\end{equation}
Putting this back into $R_{titj}$ we have
\begin{eqnarray}
R_{titj} &=& c \; c_{:ij} - v^k[c K_{ij}]_{:k} + v^k \; v^l \; ^{(3)} R_{kilj} - c^2 (K^2)_{ij} + c (K \Omega - \Omega K)_{ij}
\nonumber
\\
&& -v^m K_{mj} \; v^n K_{ni} + v^m \; c_{,m} K_{ij} + (v^k v^l \; K_{kl}) \; K_{ij}
\end{eqnarray}
which simplifies to
\begin{eqnarray}
R_{titj} &=& c \;c_{:ij} - v^k c K_{ij:k} - c^2 (K^2)_{ij} + c (K \Omega - \Omega K)_{ij}
\nonumber
\\
&& - v^m \; K_{mj} \; v^n \; K_{ni} + (v^k v^l \; K_{kl}) \; K_{ij} + v^k v^l \;^{(3)} R_{kilj}
\end{eqnarray}
We can rewrite the last line in terms of the 4-curvature as
\begin{eqnarray}
R_{titj} &=& c\;c_{:ij} - v^k \; c K_{ij:k} - c^2 (K^2)_{ij} + c (K \Omega - \Omega K)_{ij}
\nonumber
\\
&& + v^k v^l \; R_{kilj}
\end{eqnarray}
This is the simplest expression we can get in a coordinate basis.
\bigskip
\\
${\bf R_{ninj}:}$ Now define in a ``non-coordinate basis''.
\begin{eqnarray}
R_{ninj} &\equiv& n^a \; n^b \; R_{aibj} = \frac{1}{c^2} \; (1;v^i)^a \; (1;v^l)^b \; R_{aibj}
\nonumber
\\
&=& \; \frac{1}{c^2} \; \left[ R_{titj} + v^k \; R_{tikj} + v^k R_{kitj} + v^k v^l R_{kilj} \right]
\nonumber
\\
&=& \; \frac{1}{c^2} \; \left[R_{titj} - c v^k [K_{ik:j} - K_{ij:k}] - c v^k [K_{jk:i} - K_{ij:k}] - v^k v^l \; R_{kilj}\right]
\nonumber
\\
&=& \; \frac{1}{c^2} \; \left[ R_{titj} - c v^k [K_{ik:j} + K_{jk:i} - 2 K_{ij:k}] - v^k v^l \; R_{kilj}\right]
\end{eqnarray}
Then
\begin{eqnarray}
c^2 \; R_{ninj} &=& c \; c_{:ij} - v^k \; c \; K_{ij:k} - c^2 (K^2)_{ij} + c (K \Omega - \Omega K)_{ij}
\nonumber
\\
&& - c \; v^k [K_{ik:j} + K_{jk:i} - 2 K_{ij:k}]
\end{eqnarray}
There are now some simplifications
\begin{eqnarray}
c^2 \; R_{ninj} &=& c \; c_{:ij} + c \; v^k \; K_{ij:k} - c^2 (K^2)_{ij} + c (K \Omega - \Omega K)_{ij}
\nonumber
\\
&& - c \; v^k [K_{ik:j} + K_{jk:i}]
\end{eqnarray}
To simplify this we consider the Lie derivative
\begin{equation}
\mathcal{L}_{v} K_{ij} = v^k \; \partial_{k} K_{ij} + v^k{}_{,i} \; K_{kj} + v^k{}_{,j} \; K_{ik}
\end{equation}
which can also be written in terms of 3-covariant derivatives as
\begin{equation}
\mathcal{L}_v K_{ij} = v^k \; K_{ij:k} + v^k{}_{:i} \; K_{kj} + v^k{}_{;j} \; K_{ik}
\end{equation}
so that
\begin{equation}
\mathcal{L}_{v} K_{ij} = v^k \; K_{ij:k} + h^{kl} [(-c K_{ki} + \Omega_{ki}) K_{lj} + (-c K_{kj} + \Omega_{kj}) K_{il}]
\end{equation}
and
\begin{equation}
\mathcal{L}_{v} K_{ij} = v^k \; K_{ij:k} - 2 c \; (K^2)_{ij} + (K \Omega - \Omega K)_{ij}
\end{equation}
Therefore
\begin{equation}
c^2 \; R_{ninj} = c \; c_{:ij} + c \mathcal{L}_{v} K_{ij} + c^2 \; (K^2)_{ij} - c \; v^k [K_{ik:j} + K_{jk:i}]
\end{equation}
\\
{\bf {Summary
:}}

For an arbitrary stationary spacetime. taking
\[
K_{ij} = - \frac{1}{c} \; v_{(i:j)}
\]
we have 
\begin{eqnarray}
R_{ijkl} &=& \; ^{(3)} R_{ijkl} - K_{il} \; K_{jk} + K_{ik} \; K_{jl}
\\
R_{tijk} &=& c \; [K_{ik:j} - K_{ij:k}] - v^m \; R_{mijk}
\\
R_{titj} &=& c \; c_{:ij} -  v^k \; c \; K_{ij:k} - c^2 (K^2)_{ij} + c \; (K \Omega - \Omega K)_{ij} + v^k v^l \; R_{kilj}
\end{eqnarray}
We can also write this as
\begin{equation}
R_{nijk} \equiv n^a \;  R_{aijk} = K_{ij:k} - K_{ik:j}.
\end{equation}
and
\begin{equation}
R_{ninj} \equiv n^a \; n^b \; R_{aibj} = c \; c_{:ij} + c \mathcal{L}_{v} \; K_{ij} + c^2 \; (K^2)_{ij} - c \; v^k [K_{ik:j} + K_{jk:i}].
\end{equation}

   \subsection{Completely general ADM decomposition}
For the general ADM case coordinate techniques are too messy to contemplate. The easy results are on page 514 of \cite{Misner} and boil down to
\begin{equation}
K_{ij} = - \frac{1}{c} \left[ \frac{1}{2} \partial_{t} \, g_{ij} + v_{(i:j)}\right]
\end{equation}
note that $K_{ij}$ is now a combination of the zero-shift and stationary case.

Then
\begin{equation}
R_{ijkl} = \; ^{(3)} R_{ijkl} - K_{il} \; K_{jk} + K_{ik} \; K_{jl}
\end{equation}
and
\begin{equation}
R_{nijk} \equiv n^a \; R_{aijk} = K_{ij:k} - K_{ik:j}
\end{equation}
The difficult one is then encoded in page 518, \cite{Misner} which seems to assert that
\begin{equation}
R_{nink} \equiv n^a \; n^b \; R_{aibj} = [\mathcal{L}_{n} \; K]_{ij} + [K^2]_{ij} + a_{(i;j)} + a_i \; a_j
\end{equation}
where $a_i$ is the space projection of the 4-acceleration {\bf a = (n $ \cdot  \; \nabla$) n}.
   \subsection{The inverse-ADM decomposition}
We can decompose the metric as follows
\begin{equation}
g_{\mu\nu} =
\left[\begin{array}{c|c}
-1 / N^2   &v_{i} / N^2\\
\hline
v_{i} / N^2&h_{ij}-v_{i}v_{j} / N^2
\end{array}
\right]
\end{equation}
with
\[
N = N(t,x,y,z), \; v_{i} = v_{i}(t,x,y,z), \text{and} \; h_{ij} = h_{ij}(t,x,y,z).
\]
The inverse metric is now
\begin{equation}
g^{\mu\nu} =
\left[\begin{array}{c|c}
-[N^2 - h^{ij}\,v_{i}\,v_{j}]&v^{i}\\
\hline
 v^{i}                   &h^{ij}
\end{array}
\right]
\end{equation}
where $v^{i} = h^{ij}\,v_{j}$ and $h^{ij} = [h_{ij}]^{-1}$

This is the reverse of ADM formulation
\begin{equation}
g_{\mu\nu} =
\left[
\begin{array}{c|ccc}
  g_{11}&g_{12}&g_{13}&g_{14}\\
\hline
  g_{21}&g_{22}&g_{23}&g_{24}\\
  g_{31}&g_{32}&g_{33}&g_{34}\\
  g_{41}&g_{42}&g_{43}&g_{44}
\end{array}
\right]
= \left[
\begin{array}{c|ccc}
-1 / N^2   & v_{2} / N^2 &v_{3}/ N^2& v_{4} / N^2\\
\hline
v_{2} / N^2&h_{22}- v_{2}v_{2} / N^2 &h_{23}- v_{2}v_{3} / N^2 &h_{24}-v_{2}v_{4} / N^2\\
v_{3} / N^2&h_{32}-v_{3}v_{2} / N^2 &h_{33}-v_{3}v_{3} / N^2 &h_{34}-v_{3}v_{4} / N^2\\
v_{4} / N^2 &h_{42}-v_{4}v_{2} / N^2 &h_{43}-v_{4}v_{3} / N^2 & h_{44}- v_{4}v_{4} / N^2
\end{array}
\right]
\end{equation}
and
\begin{equation}
g^{\mu\nu} =
\left[\begin{array}{c|ccc}
-[N^2 - \sum_{i,j = 1}^{4}h^{ij}v_{i}v_{j}]&v^{2} &v^{3} &v^{4}\\
\hline
 v^{2}                                     &h^{22}&h^{23}&h^{24}\\
 v^{3}                                     &h^{32}&h^{33}&h^{34}\\
 v^{4}                                     &h^{42}&h^{43}&h^{44} 
\end{array}
\right]
\end{equation}
For the time being, let $v_{i} = 0$ then
\begin{equation}
g_{\mu\nu} =
\left[
\begin{array}{c|ccc}
  g_{11}&g_{12}&g_{13}&g_{14}\\
\hline
  g_{21}&g_{22}&g_{23}&g_{24}\\
  g_{31}&g_{32}&g_{33}&g_{34}\\
  g_{41}&g_{42}&g_{43}&g_{44}
\end{array}
\right]
=\left[
\begin{array}{c|ccc}
- 1 / N^2&0     &0     &0\\
\hline
0             &h_{22}&h_{23}&h_{24}\\ 
0             &h_{32}&h_{33}&h_{34}\\
0             &h_{42}&h_{43}&h_{44}
\end{array}
\right]
\end{equation}

and

\begin{equation}
g^{\mu\nu} =
\left[
\begin{array}{c|ccc}
  g^{11}&g^{12}&g^{13}&g^{14}\\
\hline
  g^{21}&g^{22}&g^{23}&g^{24}\\
  g^{31}&g^{32}&g^{33}&g^{34}\\
  g^{41}&g^{42}&g^{43}&g^{44}
\end{array}
\right]
=
\left[
\begin{array}{c|ccc}
-N^2&0     &0     &0\\
\hline
0   &h^{22}&h^{23}&h^{24}\\
0   &h^{32}&h^{33}&h^{34}\\
0   &h^{42}&h^{43}&h^{44}
\end{array}
\right]
\end{equation}

The 4 dimensional metric can be spilt up into one $ 3 \times 3$ block, and 
two $1 \times  3$ blocks with two of them trivial (zero).
That is
\begin{equation}
g_{\mu\nu} =
\left[\begin{array}{c|c}
-1 / N^2 & 0\\
\hline
 0            &h_{ij}
\end{array}
\right]
\end{equation}
The inverse tensor is very simple
\begin{equation}
g^{\mu\nu} =
\left[\begin{array}{c|c}
-N^2&0\\
\hline
 0            &h^{ij}
\end{array}
\right]
\end{equation}
where $-N^2$ and $h^{ij}$ are two dimensional inverses of the $1 \times 1$ 
matrix and $3 \times 3$ matrices $-1 / N^2$ and $h_{ij}$, respectively.

The Christoffel symbols can be computed from
\begin{equation}
\label{Christ_reverse}
\Gamma^\mu{}_{\sigma \rho} = 
g^{\mu\nu}\{g_{\nu(\sigma,\rho)}-\frac{1}{2}g_{\sigma\rho,\nu}\} 
\end{equation}
and you can spilt them up into little sub-blocks
\begin{equation}
\label{block1}
\Gamma^\mu{}_{\sigma\rho} =
\left[
\begin{array}{c|ccc}
\Gamma^\mu{}_{11}&\Gamma^\mu{}_{12}&\Gamma^\mu{}_{13}&\Gamma^\mu{}_{14}\\
\hline
\Gamma^\mu{}_{12}&\Gamma^\mu{}_{22}&\Gamma^\mu{}_{23}&\Gamma^\mu{}_{24}\\
\Gamma^\mu{}_{13}&\Gamma^\mu{}_{23}&\Gamma^\mu{}_{33}&\Gamma^\mu{}_{34}\\
\Gamma^\mu{}_{14}&\Gamma^\mu{}_{24}&\Gamma^\mu{}_{34}&\Gamma^\mu{}_{44}
\end{array}
\right]
\end{equation}
\newpage

\subsection{Christoffel symbols of the 2 metrics $-1 / N^2$ and $h_{ij}$}
\label{block2}
Some of these blocks should have a simple interpretation in terms of
Christoffel symbols of the lower dimensional metrics $-1 / N^2$ and $h_{ij}$.\\

When $t$ = time and $i,j$ = space,
for example
\noindent
\begin{eqnarray}
\Gamma^t{}_{t t} &=& g^{t\nu}\{g_{\nu(t,t)}-\frac{1}{2}g_{tt,\nu}\}
=g^{tt}\{g_{t(t,t)}-\frac{1}{2}g_{tt,t}\}
\nonumber
\\
&=& g^{tt}\{\frac{1}{2}g_{tt,t}+\frac{1}{2}g_{tt,t}- \frac{1}{2}g_{tt,t}\}
\nonumber
\\
&=&g^{tt}\{\frac{1}{2}g_{tt,t}\}
=-\frac{1}{2}N^2\{(-\frac{1}{N^2})_{,t}\} = - \frac{\dot{N}}{N}
\\
\Gamma^t{}_{i j}
&=& g^{t\nu}\{g_{\nu(i,j)}-\frac{1}{2}g_{ij,\nu}\}
=g^{tt}\{g_{t(i,j)}-\frac{1}{2}g_{ij,t}\}
\nonumber
\\
&=& g^{tt}\{0-\frac{1}{2}h_{ij,t}\}
=N^2\{\frac{1}{2}h_{ij,t}\}
=\frac{1}{2}N^2\{h_{ij,t}\}
\\
\Gamma^t{}_{t i}
&=& g^{t\nu}\{g_{\nu(t,i)}-\frac{1}{2}\,g_{ti,\nu}\}
=g^{tt}\{g_{t(t,i)}-\frac{1}{2}\,g_{ti,t}\}
\nonumber
\\
&=& g^{tt}\{\frac{1}{2}\,g_{tt,i}+\frac{1}{2}\,g_{ti,t}- \frac{1}{2}\,g_{ti,t}\}
\nonumber
\\
&=& g^{tt}\{\frac{1}{2}\,g_{tt,i}\}
=\frac{1}{2}\,N^2\{(-\frac{1}{N^2})_{,i}\} = \frac{\partial_{i} \, N}{N}
\\
\Gamma^i{}_{t t}
&=& g^{i\nu}\{g_{\nu(t,t)}-\frac{1}{2}\,g_{tt,\nu}\}
=g^{ij}\{g_{j(t,t)}-\frac{1}{2}\,g_{tt,j}\}
\nonumber
\\
&=& g^{ij}\{0-\frac{1}{2}\,h_{tt,j}\}
=-\frac{1}{2}\,h^{ij}\{(-\frac{1}{N^2})_{,j}\} = - h^{ij} \, \frac{\partial_{j} \, N}{N}
\\
\Gamma^i{}_{j k}
&=& g^{i\nu}\{g_{\nu(j,k)}-\frac{1}{2}\,g_{jk,\nu}\}
=g^{il}\{g_{l(j,k)}-\frac{1}{2}\,g_{jk,l}\}
\nonumber
\\
&=& h^{il}\{h_{l(j,k)}-\frac{1}{2}\,h_{jk,l}\}
=[\Gamma(h)]^i{}_{j k}
\\
\Gamma^i{}_{j t}
&=& g^{i\nu}\{g_{\nu(j,t)}-\frac{1}{2}\,g_{jt,\nu}\}
=g^{ik}\{g_{k(j,t)}-\frac{1}{2}\,g_{jt,k}\}
\nonumber
\\
&=& g^{ik}\{\frac{1}{2}\,g_{kj,t}+\frac{1}{2}\,g_{kt,j}- \frac{1}{2}\,g_{jt,k}\}
\nonumber
\\
&=&\frac{1}{2}\,g^{ik}\,g_{kj,t} = \frac{1}{2}\,h^{ik}\,h_{kj,t} 
\end{eqnarray}

   \subsection{Collected results for the connection}

There are a total of 6 sub-blocks to deal with:
\begin{eqnarray}
\Gamma^t{}_{t t}
&=& -\frac{1}{2}N^2\{(-\frac{1}{N^2})_{,t}\} = - \frac{\dot{N}}{N} 
\\
\Gamma^t{}_{i j} 
&=& N^2\{\frac{1}{2}h_{ij,t}\}
\\
\Gamma^t{}_{t i}
&=& \frac{1}{2}N^2\{(-\frac{1}{N^2})_{,i}\} = \frac{\partial_{i} \, N}{N}
\\
\Gamma^i{}_{t t}
&=& -\frac{1}{2}h^{ij}\{(-\frac{1}{N^2})_{,j}\} = -h^{ij} \, \frac{\partial_{i} \, N}{N}
\\
\Gamma^i{}_{j k}
&=& [\Gamma(h)]^i{}_{j k}
\\
\Gamma^i{}_{j t}
&=& \frac{1}{2}h^{ik}h_{kj,t}
\end{eqnarray}
\section{Conclusion}
In this section, we analysed a generic spacetime using both ADM and
reverse-ADM (Kaluza-Klein) decompositions which split ``spacetime"
into ``space" plus ``time" in a natural way. We use the split to describe
spacetime curvature in terms of the curvature of space and some additional
structure. Furthermore, this decomposition encodes the four-metric of 
spacetime
in terms of a time-varying three-metric on a three-dimensional domain 
together
with the lapse and shift.

In addition, we used the Gauss-Codazzi and Gauss-Weingarten equations to
decompose the (3+1)-dimensional spacetime curvature tensor in terms of the
3-dimensional spatial curvature tensor, the extrinsic curvature of the time
slice [zero], and the gravitational potential. The ADM and reverse-ADM
decompositions are identical in any static spacetime. In addition, we 
derived
the standard results in equation (\ref{static1}) to (\ref{static3}) for the
spacetime geometry in terms of the three geometry of space on a constant 
time
slice, in which this three geometry is embedded into the
spacetime. Finally, we derived equation
(\ref{Einstein1}) to (\ref{Einstein3}) for the spacetime Einstein tensor.

We consider time dependent but block diagonal metrics in the ADM and
reverse-ADM decompositions. When the ``shift vector" is zero, the ADM and
reverse-ADM decompositions are identical. Futhermore, we defined a quantity
called the extrinsic curvature in equation (\ref{extrinsic}). We are able to
calculate the Christoffel symbols of the first kind and second kind from section
(\ref{systematic}).
Furthermore, we calculated the Riemann tensor by using the formula from
(\ref{Riemann1}) to (\ref{Riemann3}). Whenever the metric $g_{ab}$ block
diagonalizes in the manner $g_{ab} = -c^2\bigoplus h_{ij}$, then even if we
allow arbitrary time dependence we have the relatively simple results as
equation (\ref{result1}) to (\ref{result3}).

For a general stationary spacetime we need to use the full ADM 
decomposition.
However, we have the simplification that all quantities are time
independent. Indeed, the case for ADM decomposition is quite distinct from
the reverse-ADM decomposition. The Christoffel symbols first kind are given in
section (\ref{part2_christ}).
We derived the simplest expression for the Riemann tensor in terms of a coordinate basis.

For the general ADM case coordinate techinques are too complicated to
contemplate (page 514 of \cite{Misner}). Using the inverse-ADM decomposition we can calculate the Christoffel symbols from
equation (\ref{Christ_reverse}). Furthermore, we can split them up into
sub-blocks in equation (\ref{block1}). Some of these blocks can
interpreted in terms of Christoffel symbols of the 2 metrics
$-1 / N^2$ and $h_{ij}$ see (\ref{block2}).

\chapter{Perfect fluid spheres in general relativity}
\label{part_2}
\section{Introduction}
   Perfect fluid spheres are interesting because they are first approximations to any attempt at building a realistic model for a general relativistic star ~\cite{Delgaty,Skea,Martin0,exact}.

Although they represent a real and important astrophysical situation, explicit and full general solutions of the perfect fluid constraints in general relativity have only very recently been developed.

The first static spherically symmetric perfect fluid solution with constant density was found by Schwarzschild in 1918.

He found two exact solutions to Einstein's field equation --- the ``exterior solution" (relevant outside the star) and the ``interior solution" (an approximation to what goes an inside the star). The so-called Schwarzschild solution is amazingly simple, especially in view of the field equations which are very complicated. Indeed, the exterior Schwarzschild solution is not a degenerate case for over-simplified situations but physically most meaningful. It is this solution by means of which one can explain most general relativistic effects in the planetary system. The reason is that it describes the gravitational field outside of a spherically symmetric body like the planets and the sun.

Over the past 90 years a confusing tangle of specific perfect fluid spheres has been discovered, with most of these examples seemingly independent from each other.

Many ``ad hoc" tricks have been explored to solve this differential equation, often by picking special coordinate systems, or making simple ansatze for one or other of the metric components. The big change over the last several years has been the introduction of ``algorithmic" techniques that permit one to generate large classess of perfect fluid spheres in a purely mechanical way. Perfect fluid spheres may be ``simple", but they still provide a surprisingly rich mathematical and physical structure.

In this chapter we will extend these algorithmic ideas, by proving several solution-generating theorems of varying levels of complexity. We shall then explore the formal properties of these solution-generating theorems and then will use these theorems to classify some of the previously known exact solution, and additionally will generate several previously unknown perfect fluid solutions.

\section{Six theorems for generating perfect fluid spheres}
   Perfect fluid spheres, either Newtonian or relativistic, are the first approximations in developing realistic stellar models.
Because of the importance of these models,
fully general solutions of the perfect fluid constraint in general relativity have been developed over the last several years.

The central idea is to start solely with spherical symmetry, which implies that in orthonormal components the stress energy tensor takes the form
\begin{equation}
T_{\hat a\hat b} = \left[ \begin{array}{cccc}
\rho&0&0&0\\ 0&p_r&0&0\\ 0&0&p_t&0\\ 0&0&0&p_t \end{array}\right]
\end{equation}
and then use the perfect fluid constraint $p_r=p_t$. This is making the radial pressure equal to the transverse pressure.
By using the Einstein equations, plus spherical symmetry, the equality  $p_r=p_t$ for the pressures becomes the statement
\begin{equation}
G_{\hat\theta\hat\theta} = G_{\hat r\hat r} = G_{\hat\phi\hat\phi}.
\end{equation}
In terms of the metric components, this leads to an ordinary differential equation [ODE], which then constrains the spacetime geometry, for \emph{any} perfect fluid sphere.

Over the last 90 years, many  ``ad hoc''  approaches to solving this differential equation have been explored. We often solve it by picking special coordinate systems, or making simple ansatze for one or other of the metric components~\cite{Hojman-et-al,Bondi1,Bondi2,Buchdahl1,Buchdahl2,Wyman}. (For recent overviews see ~\cite{Delgaty,Skea,exact}. For a table of ``well-known" perfect fluid solutions see Table \ref{T:well-known} on page \pageref{T:well-known}.)
The big recent change has been the discovery of ``algorithmic" techniques that permit one to generate large classes  of perfect fluid spheres in a purely mechanical way~\cite{Lake,Martin,Rahman}. In addition, an alternative algorithmic approach is used to generate any spherically symmetric static solution of Einstein's equation without recourse to the details of the material that makes it up.



\section*{Solution generating theorems}

\subsection*{Algorithmic techniques}
We would like to present ``Algorithmic" techniques \cite{Martin} that permit one to generate large classes of perfect fluid spheres in a purely mechanical way.
The condition of hydrostatic eqilibrium for \emph{Newtonian} gravity is
\[\frac{\d P_{r}}{\d r} = -g(r)\rho(r) = -\frac{m(r)\rho(r)}{r^2}\] 
where $g(r)$ is the local acceleration due to gravity,
$m(r)$ is the mass contained within a radius $r$
and $\rho(r)$ is the density.

We can simply calculate the mass within a radius $r$ in terms of the density:
\[m(r) = 4\pi \int_0^{r} \overline{r}^2 \rho(\overline{r})\,\d \overline{r}.\]

When we used $\rho(r)$ as a known function, we have to solve a second--order differential equation:
\[\frac{\d ^2 P_{r}}{\d r^2} + \left\{\frac{2}{r}-\frac{1}{\rho}\, \frac{\d \rho}{\d r}\right\}\, \frac{\d P_{r}}{\d r} + 4\pi \rho^2 = 0\]
We can solve  for $P_{r}(r)$ in terms of an integral of the density:
\[ P_{r}(r) = \int \frac{C_1 - 4\pi \int \overline{r}^2\, \rho(\overline{r})\,\d \overline{r}}{r^2} \, \rho(r)\, \d r + C_2.\]

We let $g(r)$ be an arbitrary function to solve simple algebra which allows us to find $m(r)$. The differentiation allows us to find $\rho(r)$ and then $P_r(r)$, respectively:
\[m(r) = r^2 g(r) = 4 \pi \int_0^{r} \overline{r}^2 \rho(\overline{r}) \, \d \overline{r}\]
\[ \Rightarrow \rho(r) = \frac{2 r g(r) + r^2 \frac{\txt \d g}{\txt \d r}}{4 \pi r^2}\]
\[ \Rightarrow P_{r}(r) = - \frac{1}{2 \pi} \int \frac{g(r)^2}{r}\, \d r - \frac{1}{8 \pi} \, g(r)^2 + P_{r}(0).\]
In both the Newtonian and relatistic cases we have to ensure that the density is non-negative. The pressure is finite and non-negative.

\subsection*{Spherical symmetry}

Spherically symmetric means ``having the same symmetry as a sphere". Indeed,    ``sphere" means $S^2$, not spheres of higher dimension.
Rather than the metric on a differentiable manifold, we are concerned with those metrics that have such symmetries. We can characterize symmetries of the metric. By the existence of Killing vectors, which we can see from the definition below.
By demanding spherical symmetry we can simplify the equations of motion considerably. It turns out that for vaccuum spherical symmetry of the metric automatically means the metric is static and asymptotically flat. This is Birkhoff's theorem, a proof of which can be found in ~\cite{D'Inverno}. Furthermore, Spherical symmetry can be defined rigorously in terms of Killing vector fields as follows.
\begin{definition}

A space-time is said to be spherically symmetric if and only if it admits three linearly independent spacelike Killing vector fields $X^a$ whose orbits are closed (i.e. topological circles) and which satisfy
\[
[X^1,X^2] = X^3, \qquad [X^2,X^3] =X^1, \qquad [X^3,X^1] =X^2\]
\end{definition}
Then there exists a coordinate system in which the Killing vectors take on a standard form as expressed in the following result.

\begin{prop}

In a spherically symmetric space-time, there exists a coordinate system $(x^a)$ (called quasi Cartesian) in which the Killing fields $X^a$ are of the form
\[X^0 \to 0 ,\]
\[X^{\alpha} \to \omega^{\alpha}\,_{\beta}\, x^{\beta}, \qquad \omega_{\alpha\beta} = -\omega_{\beta\alpha}.\]
\end{prop}
In addition, we know that what the Killing vectors of $S^2$ are, and that there are three of them. Therefore, a spherically symmetric manifold is one that has three Killing vector fields which are the commutator of the Killing vectors which is the same in either case of those on $S^2$.

\subsection*{Schwarzschild solution}
The gravitational field of a homogenous spherically symmetric body is derived in firstly Newton's and then in Einstein's gravitational theory. The gravitational field of a simple Einsteinian model star consists of the exterior and the interior Schwarzschild solutions. They are joined together at the surface of the star.

The use of arbitrary coordinates is permitted in general relativity. Indeed, the physical significance of statements about tensor or vector components are other quantities which are not always obvious. However, there are some situations where the interpretation is almost as straightforward as in special relativity. The center point of a local inertial coordinate system is the most obvious example. Its coordinate system, where the principle of equivalence comes from, allows one to treat all local quantities exactly as in special relativity. Furthermore, Schwarzschild coordinates for a spherical system turn out to be a second example ~\cite{Misner}.

\subsection*{Schwarzschild geometry}
The geometry of a spherical symmetric vacuum, i.e. vacuum spacetime outside the spherical black hole is the ``Schwarzschild geometry" can be described in terms of the Schwarzschild metric,
\begin{equation}
\label{Schwarzschild line element}
\d s^2 = -\left(1- \frac{2 M}{r}\right)\;\d t^2 + \frac{\d r^2}{1-2 M/r} + r^2 \d \Omega^2,
\end{equation}
which was derived originally as the external field of a static star. The Schwarzschild metric seems to have a singularity at the surface where $r = 2 M$ due to its coordinates, in which space and time change their meanings.

Schwarzschild geometry illustrates clearly the highly non-Euclidean character of spacetime geometry when gravity becomes strong. Furthermore, it illustrates many of techniques one can use to analyze strong gravitational fields. When appropriately truncated, it is the spacetime geometry of a black hole and of a collapsing star as well as of a wormhole~\cite{Misner}.

However, that the Schwarzschild exterior solution is the unique spherically symmetric vacuum solution, is known as \emph{Birkhoff's theorem}. This is a theorem of general relativity which states that all spherical gravitaional fields, whether from a star or from a black hole, are indistinguishable at a large distances. A consequence of this is that purely radial changes in a spherical star do not affect its external gravitational field.

That the Schwarzschild geometry is relevant to gravitational collapse follows from  ``Birkhoff's theorem".

\emph{Birkhoff's theorem:}
\emph{Let the geometry of a given region of spacetime be spherically symmetric, and be a solution to the Einstein field equations in vacuum. Then that geometry is necessarily a piece of the Schwarzschild geometry}~\cite{Misner}.

\begin{figure}[!ht] 
\centering
\includegraphics[scale = 0.6]{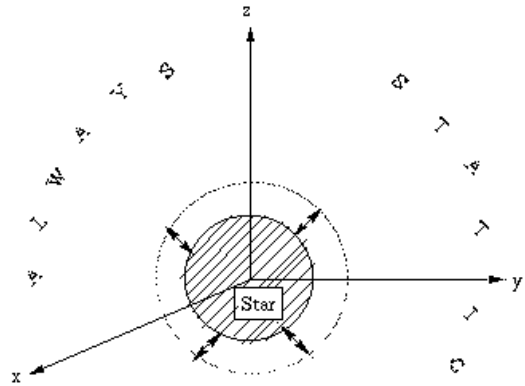}
\caption{\label{Spherical}A Pulsating spherical star cannot emit gravitational waves.}
\end{figure}

In particular, \emph{Birkhoff's theorem} implies that if a spherically symmetric source like a star changes its size, however does so always remaining spherically symmetric, then it cannot propagate any disturbances into the surrounding space. From figure \ref{Spherical}, shows that a pulsating spherically symmetric star cannot emit gravitational waves. If a spherically symmetric source is restricted to the region $r \leqslant a$ for some $a$, then the solution for $r > a$ must be the Schwarzschild solution, or can be called the Schwarzschild exterior solution. However, the converse is not true: a source which gives rise to an exterior Schwarzschild solution is not necessarily spherically symmetric. Some counter-examples are known. Therefore, in general, a source need not inherit the symmetry of its external field.
If we take the limit of equation \ref{Schwarzschild line element} as $ r \to \infty$, then the flat space Minkowski metric of special relativity in spherical polar coordinates is obtained, which is
\begin{equation}
\d s^2 = - \d t^2 + \d r^2 + r^2 ( \d \theta^2 + \sin ^2 \theta \; \d \phi^2).
\end{equation}
This shows that a spherically symmetric vacuum solution is necessarily \emph{asymptotically flat} ~\cite{D'Inverno}.

Firstly, start with some static spherically symmetric geometry in Schwarzschild (curvature) coordinates
\begin{equation} \label{line_element_new}
\d s^2 = - \zeta(r)^2 \; \d t^2 + {\d r^2\over B(r)} + r^2 \;\d\Omega^2
\end{equation}
and then assume it represents a perfect fluid sphere. That is,
$ G_{\hat\theta\hat\theta} = G_{\hat r\hat r} =G_{\hat\phi\hat\phi} $.
While $G_{\hat\theta\hat\theta} = G_{\hat\phi\hat\phi}$ is always fulfilled due to spherical symmetry, we get a nontrivial constraint from $G_{\hat\theta\hat\theta} = G_{\hat r \hat r}$.

We calculate
\begin{equation}
G_{\hat r\hat r} = - \frac{2 B r \zeta' - \zeta + \zeta B}{r^2 \zeta}
\end{equation}
and
\begin{equation}
G_{\hat\theta\hat\theta} = -\frac{1}{2} \frac{B' \zeta + 2 B \zeta' + 2 B r \zeta''  + r \zeta' B'}{r \zeta}
\end{equation}
We set $G_{\hat r\hat r} = G_{\hat\theta\hat\theta}$, which supplies us with an ODE:

\begin{equation} 
\label{ode_for_B}
[r(r\zeta)']B'+[2r^2\zeta''-2(r\zeta)']B + 2\zeta=0 \, .
\end{equation}
This reduces the freedom to choose the two functions in equation (\ref{line_element_new}) to one.

This equation is a first order-linear non-homogeneous equation in $B(r)$. Thus, once you have chosen a $\zeta(r)$ --- this equation can always be solved for $B(r)$.
Solving for $B(r)$ in terms of $\zeta(r)$ is the basis of~\cite{Lake,Martin}, (and is the basis for Theorem 1 below).

On the other hand, we can also re-group this same equation as
\begin{equation}    
\label{ode_for_zeta}
2 r^2 B \zeta'' + (r^2 B'-2rB) \zeta' +(r B'-2B+2)\zeta=0 \,,
\end{equation}
which is a linear homogeneous 2nd order ODE for $\zeta(r)$, which will become the basis for Theorem 2 below. Thus, once you have chosen a  $B(r)$  --- this equation can always be solved for $\zeta(r)$.
The question we are going to answer in this section is, how can one systematically ``deform'' this geometry while still retaining the perfect fluid property?

That is, suppose we start with the specific geometry defined by
\begin{equation}
\d s^2 = - \zeta_0(r)^2 \; \d t^2 + {\d r^2\over B_0(r)} + r^2 \d\Omega^2
\end{equation}
and assume it represents a perfect fluid sphere. In the following section, we will show how to ``deform'' this solution by applying five different transformation theorems on $\left\{ \zeta_0 , B_0  \right\}$,
such that the outcome still presents a perfect fluid sphere. The outcome of this process will depend on one or more free parameters, and so automatically generates an entire family of perfect fluid spheres of which the original starting point is only one member.

In addition, we analyze what happens if we apply these theorems more than once, iterating them in various ways.
We also try to find the connection between all six different transformation theorems.

   \subsection{The first theorem}
The first theorem we present is a variant of a result  first explicitly published in ~\cite{Martin}, though another variant of this result  can also be found in~\cite{exact}.
We first re-phrase the theorem in ~\cite{Martin}. This is slightly different formalism, and demonstrate an independent way of proving it.
\newline

\noindent{\bf Definition (Idempotence):} 
We call a transformation $\gth{}$ idempotent if $\gth{}^2 = \gth{}$. Speaking loosely, we will call a \emph{theorem} idempotent if whenever we apply the transformation more than once, no further solutions will be obtained.

We use our proof to show that the transformation in theorem 1 is (in a certain sense to be made precise below) ``idempotent''. By applying theorem 1 more than once, no new solutions will be found.
Furthermore, the idempotence property of theorem 1 and theorem 2, which we will soon establish, enables us to divide the class of perfect fluid spheres into seed metrics and non-seed metrics.
\begin{theorem} 
Suppose $\{ \zeta_0(r), B_0(r) \}$ represents a perfect fluid sphere.
Define
\begin{equation} \label{Theorem1_m_1}
\Delta_0(r)  =
 \left({ \zeta_0(r)\over  \zeta_0(r) + r  \;\zeta'_0(r)}\right)^2 \; r^2 \; 
\exp\left\{ 2 \int {\zeta'_0(r)\over  \zeta_0(r)} \; 
  { \zeta_0(r)- r\; \zeta'_0(r)\over  \zeta_0(r) + r  \;\zeta'_0(r)} \; \d r\right\}.
\end{equation}
Then for all $\lambda$, the geometry defined by holding $\zeta_0(r)$ fixed and
setting
\begin{equation}
\d s^2 = - \zeta_0(r)^2 \; \d t^2 + {\d r^2\over B_0(r)+\lambda\; \Delta_0(r) }
+ r^2 \d\Omega^2
\end{equation}
is also a perfect fluid sphere. That is, the mapping
\begin{equation}
\gth{1}(\lambda): \left\{ \zeta_0 , B_0  \right\} \mapsto 
\left\{ \zeta_0 , B_0 + \lambda\Delta_0(\zeta_0) \right\}
\end{equation}
takes perfect fluid spheres into perfect fluid spheres. Furthermore a second application of the transformation does not yield new information, $\gth{1} = \gth{1} (\sum \lambda_{i})$ is ``idempotent'', in the sense that
\begin{equation}
\gth{1}(\lambda_n) \circ \cdots \circ \gth{1}(\lambda_2) \circ \gth{1}(\lambda_1) = \gth{1} (\sum \lambda_{i}):
\left\{ \zeta_0 , B_0  \right\} \mapsto 
\left\{ \zeta_0 , B_0 + \left(\sum\nolimits_{i=1}^n \lambda_i \right)\; \Delta_0(\zeta_0) \right\}
\end{equation}
We also note that $\gth{1}$ always has an inverse
\begin{equation}
[\gth{1}(\lambda)]^{-1} = \gth{1}(-\lambda).
\end{equation}
\end{theorem}

\begin{figure}[!ht] 
\centering
\includegraphics[scale = 0.8]{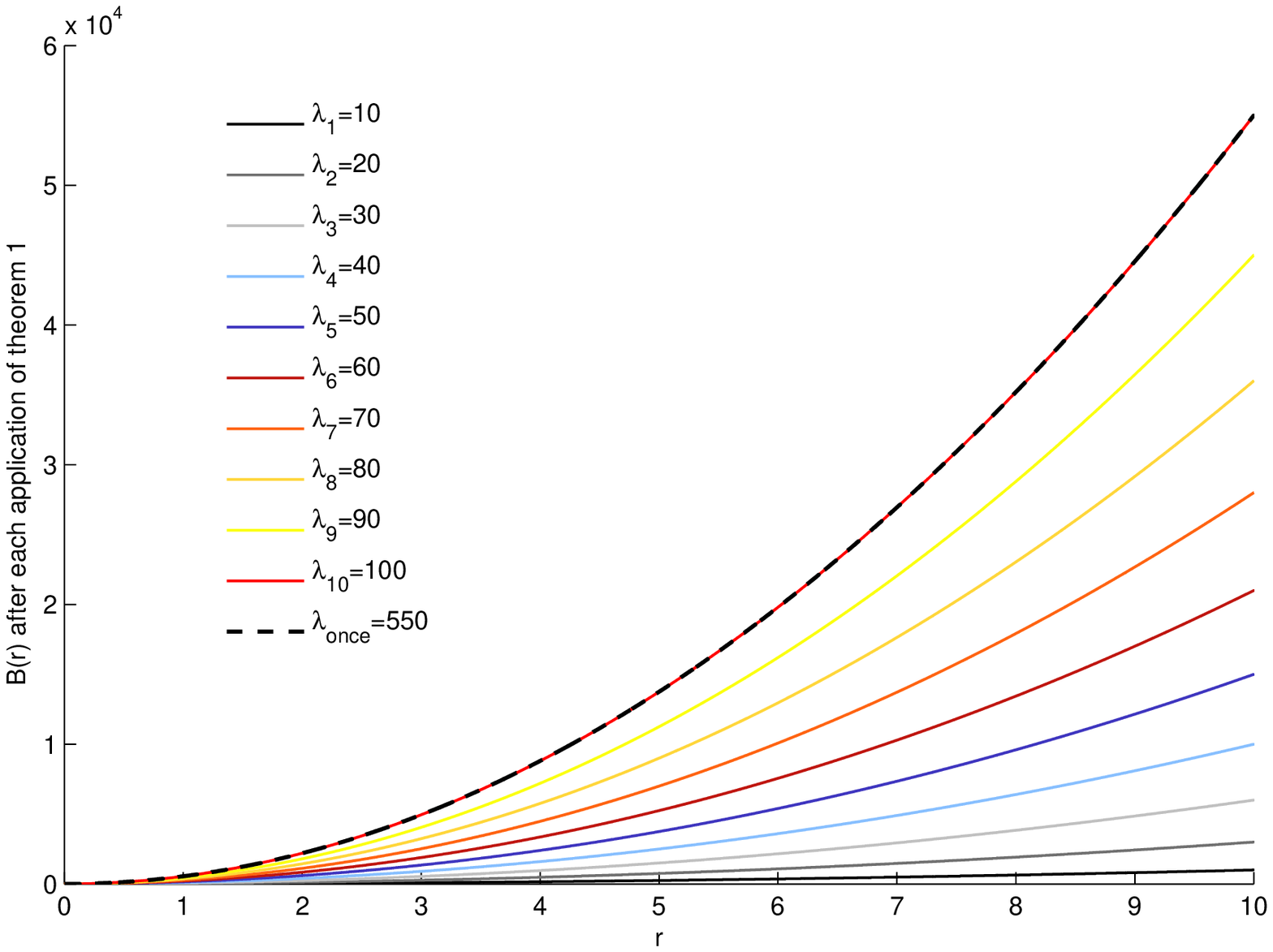}
\caption{\label{Idempotence_Theorem1}The solid lines show $B(r)$ for  $10$ reapplications of Theorem 1 onto the Minkowski metric.
The dashed line corresponds to a single application with a specific choice for $\lambda_{\mathrm{once}}$. 
It can be seen that $10$ applications of Theorem 1 can be expressed by one application.}
\end{figure}

\begin{proof}[Proof for Theorem 1]
Assume that $\left\{ \zeta_{0}(r),B_{0}(r) \right\}$ is a solution for equation (\ref{ode_for_B}).

We want to find under what conditions does $\left\{ \zeta_{0}(r),B_1(r) \right\}$  also satisfy equation (\ref{ode_for_B})?
Without loss of generality, we write
\begin{equation}
B_1(r)=B_{0}(r) + \lambda\;\Delta_{0}(r) \, .
\end{equation}
Equation ({\ref{ode_for_B}}) can now be used
to determine $\Delta_{0}(r)$.
Substitute $B_1(r)$ in equation ({\ref{ode_for_B}})
\[[r(r\zeta_0)'](B_{0} + \lambda\;\Delta_{0})'+[2r^2\zeta_0''-2(r\zeta_0)'](B_{0} + \lambda\;\Delta_{0}) + 2\zeta_0=0 \]
we can also re-group this same equation as
\[[r(r\zeta_0)']B_0'+[2r^2\zeta_0''-2(r\zeta_0)']B_0 + 2\zeta_0 + \left[ r (r \zeta_0 )' \right] \Delta_{0}'
+ \left[ 2 r^2 \zeta_{0}'' - 2 (r \zeta_{0})' \right] \Delta_{0} = 0 \]
That ordinary \emph{inhomogeneous} first-order differential equation in $B$ now
simplifies to
\begin{equation}
\label{ode_th1}
\left[ r (r \zeta_0 )' \right] \Delta_{0}'
+ \left[ 2 r^2 \zeta_{0}'' - 2 (r \zeta_{0})' \right] \Delta_{0} = 0   \, ,
\end{equation}
which is an ordinary \emph{homogeneous} first-order differential equation in $\Delta_{0}$.\\
Now we can calculate $\Delta_{0}$ by following,
\begin{equation}
\left[ r (r \zeta_0 )' \right] \Delta_{0}'
= - \left[ 2 r^2 \zeta_{0}'' - 2 (r \zeta_{0})' \right] \Delta_{0}.
\end{equation}
Rearrange it into
\begin{eqnarray}
\frac{\Delta_{0}'}{\Delta_{0}} &=& \frac{- \left[ 2 r^2 \zeta_{0}'' - 2 (r \zeta_{0})' \right]}{\left[ r (r \zeta_0 )' \right]}
\nonumber
\\
&=& \frac{-2 r \zeta_{0}''}{(r \zeta_0 )'} + \frac{2}{r}
\end{eqnarray}

A straightforward calculation, including an integration by parts, leads to
\begin{equation}
\Delta_0(r) = \frac{r^2 }{\left[ (r \zeta_{0})' \right]^2}  \; \exp\left\{{\int{\frac{4 \zeta_{0}'}{(r \zeta_{0})'} dr } }\right\} \, .
\end{equation}

We can also write this same equation as
\begin{eqnarray}
\Delta_0(r) &=& \frac{r^2 }{\left[ (r \zeta_{0})' \right]^2}  \; \exp\left\{{4\int{\frac{ \zeta_{0}'}{\zeta_{0}}\;\frac{\zeta_{0}}{(\zeta_{0} + r \zeta_{0}')} dr } }\right\} \,
\nonumber
\\
&=& \frac{r^2 }{\left[ (r \zeta_{0})' \right]^2}  \; \exp\left\{{2\int{\frac{ \zeta_{0}'}{\zeta_{0}}\;\frac{\zeta_{0}}{(\zeta_{0} + r \zeta_{0}')} dr }+2\int{\frac{ \zeta_{0}'}{\zeta_{0}}\;\frac{\zeta_{0}}{(\zeta_{0} + r \zeta_{0}')} dr }}\right\} \, .
\end{eqnarray}
Adding and subtracting 
$\pm 2 (r \zeta_{0}')/(\zeta_{0} (r \zeta_{0})')$ to the argument under the integral leads to
\begin{eqnarray}
\Delta_0(r) &=&  \frac{r^2 }{\left[ (r \zeta_{0})' \right]^2}  \; \exp\left\{2\int{\frac{ \zeta_{0}'}{\zeta_{0}}\;\frac{\zeta_{0}}{(\zeta_{0} + r \zeta_{0}')} dr }-2\int{\frac{ \zeta_{0}'}{\zeta_{0}}\;\frac{r \zeta_{0}'}{(\zeta_{0} + r \zeta_{0}')} dr } \right\}
\nonumber
\\
&&+\exp\left\{2\int{\frac{ \zeta_{0}'}{\zeta_{0}}\;\frac{\zeta_{0}}{(\zeta_{0} + r \zeta_{0}')} dr }+2\int{\frac{ \zeta_{0}'}{\zeta_{0}}\;\frac{r \zeta_{0}'}{(\zeta_{0} + r \zeta_{0}')} dr } \right\} \, .
\nonumber
\\
&=& \frac{r^2 }{\left[ (r \zeta_{0})' \right]^2}  \; \exp\left\{{2\int{\frac{ \zeta_{0}'}{\zeta_{0}}\;\frac{(\zeta_{0} - r \zeta_{0}')}{(\zeta_{0} + r \zeta_{0}')} dr }+2\int{\frac{ \zeta_{0}'}{\zeta_{0}}\;\frac{(\zeta_{0} + r \zeta_{0}')}{(\zeta_{0} + r \zeta_{0}')} dr }}
\right\}\, .
\end{eqnarray}
We can simplify to,
\[
= \frac{r^2 }{\left[ (r \zeta_{0})' \right]^2}  \; 
\exp\left\{
2\int{ \frac{ \zeta_{0}'}{\zeta_{0}}\;
\frac{(\zeta_{0} - r \zeta_{0}')}{(\zeta_{0} + r \zeta_{0}')} dr }+
2\int\frac{ \zeta_{0}'}{\zeta_{0}} \; dr
\right\} \, .
\]
Furthermore,
\[= \frac{r^2 }{\left[ (r \zeta_{0})' \right]^2}  \; \exp\left\{2\int{\frac{ \zeta_{0}'}{\zeta_{0}}\;\frac{(\zeta_{0} - r \zeta_{0}')}{(\zeta_{0} + r \zeta_{0}')} dr }\right\} \,\; \exp(\ln(\zeta_{0}^2)) .
\]
Note that, $\exp(\ln(\zeta_{0}^2)) = \zeta_{0}^2$, so we can calculate

\begin{equation}
\Delta_{0}
= \left({ \zeta_0(r)\over  \zeta_0(r) + r  \;\zeta'_0(r)}\right)^2 \; r^2 \; 
\exp\left\{ 2 \int {\zeta'_0(r)\over  \zeta_0(r)} \; 
  { \zeta_0(r)- r\; \zeta'_0(r)\over  \zeta_0(r) 
   + r  \;\zeta'_0(r)} \; \d r\right\}\, ,
\end{equation}
as previously illustrated.

If we apply this transformation a second time we obtain no additional information. To see this, consider the sequence
\begin{equation}
\{\zeta_0,B_0\} \mapsto \{\zeta_0,B_1\} \mapsto \{\zeta_0,B_2\} \dots
\end{equation}
But at the second step (and all subsequent steps), since $\zeta_0$ has not changed,  so $\Delta_1(r)=\Delta_0(r)$. More generally, at all stages of the iteration $\Delta_i(r)=\Delta_0(r)$. We can write this as
\begin{equation}
\prod_{i=1}^n \gth{1}(\lambda_i) = \gth{1}\left(\sum_{i=1}^n \lambda_i \right).
\end{equation}
or in the more suggestive form
\begin{equation}
\prod_{i=1}^n \gth{1}  \SIM \gth{1}
\end{equation}
where the symbol $\SIM$ indicates ``\emph{equality up to relabelling of the parameters}''. 
That is, transformation $\gth{1}$ is ``idempotent'' up to relabelling of the parameters (see figure \ref{Idempotence_Theorem1}).  
\end{proof}

A version of Theorem 1  can also be found in~\cite{exact}. 
Specifically, after several manipulations, changes of notation, and a change of coordinate system, the transformation exhibited in equation (16.11) of~\cite{exact} can be  converted into Theorem 1 above.

Applying theorem 1 to a fixed $\{\zeta_0,B_0\}$ generates a one dimensional space of perfect fluid spheres, which leads to the corollary below:
\begin{corollary}
Let $\{\zeta_0,B_a\}$ and  $\{\zeta_0,B_b\}$ both represent perfect fluid spheres, then for all $p$
\begin{equation}
\left\{ \zeta_0, p B_a + (1-p) B_b\right\}
\end{equation}
is also a perfect fluid sphere, furthermore all perfect fluid spheres for a fixed $\zeta_0$ can be written in this form.
\end{corollary}

\begin{proof}[Proof]
The result is automatic once you note that for fixed $\zeta_0$ the ODE for $B$ is linear, (though not homogeneous, which is why the two coefficients $p$ and $1-p$ are chosen to add up to 1) .
\end{proof}
We defer extensive discussion of the application of this theorem and its corollary until section \ref{S:classify}, where we use this and our other generating theorems as a basis for classifying perfect fluid spheres. At this stage we mention, only as a very simple example, that $\gth{1}$ applied to Minkowski space results in the geometry of the Einstein static universe.
\begin{figure}[!ht] 
\centering
\includegraphics[scale = 0.5]{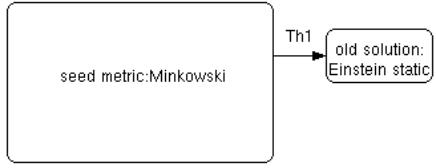}
\caption{\label{Structure graph for theorem1}This structure shows that when we apply theorem 1 to Minkowski, we get an Einstein static universe as a new metric.}
\end{figure}
   \subsection{The second theorem}
   The second theorem we present is a new transformation theorem. This is a different formalism, and demonstrates our independent way of proving it. Furthermore, we would like to show the ``idempotence" property of theorem 2 in this section.
\begin{theorem} 
Let $\{\zeta_0,B_0\}$ describe a perfect fluid sphere.
Define
\begin{equation}
Z_0(r) = \sigma +\epsilon \lint {r \; \d r\over  \zeta_0(r)^2\; \sqrt{B_0(r)} }.
\end{equation}
Then for all $\sigma$ and $\epsilon$, the geometry defined by holding $B_0(r)$ fixed and setting
\begin{equation}
\d s^2 = - \zeta_0(r)^2 \; Z_0(r)^2
\; \d t^2 + {\d r^2\over B_0(r)} + r^2 \d\Omega^2
\end{equation}
is also a perfect fluid sphere. That is, the mapping
\begin{equation}
\gth{2}(\sigma,\epsilon): \left\{ \zeta_0, B_0  \right\} \mapsto \left\{ \zeta_0 \; Z_0(\zeta_0,B_0), B_0 \right\}
\end{equation}
takes perfect fluid spheres into perfect fluid spheres. Furthermore a second application of the transformation does not yield new information, $\gth{2}$ is ``idempotent'' in the sense that
\begin{equation}
\gth{2}(\sigma_n,\epsilon_n) \circ \cdots \circ  \gth{2}(\sigma_3,\epsilon_3) \circ  \gth{2}(\sigma_2,\epsilon_2) \circ  \gth{2}(\sigma_1,\epsilon_1)  = 
\gth{2}(\sigma_n\dots \sigma_3\sigma_2\sigma_1, \; \epsilon_{n\dots321}),
\end{equation}
where
\begin{eqnarray}
\epsilon_{n\dots321} &=& (\epsilon_1\sigma_2\sigma_3 \cdots \sigma_n)
+ (\sigma_1^{-1} \epsilon_2 \sigma_3 \cdots \sigma_n)
+ (\sigma_1^{-1}\sigma_2^{-1} \epsilon_3 \cdots \sigma_n)
\nonumber
\\
&&
+ \cdots 
+ (\sigma_1^{-1} \sigma_2^{-1} \sigma_3^{-1} \cdots \epsilon_n).
\end{eqnarray}
Furthermore,  theorem 2 is invertible (as long as $\sigma\neq0$):
\begin{equation}
\left[\gth{2}(\sigma,\epsilon)\right]^{-1} = \gth{2}(1/\sigma,-\epsilon).
\end{equation}

\end{theorem}

\noindent{\bf Reduction of order:}

This is a method to find a general solution to a linear differential equation, provided you already have one particular solution. In this method, we start with an $n^{th}$--order linear differential equation.
This method is especially useful in solving a second--order linear differential equation. It reduces the problem to one for solving a first--order linear differential equation.
\begin{proof}[Proof for Theorem 2]
The proof of theorem 2 is based on the technique of ``reduction in order''. 
Assuming that $\left\{ \zeta_{0}(r),B_{0}(r) \right\}$  solves equation (\ref{ode_for_zeta}),
write 
\begin{equation}
\zeta_1(r)=\zeta_{0}(r) \; Z_0(r) \, .
\end{equation}
and demand that $\left\{ \zeta_{1}(r),B_{0}(r) \right\}$ also solves equation (\ref{ode_for_zeta}).
We find
\begin{equation}
2 r^2 B_0 (\zeta_0 \; Z_0)'' + (r^2 B_0'-2rB_0) (\zeta_0 \; Z_0)' +(r B_0'-2B_0+2)(\zeta_0 \; Z_0)=0
\end{equation}
we can expand the above equation to
\begin{eqnarray}
&&2 r^2 B_0 (\zeta_0'' \; Z_0 + 2 \zeta_0' \; Z_0' + \zeta_0 \; Z_0'') + (r^2 B_0'-2rB_0) ( \zeta_0' \; Z_0 + \zeta_0 \; Z_0')
\nonumber
\\
&&+(r B_0'-2B_0+2)(\zeta_0 \; Z_0)=0
\end{eqnarray}
we can also re-group this same equation as
\begin{eqnarray}
&&\left\{2 r^2 B_0 \zeta_0'' + (r^2 B_0'-2rB_0) \zeta_0' +(r B_0'-2B_0+2)\zeta_0 \right\} Z_0 
\nonumber
\\
&&+ (r^2 \zeta_0 B_0' + 4 r^2 \zeta_0' B_0 - 2 r \zeta_0 B_0) Z_0' + (2 r^2 \zeta_0 B_0) Z_0''  =0
\end{eqnarray}
This linear homogeneous 2nd order ODE for $Z_0$ now simplifies to
\begin{equation}
\label{ode_th2}
(r^2 \zeta_0 B_0' + 4 r^2 \zeta_0' B_0 - 2 r \zeta_0 B_0) Z_0' + (2 r^2 \zeta_0 B_0) Z_0''  =0 \, ,
\end{equation}
which is
an ordinary homogeneous second-order differential equation, depending only on $Z_0'$
and $Z_0''$. (So it can be viewed as a first-order homogeneous order differential equation in $Z'$, which is solvable.)
Separating the unknown variable to one side,
\begin{equation} \label{de_for_zetaprime}
\frac{Z_0''}{Z_0'}=-\frac{1}{2} \frac{B_0'}{B_0} - 2 \frac{\zeta_0'}{\zeta_0} + \frac{1}{r} \, .
\end{equation}
Re-write $Z_0''/ Z_0' = \d\ln(Z_0')/\d t$, and integrate twice over both sides of
equation (\ref{de_for_zetaprime}), to obtain
\begin{equation} \label{eq_for_zeta_1}
Z_0=\sigma + \epsilon \int{\frac{r \, dr}{\zeta_0(r)^2 \;\sqrt{B_0(r)} }}  \, ,
\end{equation} 
depending on the old solution $\left\{ \zeta_0 (r) , B_0 (r)  \right\}$, and two
arbitrary integration constants $\sigma$ and $\epsilon$.

To see that the transformation $\gth{2}$ defined in Theorem 2 is ``idempotent'' we first show
\begin{equation}
\gth{2} \circ\gth{2} \SIM \gth{2},
\end{equation}
and then iterate.
The precise two-step composition rule is
\begin{equation}
\gth{2}(\sigma_2,\epsilon_2)\circ\gth{2}(\sigma_1,\epsilon_1) = 
\gth{2}\left(\sigma_2\sigma_1, \;\epsilon_1\sigma_2+{\epsilon_2\over\sigma_1} \right).
\end{equation}

In this section we consider a two-step composition rule to find idempotence. In particular, the most outstanding features of these steps are:
\begin{itemize}
\item 
To see ``idempotence'', note that for fixed $B_0(r)$ equation (\ref{ode_for_zeta}) has a solution space that is exactly two dimensional.
\item
Since the first application of  $\gth{2}$  takes any specific solution and maps it into the full two-dimensional solution space, any subsequent application of  $\gth{2}$ can do no more than move one around inside this two dimensional solution space --- physically this corresponds to a relabelling of parameters describing the perfect fluid metric you are dealing with, not the generation of new solutions.
\end{itemize}
To be more explicit about this note that at step one
\begin{equation}
\zeta_0 \to \zeta_1 = \zeta_0 \; \left\{
\sigma_1 +\epsilon_1 \lint {r \; \d r\over  \zeta_0(r)^2\; \sqrt{B_0(r) }}
\right\},
\end{equation}
while at the second step
\begin{equation}
\zeta_1 \to \zeta_2 = \zeta_1 \; \left\{
\sigma_2 +\epsilon_2 \lint {r \; \d r\over  \zeta_1(r)^2\; \sqrt{B_0(r) }}
\right\}.
\end{equation}
That is:
\begin{eqnarray}
\zeta_2 &=& \zeta_0 \;  \left\{
\sigma_1 +\epsilon_1 \lint {r \; \d r\over  \zeta_0(r)^2\; \sqrt{B_0(r)}} 
\right\}
\\
&&
\times \left\{
\sigma_2 +\epsilon_2 \lint {r \; \d r\over  \zeta_0(r)^2\; \sqrt{B_0(r)} \;  \;
[
\sigma_1 +\epsilon_1 \int {r \; \d r /( \zeta_0(r)^2\; \sqrt{B_0(r) })}
]^2  }
\right\}.
\nonumber
\end{eqnarray}
But this can be rewritten as
\begin{eqnarray}
\zeta_2 &=& \zeta_0 \; Z_0 \; 
\left\{ \sigma_2 + {\epsilon_2\over\epsilon_1} \lint { \d Z_0\over Z_0^2 } \right\}
\nonumber
\\
&=&
\zeta_0 \; Z_0 \; 
\left\{ \sigma_2 - {\epsilon_2\over\epsilon_1} \int { \d \left({1\over Z_0}\right) } \right\}
\nonumber
\\
&=&
\zeta_0 \; Z_0 \; 
\left\{ \sigma_2 - {\epsilon_2\over\epsilon_1}  \left[{1\over Z_0 }-{1\over \sigma_1}\right] \right\}.
\end{eqnarray}
Therefore
\begin{equation}
\zeta_2
=
\zeta_0 \;  \left\{  - {\epsilon_2\over\epsilon_1}  + 
\left[\sigma_2+{\epsilon_2\over\epsilon_1} \;{1\over \sigma_1}\right] \; Z_0 \right\}.
\end{equation}
That is
\begin{eqnarray}
Z_1 &=&  - {\epsilon_2\over\epsilon_1}  
+ \left[\sigma_2+{\epsilon_2\over\epsilon_1} \;{1\over \sigma_1}\right]  \; Z_0,
\\
&=& \sigma_2 \sigma_1 + \left( \sigma_2 \epsilon_1 +{\epsilon_2 \over \sigma_1}\right) \lint {r \, \d r \over \zeta_{0}(r)^2 \sqrt{B_0(r)}}
\end{eqnarray}
from which the composition law 
\begin{equation}
\gth{2}(\sigma_2,\epsilon_2)\circ\gth{2}(\sigma_1,\epsilon_1) = 
\gth{2}\left(\sigma_2\sigma_1, \;\epsilon_1\sigma_2+{\epsilon_2\over\sigma_1} \right)
\end{equation}
follows immediately. (Note that the composition law for $\gth{2}$ is actually a statement about applying reduction of order to 2nd-order ODEs, it is not specifically a statement about perfect fluid spheres, though that is how we will use it in the thesis).
The general composition law then follows by induction. To find the inverse transformation we choose $\sigma_2=1/\sigma_1$ and $\epsilon_1=-\epsilon_2$, for which
\begin{equation}
\label{E:composition2}
\gth{2}(1/\sigma_1,-\epsilon_1)\circ\gth{2}(\sigma_1,\epsilon_1)= 
\gth{2}\left(1, \;0\right) =  \mathbf{I}.
\end{equation}
\end{proof}

\begin{figure}[!ht]
\centering
\includegraphics[scale = 0.65]{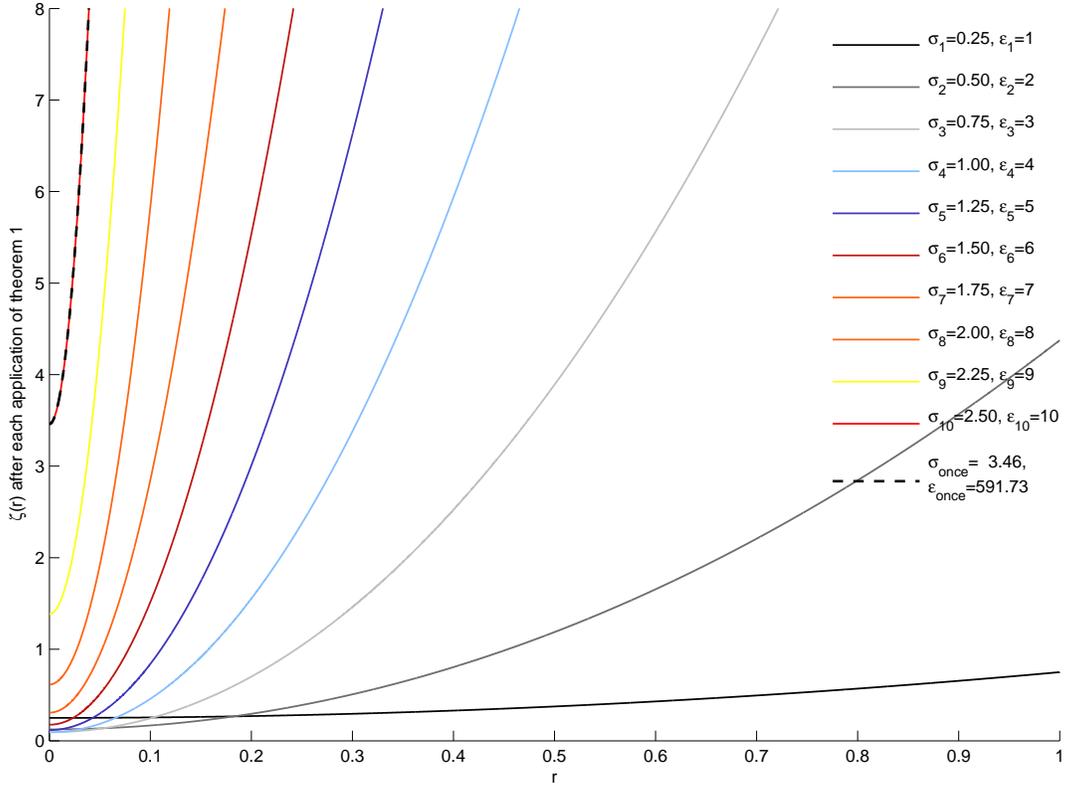}
\caption{\label{Idempotence_Theorem2}The solid lines show $\zeta(r)$ for  $10$ reapplications of Theorem 2 onto the Minkowski metric.
The dashed line corresponds to a single application with a specific choice for $\sigma_{\mathrm{once}}$ and $\epsilon_{\mathrm{once}}$. 
It can be seen that $10$ applications of Theorem 2 can be expressed by one application.}
\end{figure}

Comment: As other special cases of the composition law we will also mention the results that
\begin{equation}
\prod_{i=1}^n \gth{2}(1,\epsilon_i)= \gth{2}\left(1,\sum_{i=1}^n \epsilon_i \right),
\end{equation}
and
\begin{equation}
\gth{2}(\sigma,\epsilon)^n= 
\gth{2}\left(\sigma^n,\epsilon\;
\left[\sigma^{n-1}+\sigma^ {n-3} \cdots +\sigma^{-(n-3)}+\sigma^{-(n-1)}\right] 
\right).
\end{equation}
Now as long as $\sigma>1$ then for sufficiently large $n$ we see
\begin{equation}
\gth{2}(\sigma,\epsilon)^n \approx
\gth{2}\left(\sigma^n, \sigma^{n-1} \epsilon \right) = \sigma^{n-1} \; \gth{2}(\sigma,\epsilon) \SIM \gth{2}(\sigma,\epsilon),
\end{equation}
where at the last step we have used the fact that the overall multiplicative factor $\sigma^{n-1}$ can simply be reabsorbed into a redefinition of the time coordinate. Because of this result, we see that for fixed $\sigma>1$ (and fixed but arbitrary $\epsilon$) repeated numerical applications of $\gth{2}(\sigma,\epsilon)$ will have a well-defined limit. In figure \ref{Idempotence_Theorem2} we have tested the composition law numerically.

A strictly limited version of theorem 2,  with little comment on its importance, can be found in~\cite{exact}.
Specifically, after several manipulations, changes of notation, and a change of coordinate system, the transformation exhibited in equation (16.12) of~\cite{exact} can be seen to be equivalent to the sub-case $\sigma=0$, $\epsilon=1$ of theorem 2 above.

For some purposes it is more useful to rephrase theorem 2 as below:
\begin{corollary}
Let $\{\zeta_a,B_0\}$ and  $\{\zeta_b,B_0\}$ both represent perfect fluid spheres, then for all $p$ and $q$
\begin{equation}
\left\{ p \;\zeta_a + q \;\zeta_b, B_0\right\}
\end{equation}
is also a perfect fluid sphere. Furthermore, for fixed $B_0$ all perfect fluid spheres can be written in this form.
\end{corollary}

\begin{proof}[Proof]
The result is automatic once you note that for fixed $B_0$ the 2nd order ODE for $\zeta$ is linear and homogeneous.
\end{proof}

We defer extensive discussion about the application of these theorems and corollary until section \ref{S:classify}, which cover the following:
\begin{itemize}
\item
A simple example involving $\gth{2}$ applied to either the Einstein static universe or the anti-de Sitter universe results in Schwarzschild's stellar solution (position-independent density).
\item
Similarly, corollary 2 applied to a combination of the Einstein static universe and anti-de Sitter space is another way of obtaining Schwarzschild's stellar solution.
\end{itemize}
Note that when comparing the results of our computations with the extant literature we shall adopt the naming conventions of the review article by Delgaty and Lake \cite{Delgaty}. For example, some of the less standard names are:
\begin{itemize}
\item
K-O III: The third spacetime considered by Korkina and Orlyanskii.
\item
M-W III: The third spacetime considered by Matese and Whitman.
\item
B-L: The spacetime considered by Buchdahl and Land.
\item
Kuch: One of several spacetimes considered by Kuchowicz.
\item
Heint: One of several spacetime considered by Heintmann.
\end{itemize}
References to the primary literature can be found in Delgaty and Lake \cite{Delgaty}. We also add several new abbreviations:
\begin{itemize}
\item
Martin: One of several spacetimes considered by Martin and Visser.
\item
S1: A special name for a particularly simple spacetime.
\item
P1-P8: Special names for several apparently new perfect fluid spacetimes.
\end{itemize}
See in particular tables (3.2) and (3.3) for the spacetime metrics corresponding to these geometries.
\begin{figure}[!ht] 
\centering
\includegraphics[scale = 0.5]{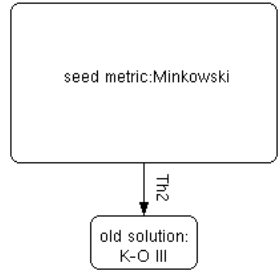}
\caption{\label{Structure graph for theorem2}This structure shows that when we apply theorem 2 to Minkowski, we get the K-O III geometry.}
\end{figure}
   \subsection{The third and fourth theorems}
Having now found the first and second generating theorems, is now useful to define two new theorems by composing them. What we are going to show in the following section is how to deform the geometry while still retaining the perfect fluid property.
To further illustrate the formalism, we will show how to classify perfect fluid spheres by using two new generating theorems.
We present theorem 3 by taking a perfect fluid sphere solution $\left\{ \zeta_0, B_0  \right\}$ and applying theorem 1 onto it which gives us a new perfect fluid sphere $\left\{ \zeta_0, B_1 \right\}$. The new $B_1$ is given in
equation (\ref{Theorem1_m_1}). We now continue with applying theorem 2, again we get a new solution $\{ \tilde\zeta, B_1 \}$, where $\tilde{\zeta}$ now depends on the new $B_1$.
All together we can consider this as a single process by introducing the following theorem:\\
\begin{theorem}
If $\{\zeta_0,B_0\}$ denotes a perfect fluid sphere, then for all $\sigma$,  $\epsilon$, and $\lambda$, the three-parameters geometry defined by
\begin{equation}
\d s^2 = - \zeta_0(r)^2 
\left\{
\sigma+\epsilon \lint {r \; \d r\over  \zeta_0(r)^2\; \sqrt{B_0(r)+\lambda \;\Delta_0(r)}} 
\right\}^2
\; \d t^2 + {\d r^2\over B_0(r) +\lambda \Delta_0(r)   }
+ r^2 \d\Omega^2
\end{equation}
is also a perfect fluid sphere, where $\Delta_0$ is 
\begin{equation}
\Delta_0(r) =
\left({ \zeta_0(r)\over  \zeta_0(r) + r  \;\zeta'_0(r)}\right)^2 \; r^2 \; 
\exp\left\{ 2 \int {\zeta'_0(r)\over  \zeta_0(r)} \; 
  { \zeta_0(r)- r\; \zeta'_0(r)\over  \zeta_0(r) + r  \;\zeta'_0(r)} \; \d r\right\} \, .
\end{equation}
That is
\begin{eqnarray}
\gth{3} &=& \gth{2}\circ\gth{1}: \{\zeta_0,B_0\} \mapsto \{\zeta_0,B_0+\lambda\;\Delta_0(\zeta_0)\} 
\nonumber
\\
&\mapsto&
\{\zeta_0 \; Z_0(\zeta_0,B_0+\lambda\Delta_0(\zeta_0)),B_0+\lambda\Delta_0(\zeta_0)\}. 
\end{eqnarray}
\end{theorem}

Now we would like to present another new theorem. Instead of starting with theorem 1 we could first apply theorem 2 on $\left\{ \zeta_0, B_0  \right\}$.
This gives us a new perfect fluid sphere $\left\{ \zeta_1, B_0  \right\}$, 
where $\zeta_1=\zeta_0 \; Z_0(\zeta_0,B_0)$ is given by equation (\ref{eq_for_zeta_1}). We now continue with theorem 1 which leads to $\{ \zeta_1, \tilde{B}  \}$ where $\tilde{B}$ depends on the new $\zeta_1$. Again, we can consider this as a single process by introducing the following theorem:\\

\begin{theorem}
If $\{\zeta_0,B_0\}$ denotes a perfect fluid sphere, then for all $\sigma$,  $\epsilon$, and $\lambda$, the three-parameters geometry are defined by
\begin{equation}
\d s^2 = - \zeta_0(r)^2 
\left\{
\sigma+\epsilon \lint {r \; \d r\over  \zeta_0(r)^2\; \sqrt{ B_0(r)   }} 
\right\}^2
\; \d t^2 + {\d r^2\over B_0(r) + \lambda \Delta_0(\zeta_1,r)  }
+ r^2 \d\Omega^2
\end{equation}
is also a perfect fluid sphere, where $\Delta_0(\zeta_1,r)$ is defined as
\begin{equation}
\Delta_0(\zeta_1,r)=
\left({ \zeta_1(r)\over  \zeta_1(r) + r  \;\zeta_1'(r)}\right)^2 \; r^2 \; 
\exp\left\{ 2 \int {\zeta_1'(r)\over  \zeta_1(r)} \; 
  { \zeta_1(r)- r\; \zeta_1'(r)\over  \zeta_1(r) + r  \;\zeta'_1(r)} \; \d r\right\} \, ,
\end{equation}
depending on $\zeta_1 = \zeta_0\;  Z_0$, where as before
\begin{equation}
Z_0(r) = 
\sigma+\epsilon \lint {r \; \d r\over  \zeta_0(r)^2\; \sqrt{ B_0(r) }}  \, .
\end{equation}
That is
\begin{eqnarray}
\gth{4} &=& \gth{1}\circ\gth{2}: \{\zeta_0,B_0\} \mapsto \{\zeta_0 \;Z_0(\zeta_0,B_0),B_0\}
\nonumber
\\
&\mapsto&
\{\zeta_0 \; Z_0(\zeta_0,B_0),B_0+\lambda \Delta_0(\zeta_0 \; Z_0(\zeta_0,B_0))\}.
\end{eqnarray}

\end{theorem}
In addition, we analyze what happens, if we apply theorems 3 or 4 more than once.
Some general comments about both new transformation theorems:
\begin{itemize}
\item
Theorem 3 and theorem 4 are in general distinct, which can be traced back to the fact that theorem 1 and theorem 2 do not in general commute.
\item
Theorem 3 and theorem 4 are in general not idempotent. That means when we apply theorem 3 more than once further distinct solutions will be obtained. Similarly for theorem 4.
\end{itemize}
In addition, these comments can be described in terms of the equations below.
\begin{equation}
\gth{3} \not\SIM \gth{4}; \qquad \gth{3} \circ \gth{3} \not\SIM \gth{3}; \qquad  \gth{4} \circ \gth{4} \not\SIM \gth{4}.
\end{equation}
The best way to verify this is to try a few specific examples. There may be some specific and isolated special metrics for which theorem 3 and theorem 4 happen to be degenerate, or idempotent, and finding such metrics is important for our classification programme.
\begin{figure}[!ht] 
\centering
\includegraphics[scale = 0.6]{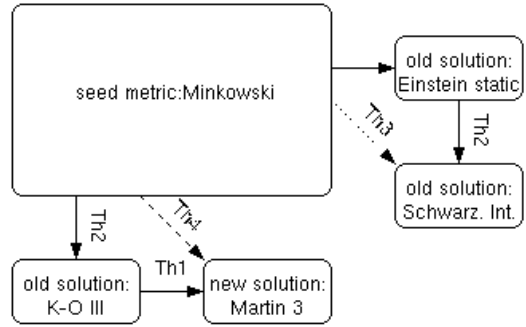}
\caption{\label{Structure graph for theorem3 and 4}This structure shows that Theorem 3 and theorem 4 are in general distinct. When we apply theorem 3 to Minkowski, we get the Schwarzschild Interior geometry, while applying theorem 4 to the Minkowski, we get Martin 3 .}
\end{figure}
\subsection{Formal properties of the generating theorems}
The solution generating theorems we have developed interact in a number of interesting ways and exhibit numerous formal properties that will be useful when classifying generic perfect fluid spheres.
To start with, theorem 3 $(\gth{3})$ and theorem 4 $(\gth{4})$ can be expressed in terms of theorem 1 $(\gth{1})$ 
and theorem 2 $(\gth{2})$:
\begin{equation}
\begin{split}
\gth{3} (g) :=& \left( \gth{2} \circ \gth{1} \right) g = \gth{2}\left(\gth{1}(g) \right) \, ; \\
\gth{4} (g) :=& \left( \gth{1} \circ \gth{2} \right) g = \gth{1}\left(\gth{2}(g) \right) \, .\\
\end{split} 
\end{equation}
where $g$ is a metric representing a perfect fluid sphere. 
Having this in mind, some of the new solutions generated by starting from some specific solution can be identified. For example:
\begin{equation}
\gth{4} \circ \gth{1}
\equiv \gth{1} \circ \gth{2} \circ \gth{1}
 \equiv \gth{1} \circ \gth{3}  \, ,
\end{equation}
or
\begin{equation}
\gth{3} \circ \gth{3}
\equiv \gth{2} \circ \gth{1} \circ \gth{2} \circ \gth{1}
 \equiv \gth{2} \circ \gth{4} \circ \gth{1}  \, .
\end{equation}
The idempotence of $\gth{1}$ and $\gth{2}$ in this formalism is:
\begin{equation}
\begin{split}
\left( \gth{1} \circ \gth{1} \right) g =& \gth{1}\left(\gth{1}(g) \right) \SIM \gth{1}(g) \, ;\\
\left( \gth{2} \circ \gth{2} \right) g =& \gth{2}\left(\gth{2}(g) \right) \SIM \gth{2}(g) \, .\\
\end{split} 
\end{equation}
Taken together, it is possible to simplify all formulae wherever $\gth{1}$ and $\gth{2}$
appear next to each other more than once. The following examples should demonstrate how this works:
\begin{equation}
\begin{split}
\left( \gth{2} \circ \gth{3} \right) g &\equiv \left( \gth{2} \circ \gth{2} \circ \gth{1} \right) g 
\SIM \left( \gth{2} \circ \gth{1} \right) g
\equiv \gth{3} (g) \, ; 
\\
\left( \gth{1} \circ \gth{4} \right) g &\equiv \left( \gth{1} \circ \gth{1} \circ \gth{2} \right) g 
\SIM  \left( \gth{1} \circ \gth{2} \right) g
\equiv \gth{4} (g) \, , \\
\end{split}
\end{equation}
and in the same way
\begin{equation}
\begin{split}
 \gth{4} \circ \gth{3}  &\equiv  \gth{1} \circ \gth{2} \circ \gth{2} \circ \gth{1}  
\SIM  \gth{1} \circ \gth{2}  \circ \gth{1} 
\equiv \gth{1} \circ \gth{3}  \equiv \gth{4} \circ \gth{1} \, ; 
\\
\gth{3} \circ \gth{4}  &\equiv  \gth{2} \circ \gth{1} \circ \gth{1} \circ \gth{2}  
\SIM \gth{2} \circ \gth{1}  \circ \gth{2} \equiv \gth{2} \circ \gth{4}  \equiv \gth{3} \circ \gth{2} \, . \\
\end{split}
\end{equation}
These relationships can be used to structure the solution set generated starting from any particular perfect fluid sphere, and
moreover to classify which metrics can be produced by our theorems, and which ones cannot.
For example, the idempotence property of theorem 1 and theorem 2 enables us to divide the class of perfect fluid spheres into seed metrics and non-seed metrics. Seed metrics can never be generated by using one of the two theorems $\gth{1}$ or $\gth{2}$,
while non-seed metrics are connected to other simpler metrics via one of these theorems.
We formalize this in the following subsection.

\subsection{Seed and non-seed metrics}

\noindent{\bf Definition (Seed metric):} 
Take a metric $g$ (or a parameterized class of metrics) and apply theorem 1 or theorem 2 on it.
Three different cases are possible:
\begin{itemize}
\item 
Each of the applications supplies us with a new solution. [$\gth{1}(g)\not\SIM g \not\SIM \gth{2}(g)$.] We define a metric with this behaviour as a seed metric.  (We shall soon see several examples of this behaviour.)
For example, figure 2.5 show that Minkowski is a seed metric. When we apply theorem 1 to it, we get Einstein static. While applying theorem 2 to Minkowski, we get K-O III instead.
\item 
Only one of the applications supplies us with a new solution, while the other one gives us the same metric we started with.  [$\gth{1}(g)\SIM g$ or $\gth{2}(g)\SIM g$.] These metrics are non-seed metrics.  (We shall soon see several examples of this behaviour.) 
For example, figure 2.5 show that Einstein static and K-O III are both non-seed metric. The reason is only one of the applications gives us a new solution, while the other one still give us the same metric which are Einstein static and K-O III, respectively.
\item 
Both applications give us the same metric we started with. [$\gth{1}(g)\SIM g \SIM \gth{2}(g)$.]  Metrics of this type are fixed points of the transformation process and we then also have $\gth{3}(g)\SIM g \SIM \gth{4}(g)$. While we have encountered numerical examples that seem to exhibit this behaviour, we have no analytic proof for the existence of non-obvious fixed-point metrics. There is one obvious but not particularly useful example of a fixed point class of metrics. If we take the ODE in equation (\ref{ode_for_B}), and write down its most general solution as a function of the arbitrary parameters $\zeta(r)$,  then any of our solution generating theorems applied to this most general solutions will at most move us around in the parameter space characterizing the most general solution --- the most general solution of equation (\ref{ode_for_B}), or  equivalently equation (\ref{ode_for_zeta}), is thus an infinite-parameter fixed point of the generating theorems. But apart from this obvious example, it is unclear whether other fixed point classes of metric exist.
\end{itemize}

Classifying seed and non-seed metrics are very important. In particular, the most outstanding features of this algorithm are:
\begin{itemize}
\item
We developed a tool to generate new solutions for a perfect fluid sphere, which does  not require to solve the Einstein equations.
\item
We also established the relationships among the generating theorems. Before all metrics seem to have nothing more in common than presenting a perfect fluid sphere.
\item
We invented the concept of a seed metric, which cannot be generated by our theorems. Starting with such a metric, a class of non-seed metric can be produced.
\item
This can be used to structure the solutions generated by a perfect fluid sphere and classify which metrics can and cannot be produced by our theorems.
\end{itemize}
\begin{table}[!ht]
\centerline{Examples of some seed and non-seed metrics}
\bigskip
\begin{center}
\begin{tabular}{| l | l |}
\hline
\hline
Seed metric & Non-Seed metric \\
\hline
\hline
Minkowski & B-L
\\
Exterior Schwarzschild & Interior Schwarzschild
\\
de Sitter & Einstein static
\\
\hline
\vphantom{\Big|}Tolman IV ($A = 0$) & Tolman IV , Tolman V ($n = + 1$) , Tolman VI
\\
M-W III & K-O III
\\
Heint IIa (C=0) & Heint IIa
\\
S1 & Wyman IIa, Wyman III
\\
\hline
\end{tabular}
\end{center}
\caption{This table shows several seed and non-seed metrics, which satisfied the definition of seed and non-seed metrics (see Table 3.2 for further details).}
\end{table}
We can nevertheless develop several formal lemmata regarding fixed-point metrics.  For instance
\begin{lemma}
Suppose we have a metric such that $\forall \; \sigma$, $\epsilon$, $\lambda$
\begin{equation}
\gth{3}(\sigma,\epsilon,\lambda)\; g \SIM g,
\end{equation}
where we recall that $\SIM$ denotes equality up to redefinition of parameters.
Then in particular
\begin{equation}
\gth{1}(\lambda)\; g \SIM g \SIM \gth{2}(\sigma,\epsilon)\; g,
\end{equation}
and so
\begin{equation}
\gth{4}(\sigma,\epsilon,\lambda)\; g \SIM g.
\end{equation}

Conversely, suppose we have a metric such that
\begin{equation}
\gth{4}(\sigma,\epsilon,\lambda)\; g \SIM g,
\end{equation}
then
\begin{equation}
\gth{1}(\lambda)\; g \SIM g \SIM \gth{2}(\sigma,\epsilon)\; g,
\end{equation}
and so
\begin{equation}
\gth{3}(\sigma,\epsilon,\lambda)\; g \SIM g.
\end{equation}
\end{lemma}

\begin{proof} Trivial, note that $\gth{3}(\sigma,\epsilon,\lambda=0)=\gth{2}(\sigma,\epsilon)$ and $\gth{3}(\sigma=0,\epsilon=0,\lambda)=\gth{1}(\lambda)$. Then recall $\gth{4} = \gth{1}\circ\gth{2}$. Similarly for the converse.
\end{proof}

\begin{lemma}
Suppose we have a metric $g$ such that
\begin{equation}
\gth{4}\; g \SIM \gth{3}\; g
\end{equation}
and then define $g'$ by
\begin{equation}
\gth{4}\; g \SIM g' \SIM \gth{3}\; g.
\end{equation}
Then we have
\begin{equation}
\gth{4}\; g' \SIM g' \SIM \gth{3}\; g'
\end{equation}
so that $g'$ is a ``fixed point'' of both $\gth{3}$ and $\gth{4}$.
\end{lemma}

\begin{proof}
Note
\begin{equation}
\gth{4}\; g' \SIM \gth{4} \circ \gth{3}\; g \SIM \gth{1}\circ\gth{2}\circ\gth{1}\; g 
\SIM \gth{1}\circ\gth{3}\; g \SIM \gth{1}\circ\gth{4}\; g \SIM \gth{4}\; g = g'
\end{equation}
and similarly for $\gth{3}$:
\begin{equation}
\gth{3}\; g' \SIM \gth{3} \circ \gth{4}\; g \SIM \gth{2}\circ\gth{1}\circ\gth{2}\; g 
\SIM \gth{2}\circ\gth{4}\; g \SIM \gth{2}\circ\gth{3}\; g \SIM \gth{3}\; g = g'
\end{equation}
\end{proof}

Several other formal theorems along these lines can be constructed, but these seem the most important results.

Finally, among the formal properties enjoyed by the generating theorems, we mention the fact that theorems 3 and 4 are  ``conjugate'' to each other in the following sense
\begin{equation}
\gth{4} \equiv \gth{1}\circ\gth{2} =  \gth{1}\circ\gth{2} \circ \gth{1}\circ[\gth{1}]^{-1} =
 \gth{1}\circ\gth{3} \circ [\gth{1}]^{-1},
\end{equation}
and similarly (when the appropriate inverse $[\gth{2}]^{-1}$ exists)
\begin{equation}
\gth{3} \equiv \gth{2}\circ\gth{1} =  \gth{2}\circ\gth{1} \circ \gth{2}\circ[\gth{2}]^{-1} =
 \gth{2}\circ\gth{4} \circ [\gth{2}]^{-1}.
\end{equation}
We can write this as
\begin{equation}
\gth{3}(\sigma,\epsilon,\lambda)\sim\gth{4}(\sigma,\epsilon,\lambda),
\end{equation}
where $\sim$ denotes the concept of ``similarity'' under conjugation by invertible generating theorems.

In these properties we would like to ensure that
\begin{itemize}
\item
We wish to emphasise that similarity $\sim$ is a statement that holds for particular and fixed values of the parameters  $(\sigma,\epsilon,\lambda)$, as opposed to $\SIM$ which denotes equivalence under redefinition of parameters. 
\item
If one is working numerically, it is much easier to ask questions involving similarity $\sim$. For analytic work, it is typically easier to ask questions involving equivalence $\SIM$.
\end{itemize}

   \subsection{The fifth and sixth theorems}
\subsection*{Two linking theorems}

The last two solution generating theorems we shall present are slightly different from those developed so far:
Consider a perfect fluid sphere solution $\left\{ \zeta_{0},B_{0}\right\}$ and extend it
to a new perfect fluid sphere $\left\{ \zeta_{0} Z_{0},B_{0} + \Delta \right\}$.  Previously, we had either set $Z_{0}=1$ and obtained theorem 1,
or we had set $\Delta=0$ and obtained theorem 2. In other words, we only changed one metric component at a time.
(From this point of view theorem 3 and theorem 4 are, strictly speaking, not new theorems, in that they are replaceable by
iterations of theorem 1 and theorem 2 and vice versa.)
We now investigate what happens if we place no \emph{a priori} restrictions on $Z$ and $\Delta$, and allow both metric components to vary simultaneously. The differential equation
(\ref{ode_for_B}) for this problem now becomes
\begin{eqnarray} 
\label{general_ode_for_B}
&\left[ r (r \zeta_0 Z_{0} )' \right] \Delta'&
+ \left[ 2 r^2 (\zeta_{0} Z_{0} )'' - 2 (r \zeta_{0} Z_{0})' \right] \Delta 
\nonumber
\\
&&+ \left[ r^2 \zeta_0  B_0' + 4 r^2 \zeta_0' B_0 - 2 r \zeta_0 B_0\right] Z_0' + 2 r^2 \zeta_0 B_0 Z_0''  =0.
\end{eqnarray}
Note that if $\Delta=0$ this becomes equation (\ref{ode_th1}), while if $Z_0=1$ this becomes equation (\ref{ode_th2}). 
In general, this ODE of first-order in $\Delta$, and --- as long $Z$ is not a constant --- inhomogeneous. In terms of
$\Delta$ this ODE can be solved explicitly and the result stated as a new theorem:
\begin{theorem}
Suppose $\left\{ \zeta_{0} , B_{0} \right\} $ describes a perfect fluid sphere, and let $Z_0(r)$ be arbitrary.
Define
\begin{eqnarray}
\Delta(\lambda,r) &=& \Delta_{0}(r)
\Bigg\{ \lambda -\int \frac{\left[
\left(4r^2\zeta_{0}'B_{0}+r^2 \zeta_{0} B_{0}'-2r\zeta_{0}B_{0}\right) Z_{0}'
+ 2 r^2 \zeta_0 B_0 Z_0'' \right] 
\left\{ \zeta_0 + r \zeta_0' \right\}^2 
} { r^3 \;\left( r \zeta_{0} Z_{0}  \right)' \; \zeta_0^2\; Z_0^2 
}
\; 
\nonumber\\
&&
\qquad\qquad
\exp\left\{ -2 \int \frac{(\zeta_0 Z_{0})'}{ \zeta_0 Z_{0}} \; 
  \frac{ \zeta_0 Z_{0}- r\; (\zeta_0 Z_{0})' }{\zeta_0 Z_{0} + r  \;(\zeta_0 Z_{0})'} \; \d r\right\}
\d r    \Bigg\} ,
\end{eqnarray}
where
\begin{equation} 
\Delta_0(r)  =
 \left(    \frac{ \zeta_0 Z_{0}}{  \zeta_0 Z_{0} + r  \; (\zeta_0 Z_{0})'}\right)^2 \; r^2 \; 
\exp\left\{ 2 \int \frac{(\zeta_0 Z_{0})'}{ \zeta_0 Z_{0}} \; 
  \frac{ \zeta_0 Z_{0}- r\; (\zeta_0 Z_{0})' }{\zeta_0 Z_{0} + r  \;(\zeta_0 Z_{0})'} \; \d r\right\}.
\end{equation}
Then for all $\lambda$, the geometry defined by an arbitrary chosen $Z_0(r)$ and
setting
\begin{equation}
\d s^2 = - \zeta_0(r)^2 Z_{0}(r)^2 \; \d t^2 + {\d r^2\over B_0(r)+\Delta(\lambda,r) }
+ r^2 \d\Omega^2
\end{equation}
corresponds to a perfect fluid sphere. That is, the mapping
\begin{equation}
\gth{5}(\lambda): \left\{ \zeta_0 , B_0  \right\} \mapsto 
\left\{ \zeta_0 \; Z_0, B_0 + \Delta(\lambda,\zeta_0) \right\}
\end{equation}
takes perfect fluid spheres into perfect fluid spheres.
\end{theorem}

Note that if $Z_0(r)=1$ this simply reduces to theorem 1.
Re-arranging equation (\ref{general_ode_for_B}) in terms of $Z_{0}$ leads to
a second-order inhomogenous differential equation,  which cannot in general be solved for a prescribed $\Delta$, \emph{unless one imposes further constraints}. So further exploration in that direction is moot.
There is however a related theorem (which may be easier to understand) in terms of  parametric derivatives:
\begin{theorem}
Let $\{\zeta(\mu), B(\mu)\}$ denote a one-parameter class of perfect fluid spheres, so that the differential equation  (\ref{ode_for_B}) is satisfied for all $\mu$. Then
\begin{eqnarray}
[r(r\zeta)'] \left({\d B\over\d\mu}\right)'+[2r^2\zeta''-2(r\zeta)']\left({\d B\over\d\mu}\right) 
+ 2 r^2 B \left({\d\zeta\over\d\mu}\right)'' 
\nonumber
\\
+ (r^2 B'-2rB)  \left({\d\zeta\over\d\mu}\right)' +(r B'-2B+2)  \left({\d\zeta\over\d\mu}\right)=0.
\end{eqnarray}
In particular if $\d\zeta/\d \mu=0$ this reduces to the ODE (\ref{ode_th1}), while if $\d B/\d\mu=0$ this reduces to the 
ODE (\ref{ode_for_zeta}).
\end{theorem}
This is simply an alternative viewpoint on the previous theorem, emphasising the differential equation to be solved.

We again defer extensive discussion to the next section, but that this stage point out that if we invoke theorem 5 and apply it to Minkowski space, then making the choice $Z_0=1+r^2/a^2$ leads to the general Tolman IV metric --- that is:
\begin{equation}
\gth{5}(\hbox{Minkowski}; \; Z_0=1+r^2/a^2) = (\hbox{Tolman IV}).
\end{equation}
Even before we systematically start our classification efforts, it is clear that the solution generating theorems we have established will inter-relate \emph{many} of the standard perfect fluid spheres.

\begin{figure}[!ht] 
\centering
\includegraphics[scale = 0.6]{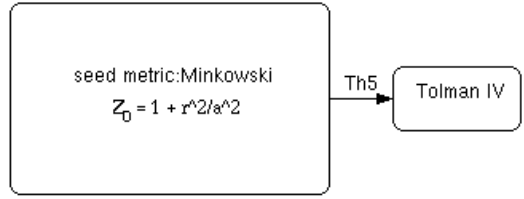}
\caption{\label{Structure graph for theorem 5}This structure shows that when we invoke theorem 5 and apply it to Minkowski space, then making the choice $Z_0=1+r^2/a^2$ leads to the general Tolman IV metric.}
\end{figure}
\section{Formal properties of the linking theorems}

Before turning to issues of systematic classification of perfect fluid metrics, we wish to establish a few formal properties of the linking theorems. To simplify the notation,  let us define the differential expression
\begin{eqnarray}
D(\zeta,B) &\equiv&
[r(r\zeta)']B'+[2r^2\zeta''-2(r\zeta)']B + 2\zeta 
\nonumber
\\
&=&
2 r^2 B \zeta'' + (r^2 B'-2rB) \zeta' +(r B'-2B+2)\zeta.
\end{eqnarray}
Then the condition for a perfect fluid sphere is simply
\begin{equation}
D(\zeta,B)=0
\end{equation}
Now define
\begin{equation}
D(\zeta_0,B_0+\Delta_0) = D(\zeta_0,B_0) + D_1(\zeta_0;\Delta_0)
\end{equation}
Then it is easy to check that
\begin{equation}
D_1(\zeta_0;\Delta_0) =  \left[ r (r \zeta_0 )' \right] \Delta_{0}'
+ \left[ 2 r^2 \zeta_{0}'' - 2 (r \zeta_{0})' \right] \Delta_{0}
\end{equation}
The ODE for theorem 1, where we assume $\{\zeta_0,B_0\}$ is a perfect fluid sphere, is then
\begin{equation}
D_1(\zeta_0;\Delta_0)=0.
\end{equation}
Now define
\begin{equation}
D(\zeta_0 Z_0,B_0) = Z_0 D(\zeta_0,B_0) + D_2(\zeta_0,B_0;Z_0),
\end{equation}
then it is easy to check that
\begin{equation}
D_2(\zeta_0,B_0;Z_0) = 
(r^2 \zeta_0 B_0' + 4 r^2 \zeta_0' B_0 - 2 r \zeta_0 B_0) Z_0' + (2 r^2 \zeta_0 B_0) Z_0'' .
\end{equation}
The ODE for theorem 2, where we assume $\{\zeta_0,B_0\}$ is a perfect fluid sphere, is then
\begin{equation}
D_2(\zeta_0,B_0;Z_0)=0.
\end{equation}
Now let us consider any simultaneous shift in $\zeta$ and $B$, as considered in theorem 5. We have
\begin{eqnarray}
D(\zeta_0 Z_0, B_0+\Delta_0) &=&
D(\zeta_0 Z_0,B_0)+D_1(\zeta_0 Z_0;\Delta_0)
\\
&=&
Z_0 D(\zeta_0,B_0) + D_2(\zeta_0,B_0;Z_0) + D_1(\zeta_0 Z_0;\Delta_0).
\nonumber
\end{eqnarray}
But now let us write
\begin{equation}
D_1(\zeta_0 Z_0;\Delta_0) = Z_0 D_1(\zeta_0;\Delta_0) + D_{12}(\zeta_0;Z_0,\Delta_0),
\end{equation}
where a brief computation yields
\begin{equation}
D_{12}(\zeta_0;Z_0,\Delta_0) = r^2 \zeta_0 Z_0' \Delta_0' + [2r^2\zeta_0 Z_0''+4r^2\zeta_0'Z_0' - 2 r \zeta_0 Z_0' ] \Delta_0.
\end{equation}
Then all in all
\begin{eqnarray}
D(\zeta_0 Z_0, B_0+\Delta_0) &=& Z_0 D(\zeta_0,B_0) + Z_0 D_1(\zeta_0;\Delta_0) 
\nonumber
\\
&&
+  D_{12}(\zeta_0;Z_0,\Delta_0) +  D_2(\zeta_0,B_0;Z_0).
\end{eqnarray}
So if $\{\zeta_0,B_0\}$ and $\{\zeta_0 Z_0,B_0+\Delta_0\}$ are both perfect fluid spheres we must have
\begin{equation}
Z_0 D_1(\zeta_0;\Delta_0) +  D_{12}(\zeta_0;Z_0,\Delta_0) +  D_2(\zeta_0,B_0;Z_0)
= 0.
\end{equation}
 This is
 \begin{equation}
 Z_0\times(\hbox{ODE for theorem1}) + (\hbox{cross term}) + (\hbox{ODE for theorem2}) = 0.
\end{equation}
The cross term vanishes if either $Z_0=\hbox{constant}$ or $\Delta_0=0$ in which case we recover the usual theorem 1 and theorem 2. If we do things in the opposite order
 \begin{equation}
 D(\zeta_0 Z_0, B_0+\Delta_0) 
 = Z_0 D(\zeta_0,B_0+\Delta_0)+D_2(\zeta_0, B_0+\Delta_0;Z_0);
 \end{equation}
 \begin{equation}
 = Z_0 D(\zeta_0,B_0) + Z_0 D_1(\zeta_0,\Delta_0) + D_2(\zeta_0,B_0+\Delta_0;Z_0).
 \end{equation}
 We now have to compute
 \begin{equation}
 D_2(\zeta_0,B_0+\Delta_0;Z_0) =  D_2(\zeta_0,B_0;Z_0) + D_{21}(\zeta_0,B_0;Z_0,\Delta_0),
 \end{equation}
 and after a brief computation
 \begin{eqnarray}
 D_{21}(\zeta_0,B_0;Z_0,\Delta_0) 
&=&
2 r^2\zeta_0\Delta_0 Z_0'' +
[r^2\zeta_0\Delta_0'+4 r^2\zeta_0'\Delta_0- 2 r \zeta_0\Delta_0] Z_0'
\nonumber
\\
&=&
D_{12}(\zeta_0,B_0;Z_0,\Delta_0).
\end{eqnarray}
Thus the cross term is the same, no matter how you calculate it, and we still have the identity
\begin{eqnarray}
D(\zeta_0 Z_0, B_0+\Delta_0) &=& Z_0 D(\zeta_0,B_0) + Z_0 D_1(\zeta_0;\Delta_0) 
\nonumber
\\
&&
+  D_{12}(\zeta_0;Z_0,\Delta_0) +  D_2(\zeta_0,B_0;Z_0).
\end{eqnarray}
Provided  ${\zeta_0,B_0}$ and $\{\zeta_0 Z_0,B_0+\Delta_0\}$ are both perfect fluid spheres we again deduce
\begin{equation}
Z_0 D_1(\zeta_0;\Delta_0) +  D_{12}(\zeta_0;Z_0,\Delta_0) +  D_2(\zeta_0,B_0;Z_0)
= 0.
 \end{equation}
Now this gives us another way of looking at theorem 3 and theorem 4. For theorem 3 we first apply theorem 1 so we have the two equations
\begin{equation}
D_1(\zeta_0;\Delta_0) = 0;
\end{equation}
and
\begin{equation}
D_2(\zeta_0,B_0+\Delta_0;Z_0)\equiv D_{12}(\zeta_0;Z_0,\Delta_0) +  D_2(\zeta_0,B_0;Z_0) = 0.
 \end{equation}
Conversely, for theorem 4 where we first apply theorem 2 we see that we need to solve
\begin{equation}
D_2(\zeta_0,B_0;Z_0)= 0;
\end{equation}
and
\begin{equation}
D_1(\zeta_0 Z_0;\Delta_0) \equiv Z_0 D_1(\zeta_0;\Delta_0) +  D_{12}(\zeta_0;Z_0,\Delta_0) 
= 0.
 \end{equation}
 For theorem 5 we pick $Z_0$ arbitrarily, and solve the single ODE
 \begin{equation}
 D_1(\zeta_0 Z_0;\Delta_0)+ D_2(\zeta_0,B_0;Z_0)=0.
 \end{equation}
 This is a single first-order linear inhomogeneous ODE for $\Delta_0$, and hence solvable.
(In particular this makes it clear that theorem 5 is an \emph{inhomogeneous} version of theorem 1 with a carefully arranged ``source term'' $D_2(\zeta_0,B_0;Z_0)$. While theorem 5 is not ``idempotent'' it does satisfy the important formal property that:
\begin{lemma}
\begin{equation}
\gth{1}\circ\gth{5} \SIM \gth{5}
\end{equation}
which in particular tells us that the output from theorem 5 is never a seed metric.
\end{lemma}

\begin{proof}
Applying theorem 5 we need to solve
 \begin{equation}
 D_1(\zeta_0 Z_0;\Delta_0)+ D_2(\zeta_0,B_0;Z_0)=0
 \end{equation}
 in order to map
 \begin{equation}
 \{\zeta_0,B_0\} \to \{\zeta_0Z_0,B_0+\Delta_0\}
 \end{equation}
 Now apply theorem 1 to  $\{\zeta_0Z_0,B_0+\Delta_0\}$, this means we have to solve the homogeneous ODE
 \begin{equation}
D_1(\zeta_0 Z_0;\Delta_1)=0
 \end{equation} 
But then, using properties of first-order ODEs
\begin{equation}
\left\{  D_1(\zeta_0 Z_0;\Delta_0)+ D_2(\zeta_0,B_0;Z_0)=0 \right\} \oplus
\left\{ D_1(\zeta_0 Z_0;\Delta_1)=0 \right\} 
\end{equation}
\begin{equation}
\implies
\left\{  D_1(\zeta_0 Z_0;\Delta_0+\Delta_1)+ D_2(\zeta_0,B_0;Z_0)=0 \right\}
\end{equation}
which is the ODE from theorem 5 back again.
\end{proof}
(The net result of this observation, as we shall see in the next section, is that theorem 5 can be used to connect one seed metric with the ``descendants'' of another seed metric.)

\section{Classifying perfect fluid spheres}
\label{S:classify}
We will now see the power of these transformation theorems (solution generating theorems) by using them in a number of different ways:
\begin{itemize}
\item
To generate several new perfect fluid spheres
\item
To relate various perfect previously known fluid spheres to each other.
\item
And also to classify the geometries we encounter.
\end{itemize}
First some minor warnings: Despite comments made in~\cite{Delgaty},
Kuch~2~I $\equiv$ Tolman~V;   and Kuchb~I~b \emph{is} a perfect fluid for general values of its parameters.
Furthermore RR--I $\equiv$ Einstein static; RR--V = Tolman V ($n=-5/4$). If we had not noted these degeneracies then our tables below would have been more extensive, but would have conveyed no extra information.


\label{P:well-known}
\begin{table}[!ht]
\centerline{Some selected perfect fluid solutions}
\bigskip
\hskip -1.5 cm
{\footnotesize
\begin{tabular}{|| l | l ||}
\hline
\hline
Name & Metric \\
\hline
\hline
Minkowski & $-\d t^2 + \d r^2 + r^2\d\Omega^2$ 
\\
Einstein static & $- \d t^2 + (1- {r^2}/{R^2})^{-1} \d r^2 + r^2 \d\Omega^2$
\\
de Sitter & $-(1 - {r^2}/{R^2})\, \d t^2 + (1 - {r^2}/{R^2})^{-1}\, \d r^2 + r^2 \d\Omega ^2$ 
\\
Schwarzschild Interior & $-\left(A - B \sqrt{1 - {r^2}/{R^2}}\right)^2 \d t^2 + (1- {r^2}/{R^2})^{-1} \d r^2 + r^2 \d\Omega^2$ 
\\
Schwarzschild Exterior & $-(1-2m/r) \, \d t^2 + (1- 2m/r)^{-1} \d r^2 + r^2 \d\Omega^2$ 
\\
\hline
S1 & $- r^4 \d t^2 + \d r^2 + r^2 \d\Omega^2$
\\
K-O III & $- (A + Br^2)^2 \d t^2 + \d r^2 + r^2 \d\Omega^2$ 
\\
Kuch1 Ib & $ - (Ar + Br \ln r)^2 \d t^2+ 2 \d r^2 + r^2 \d\Omega^2$ 
\\
B--L & $- A (r^2/a^2)\d t^2 +  2(1+ r^2/a^2)^{-1} \d r^2 + r^2 \d \Omega^2$
\\
\hline
\vphantom{\Big|}Tolman IV &  
$- B^2 \left(1 + {r^2}/{a^2}\right) \d t^2 + 
\frac{1 + 2 {r^2}/{a^2}}{\left(1 - {r^2}/{b^2}\right)\left(1 + {r^2}/{a^2}\right)} \d r^2 
+ r^2 \d\Omega^2$ 
\\
\vphantom{\Big|}Tolman IV ($b\to\infty$)& 
 $- B^2 \left(1 + {r^2}/{a^2}\right) \d t^2 + \frac{1 + 2 {r^2}/{a^2}}{1 + {r^2}/{a^2}} \d r^2 + r^2 \d\Omega^2$ 
\\
\hline
Tolman V &  
$-B^2 r^{2(1+ n)}\d t^2 + (2-n^2)\; 
[1 - A r^{2(2-n^2)/(2+ n)}]^{-1} \d r^2 + r^2\d\Omega^2$
\\
Tolman V ($A\to0$) & $-B^2 r^{2(1+ n)}\d t^2 + (2-n^2)\d r^2 + r^2\d\Omega^2$
\\
Tolman VI & $-(Ar^{1-n} + Br^{1+n})^2 \d t^2 + (2-n^2)\d r^2 + r^2 \d\Omega^2$ 
\\
Tolman VII &
 $-B^2 \cos\left\{\ln\left[{\sqrt{1-2r^2/a^2+r^4/b^4}+r^2/a^2-b^2/a^2}\right]^{1/2}+\theta\right\}^2$ \\
 & \qquad 
 $+ (1-2r^2/a^2+r^4/b^4)^{-1} \d r^2 + r^2\d\Omega^2$
\\
Tolman VIII &
$ - A^2 r^{2(n-1)(n-4)/n}\left( 
{-n^2\over(n^2-4n+2)(n^2-8n+8)} + B r^{-(n^2-8n+8)/n} + C r^{-2(n^2-4n+2)/n} \right)
$
\\
\qquad 
&
$
+ \left( 
{-n^2\over(n^2-4n+2)(n^2-8n+8)} + B r^{-(n^2-8n+8)/n} + C r^{-2(n^2-4n+2)/n} \right)^{-1} 
\d r^2 + r^2\d\Omega^2$
\\
\hline
Kuch 68 II & $-\left(1-{2m/ r}\right)\d t^2 
+ \left[ \left(1-{2m/r}\right)\left(1 + C(2r-2m)^2 \right)\right]^{-1} \d r^2 + r^2\d\Omega^2$
\\
Kuch 68 I &
$ - \left(A\sqrt{1-2m/r}+
B\left[r^2/m^2+5r/m-30+15\sqrt{1-2m/r}\ln\{1-r/m-\sqrt{r(r-2m)}/m\}\right] \right)^2\d t^2$
\\
&\qquad
$+(1-2m/r)^{-1} \d r^2 + r^2 \d\Omega^2$
\\
M--W III & $-A r (r-a) +{7/4\over 1-r^2/a^2} \d r^2 + r^2 \d\Omega^2 $
\\
Kuch I b &
$ - r^2[A+B\,\arctanh(a/\sqrt{a^2+r^2})]^2\d t^2 + 2(1+r^2/a^2)^{-1}\d r^2 + r^2\d\Omega^2$
\\
Heint IIa (C=0) & $-(1+ar^2)^3 \d t^2 + [(1+ar^2)/(1-ar^2/2)] \d r^2 + r^2\d\Omega^2$
\\
Heint IIa &  $-(1+ar^2)^3 \d t^2 + \left[1-{3ar^2\over2(1+ar^2)} + {Cr^2\over(1+ar^2)\sqrt{1+4ar^2}} \right]^{-1} \d r^2 + r^2\d\Omega^2$
\\
\hline
\hline
\end{tabular}
}  
\caption{\label{T:well-known}Some well-known perfect fluid spheres and their coresponding metrics. Note that we have often reparameterized these metrics to make them easier to deal with, and so their appearance (but not the substance)  may differ from other sources~\cite{Delgaty,Skea,exact}.  Metric S1 is a special case of K--O III, Tolman V, and Tolman VI, notable for its extreme simplicity.}
\end{table}


\label{P:well-known2}
\begin{table}[!ht]
\centerline{Some apparently new perfect fluid solutions}
\bigskip
\hskip - 2 cm
{\footnotesize
\begin{tabular}{|| l | l ||}
\hline
\hline
Name & Metric \\
\hline
\hline
Martin 1 & $ - ( Ar + Br \ln r)^2 \d t^2 + 2\frac{2A + 2B \ln r + B}{2A + 2B \ln r+B-Cr^2} \d r^2 + r^2 \d\Omega^2$ 
\\
Martin 2 & $- Ar(r-a) \d t^2 + {7\over 4} \left(1-{r^2\over a^2} - B {(r-a)r^{7/3}\over(4r-3a)^{4/3}} \right)^{-1}\d r^2 
+ r^2 \d\Omega^2$ 
\\
Martin 3 &$-(1 + ar^2)^2 \d t^2 + [1-br^2/(1+3ar^2)^{2/3}]^{-1} \d r^2 + r^2 \d\Omega^2$ 
\\
\hline

P1 & 
$-(1 + ar^2)^2 
\left[A + B\int {(1+ar^2)^{-2}\over\sqrt{1-br^2/(1+3ar^2)^{2/3}}}r \d r \right]^2 \d t^2 
+ [1-br^2/(1+3ar^2)^{2/3}]^{-1} \d r^2+ r^2 \d\Omega^2$ 
\\

P2 &  
$  - ( Ar + Br \ln r)^2 
\left[\sigma+\epsilon
\int  ( Ar + Br \ln r)^{-2} 
\left(2\frac{2A + 2B \ln r + B}{2A + 2B \ln r+B-Cr^2}\right)^{-1/2} r\d r \right]^2\d t^2 
$\\
& \qquad $
+ 2\frac{2A + 2B \ln r + B}{2A + 2B \ln r+B-Cr^2} \d r^2+ r^2 \d\Omega^2$
\\

P3 & $-(1+ar^2)^3 \left[ A + B {(5+2ar^2)\sqrt{1-ar^2/2}\over(1+ar^2)^{3/2}}\right]^2\d t^2 
+ [(1+ar^2)/(1-ar^2/2)] \d r^2 + r^2\d\Omega^2$
 \\
 
P4 & 
$- r^4\left( A + B\int {\d r\over r^3 \sqrt{1+\lambda r^{2/3}}}\right)^2 \d t^2 
+ (1+\lambda r^{2/3})^{-1} \d r^2 + r^2\d \Omega^2$\\
P4 & $- r^4\left( A + B\left[{15\over16}\lambda^3\tanh^{-1}(1/\sqrt{1+\lambda r^{2/3}}) 
- {1\over16}\sqrt{1+\lambda r^{2/3}} (8 r^{-2} - 10\lambda r^{-4/3} +15\lambda^2 r^{-2/3} ) \right]\right)^2 \d t^2 $
\\
 & \qquad $
+ (1+\lambda r^{2/3})^{-1} \d r^2 + r^2\d \Omega^2$\\

P5 & $- r \left(A\sqrt{r-a}+B\sqrt{r+a}\right)^2 
+{7/4\over 1-r^2/a^2} \d r^2 + r^2 \d\Omega^2 $
\\
P6 & $- r(r-a) \left(A+B\int{\d r\over(r-a)\sqrt{ 1-{r^2\over a^2} - 
B {(r-a)r^{7/3}\over(4r-3a)^{4/3}}  }}\right)^2 
+ {7\over4}\left(1-{r^2\over a^2}- 
B {(r-a)r^{7/3}\over(4r-3a)^{4/3}} \right)^{-1} \d r^2 + r^2 \d\Omega^2 $
\\

P7 &
$
- B^2 \left(1 + {r^2}/{a^2}\right) 
\left[ A + B \int {\sqrt{a^2-2r^2}\over\sqrt{b^2-r^2}(a^2-r^2)^{3/2}} r\d r \right]^2 \d t^2 + 
\frac{1 + 2 {r^2}/{a^2}}{\left(1 - {r^2}/{b^2}\right)\left(1 + {r^2}/{a^2}\right)} \d r^2 
+ r^2 \d\Omega^2$ 
\\

P8 &
$
-(1+ar^2)^3
 \left[ A + B  \int { r \; \d r \over (1+ar^2)^2 \sqrt{1-{3ar^2\over2(1+ar^2)} + {Cr^2\over(1+ar^2)\sqrt{1+4ar^2}}} }
 \right]^2 
 \d t^2 
+ \left[1-{3ar^2\over2(1+ar^2)} + {Cr^2\over(1+ar^2)\sqrt{1+4ar^2}} \right]^{-1} \d r^2
+ r^2 \d\Omega^2$ 
\\

\hline
\hline
\end{tabular}
} 
\caption{\label{T:well-known2}Some apparently new prefect fluid spheres and their coresponding metrics. Sometimes the relevant integrals cannot be done in elementary form. When they can be done they are explicitly shown.}
\end{table}

Starting with the metric for any known perfect fluid sphere and successively applying theorem 1 and theorem 2 numerous times will supply us with endless
``new'' perfect fluid sphere solutions. There are so many types of ``new" solutions such as:
\begin{itemize}
\item
Some of these ``new'' solutions might already be findable in the literature, some of them might be truly novel.
\item
Some of these solutions can be written down in a fully explicit form. 
\item
Some solutions are explicit but not elementary, in the sense that while the metric components can be exhibited as specific and explicit integrals, these integrals cannot be done in elementary form.
\item
Some solutions are so complex that present day symbolic manipulation programs quickly bog down.
\end{itemize}
(For specific symbolic computations we have used a vanilla installation of {\sf Maple}. See appendix C for some sample worksheets).

\label{P:well-known3}
\begin{table}[!ht]
\centerline{Some seed geometries and their descendants}
\bigskip
\hskip-1cm
{\small
\begin{tabular}{|  l  |  l  |  l  |  l  | l |}
\hline
Seed & Theorem1 & Theorem2 & Theorem 3 & Theorem 4\\
\hline
Minkowski & Einstein static & K-O III & interior Schwarzschild & Martin 3\\ 
exterior Schwarzschild & Kuch68 II & Kuch 86 I  &    [integral] & [integral] \\
de Sitter& Tolman IV   & interior Schwarzschild  &  P7 & [integral] \\
Tolman V ($A=0$) & Tolman V &  Tolman VI & Wyman III & Wyman IIa\\
S1 & Tolman V ($n=+1$) & K--O III & P4 & Martin 3 \\
M--W III & Martin 2 &  P5 &  P6 & [integral] \\
Heint IIa (C=0) &  Heint IIa & P3 & P8 & [integral] \\
\hline
\hline
\end{tabular}
} 
\caption{\label{T:well-known3}Seed solutions and their generalizations derived via theorems 1--4. The notation ``[integral]'' denotes a metric so complicated that explicitly writing out the relevant integral is so tedious that it does not seem worthwhile.}
\bigskip
\end{table}

\label{P:well-known4}
\begin{table}[!ht]
\centerline{Some non-seed perfect fluid geometries and their descendants}
\bigskip
\begin{center}
\begin{tabular}{|  l  |  l  |  l  |  l  | l |}
\hline
Base & Theorem1 & Theorem2 & Theorem 3 & Theorem 4\\
\hline
Tolman IV & Tolman IV &   P7  &   P7  & [integral] \\
B-L & B-L &Kuchb I b & Kuchb I b  & [integral]  \\
Heint IIa &  Heint IIa & P8 & P8 & [integral] \\
\hline

Tolman VI& Wyman IIa &  Tolman VI & [integral] & Wyman IIa \\

Kuch1 Ib & Martin 1 & Kuch1 Ib &  P2 & Martin 1\\

K--O III & Martin 3&  K--O III &   P1  & Martin 3\\
\hline



\hline
\end{tabular}
\end{center}
\caption{\label{T:well-known4}Non-seed solutions and their generalizations.}
\bigskip
\end{table}

We present several tables and diagrams to summarize the situation. Two tables are used to provide the names and explicit metrics for the perfect fluid spheres we consider. Two other tables are used to describe the inter-relationships of these perfect fluid spheres under $\gth{1}$, $\gth{2}$, $\gth{3}$, and $\gth{4}$.  
Recall that a seed metric is one for which theorem 1 and theorem 2 both yield metrics  distinct from the seed: $\gth{1}(g)\not\SIM g\not\SIM\gth{2}(g)$. In contrast for non-seed metrics one or the other of these theorems is trivial, either $\gth{1}(g)\SIM g$ or $\gth{2}(g)\SIM g$.

In these tables, we can see the notation  ``[integral]'' means that it is definitely a novel perfect fluid solution. However, the metric components involve an explicit integral that does not appear to be do-able by elementary methods, and is so complicated that it does not seem worthwhile to even write it down.

By considering that theorem 1 and theorem 2 have ``idempotent'' properties , and the fact that theorem 3 and theorem 4 can be expressed in terms of the first two theorems, it is possible to structure, and therefore to graphically visualize the relationship between all metrics generated from a given seed metric.
We demonstrate this behaviour starting with Minkowski spacetime as seed metric where we are starting with this trivial seed to check if it is possible to create endless new perfect fluid spheres. The first few steps can be carried out explicitly, and show that Minkowski space generates several well known interesting perfect fluid models.

\begin{figure}[!ht]
\centering
\includegraphics[width = \textwidth]{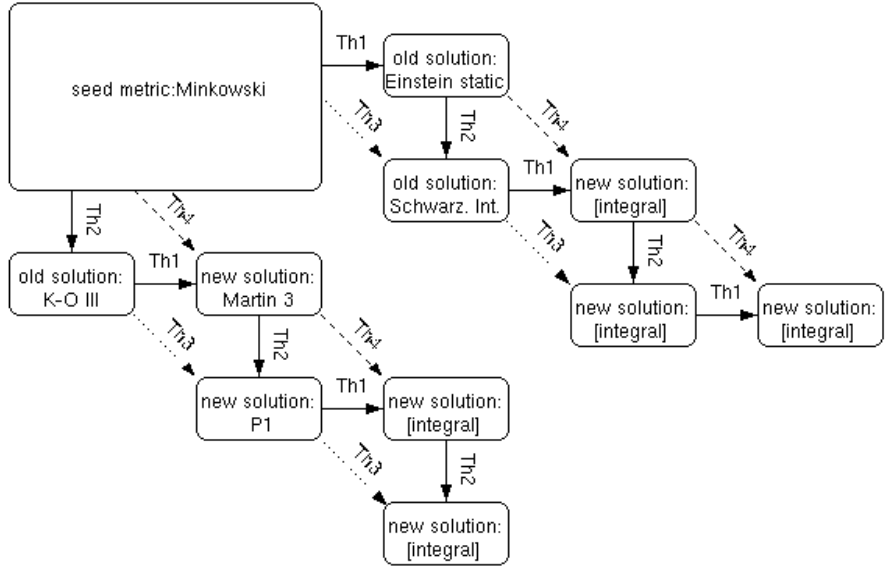}
\label{structure}
\caption{\label{Structure_Graph1}Structure graph for Minkowski space used as seed metric.}
\end{figure}

In figure \ref{Structure_Graph1} all the ideas from the previous sections are used.
Each box represents a specific metric (perfect fluid sphere) while the arrows correspond to the 
application of the different theorems. The horizontal arrows correspond 
to an application of theorem 1 and the vertical arrows to an application 
of theorem 2. In addition, the vectors pointing along the diagonals can 
either indicate an application of theorem 3 or theorem 4. A dotted arrow 
corresponds to the application of theorem 3 while the dashed arrow 
represents an application of theorem 4.

\begin{figure}[!htb]
\includegraphics[width = \textwidth]{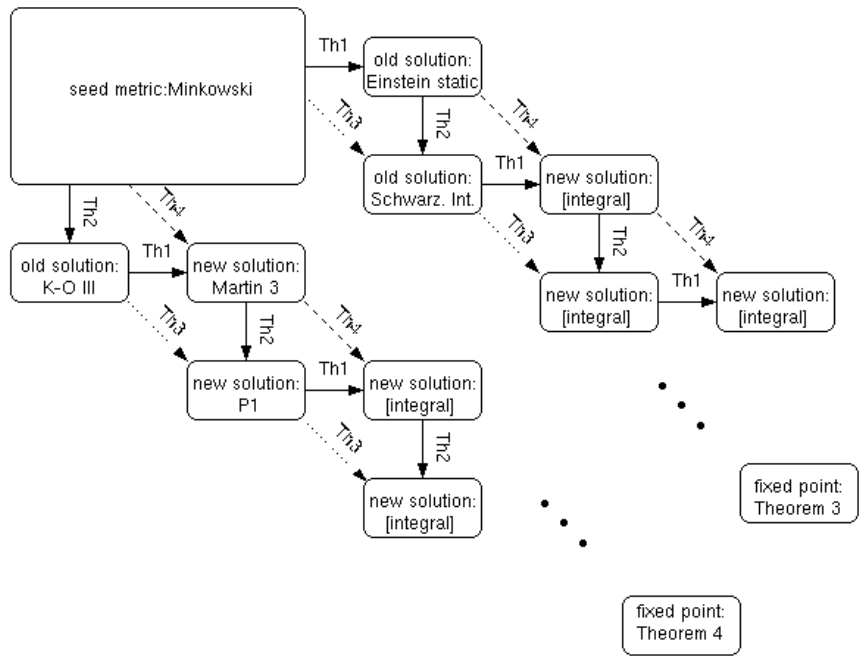}
\caption{\label{Structure_Graph2}Numerical implementations of Theorem 3 and Theorem 4 can be used to investigate the behavior for $n$ applications, if the number of applications goes to infinity.
For specific choices of parameters $\lambda_{i}$, $\sigma_{i}$, and $\epsilon_{i}$, multiple  reapplications of Theorem 3 and Theorem 4 approach a limit, in the sense that both metric components are converging towards there fixed-points.}
\end{figure}

Figure \ref{Idempotence_Theorem1} shows that after applying theorem 1 to the Minkowski seed metric, we get the Einstein static universe.  By the idempotence of theorem 1 
$n$ applications of  $\gth{1}$ to the seed metric still results in the Einstein 
static. Similary, any number of  applications of theorem 2 after the first (which leads to the K--O III solution) does not give us any further new solutions (see figure \ref{Idempotence_Theorem2}) so we can see form this figure that theorem 2 is ``idempotent". In addition, we can also see that the first and second generating theorems are not commutative.  Application of theorem 1 and theorem 2 in that order to the Minkowski seed metric results in the Schwarzschild Interior geometry, whereas application of theorem 1 after theorem 2 gives us the Martin 3 solution,
which is quite distinct from the three-parameter Schwarzschild interior solution.
Indeed, it seems as if it is possible to create endless new solutions for a perfect fluid sphere out
of the Minkowski metric (or any other of the seed metrics).
After several iterations the calculations become more complex, and  can no longer be carried out analytically. We then resort to numerical computation  to find out whether theorem 3 and theorem 4 have some sort of numerical limit, a numerical fixed point, or not. 
Depending on the choice for $(\lambda,\sigma,\epsilon)$ both theorems converge very quickly. Specifically, we used {\sf Matlab} to numerically analyze the evolution of  $\zeta(r)$ and $B(r)$ after applying theorem 3 several times to the Minkowski metric. 
Numerical results are summarized in figure \ref{Structure_Graph2}. This figure indicates that theorem 3 and theorem 4 both appear to have well  defined numerical limits, though we have no direct analytic solution for the fixed point metric.

\section{Discussion}
We have developed several transformation theorems that map perfect fluid
spheres to perfect fluid spheres using Schwarzschild coordinates. We
used these transformations as a basis for classifying different types of
perfect fluid sphere solutions.  Applying these theorems on a known perfect
fluid sphere, different solutions are often obtained. Some of the solutions
are already known in the literature but most of them are novel. Moreover, we
can classify which metrics can be produced by our theorems, and which ones
cannot. The first theorem we presented is in a slightly different formalism, 
and demonstrates an independent way of proving it. Indeed, the second theorem we
presented is a new transformation theorem. The idempotence property of
Theorem 1 and Theorem 2 can divide the class of perfect fluid spheres into seed and non-seed metrics.  Classifying seed and non-seed metrics are important because we have developed a tool to generate new solutions for a perfect fluid sphere, which does not require us to directly solve the Einstein
equations. Indeed, the whole procedure was set up in such a way that we are
implicitly and indirectly solving the Einstein equations, but the utility of
the transformation theorems is that one does not have to go back to first
principles for each new calculation.

We established two new theorems by composing first and second generating
theorems. They are in general distinct, which can be traced back to the fact
that Theorem 1 and Theorem 2 do not, in general, commute. Indeed, Theorem 3 
and Theorem 4 are, in general, not idempotent. That means when we apply Theorem 3 or Theorem 4 more than once, further solutions will in general be obtained. In addition, we can see the power of these transformation theorems as we can generate several new perfect fluid spheres and also can classify the
geometries we encountered.

Furthermore, we have also established several relationships among the 
generating theorems. Previously, all metrics seemed to have nothing more in
common than presenting a perfect fluid sphere. We have developed the concept
of a seed metric,  which is one that cannot be generated by our theorems.
Starting with a seed metric and applying our theorems, it is possible to
structure, therefore to visualize, the relationship between all metrics
generated from a given seed metric in a graph.  Based on this example it is
possible to create endless new solutions out of the Minkowski metric. We 
also used a numeric program to investigate whether Theorem 3 and Theorem 4 have
fixed point limits. Both seem to converge very quickly.
The situation is summarized graphically in figures (3.1) to (3.9), and in tabular form in tables (3.1) to (3.5).
\chapter{Solution generating theorems --- other coordinates}
Consider other coordinate systems: This chapter generalizes the theorems which we derived for Schwarzschild coordinates to a number of other coordinate systems.  This may allow us to develop extra relations between the known solutions.


In all cases we give the metric and the pressure isotropy condition $G_{\hat r\hat r} = G_{\hat\theta\hat\theta}$.
Finch and Skea \cite{Skea} estimate that about 55\% of all work on fluid spheres is carried out in Schwarzschild coordinates, that isotropic coordinates account for 35\% of research, and that the remaining 10\% is spread over multiple special cases (see below).

\bigskip
\section{General diagonal coordinates:}
Consider the metric:
\begin{equation} \label{general_diagonal_1}
\d s^2 = - \zeta(r)^2 \; \d t^2 + {\d r^2\over B(r)} + R(r)^2 \d \Omega^2
\end{equation}
and then assume it represents a perfect fluid sphere. That is,
$ G_{\hat\theta\hat\theta} = G_{\hat r\hat r} =G_{\hat\phi\hat\phi} $.
The equation $G_{\hat\theta\hat\theta} = G_{\hat\phi\hat\phi}$ is always fulfilled due to spherical symmetry.

Calculate
\begin{equation}
G_{\hat r\hat r} = \frac{\zeta B (R')^2 + 2 B R \zeta' R' - \zeta }{R^2 \zeta}
\end{equation}
and
\begin{equation}
G_{\hat\theta\hat\theta} = -\frac{1}{2}\; \frac{2 B \zeta' R' + 2 \zeta R'' B + \zeta B' R' + 2 R \zeta'' B + R \zeta' B'}{R \zeta}
\end{equation}
We set $G_{\hat r\hat r} = G_{\hat\theta\hat\theta}$, which supplies us with an ODE. This ODE reduces the freedom to choose the three functions in equation (\ref{general_diagonal_1})
to two:
\begin{equation}
\label{ODE_for_B_diagonal_1}
[R(R\zeta)']B' +[2R R'' \zeta+2R^2\zeta''-2RR'\zeta'-2(R')^2\zeta] B + 2 \zeta=0.
\end{equation}
This is first order-linear non-homogeneous in $B(r)$, and second-order linear homogeneous in $\zeta(r)$. Note in particular it is linear in both $\zeta(r)$ and $B(r)$ and the comment below equation (16.10) in ``exact solutions" \cite{exact} (while true) completely misses the mark.
There is nothing particularly nice about the $R(r)$ dependence. 

\bigskip

\section{Schwarzschild (curvature) coordinates:} 
This is the basis of the theorems which were previously derived above.

Consider the metric
\begin{equation}
\d s^2 = - \zeta(r)^2 \; \d t^2 + {\d r^2\over B(r)} + r^2 \d\Omega^2
\end{equation}
and the ODE
\begin{equation}
[r(r\zeta)']B'+[2r^2\zeta''-2(r\zeta)']B + 2\zeta=0.
\end{equation}
This is first order-linear non-homogeneous in $B(r)$, and second-order linear homogeneous in $\zeta(r)$.  Solving for $B(r)$ in terms of $\zeta(r)$ is the basis of the \cite{Martin}, (and is the basis for Theorem 2 in chapter 3).

Sometimes people prefer to write $B(r)$ in terms of $m(r)$
\begin{equation}
\d s^2 = - \zeta(r)^2 \; \d t^2 + {\d r^2\over1-{\txt2m(r)\over\txt r}} + r^2 \d\Omega^2
\end{equation}
or even $\mu(r)=m/r^3$, but there is no qualitative gain in doing so.

If we re-group things in terms of $\zeta$ we find
\begin{equation}
2 r^2 \zeta'' + (r^2 B'-2rB) \zeta' +(r B'-2B+2)\zeta=0,
\end{equation}
which is a linear homogeneous 2nd order ODE. This is the basis of Theorems 1 and 3 in the above section.

\bigskip

\section{Isotropic coordinates:} 
Consider the metric
\begin{equation} \label{Isotropic_1}
\d s^2 = - \zeta(r)^2 \; \d t^2 +{1\over\zeta(r)^2\;B(r)^2} \{ \d r^2 + r^2 \d\Omega^2 \}
\end{equation}
Calculate
\begin{equation}
G_{\hat r\hat r} = (\zeta')^2 B^2 - (B')^2 \zeta^2 + 2 B' B \zeta^2 /r
\end{equation}
and
\begin{equation}
G_{\hat\theta\hat\theta} = -(B')^2 \zeta^2 + \zeta^2 B B'' - (\zeta')^2 B^2 + B' B \zeta^2 /r
\end{equation}
We set $G_{\hat r\hat r} = G_{\hat\theta\hat\theta}$, which supplies us with an ODE. This ODE reduces the freedom to choose the two functions in equation (\ref{Isotropic_1})
to one:
\begin{equation} \label{ode_for_iso_zeta}
\left({\zeta'\over\zeta}\right)^2 ={B''-B'/r\over2B}.
\end{equation}
There are several ways of improving this. If we write $\zeta(r)=\exp(\int g(r) \d r)$ then we have an algebraic equation for $g(r)$:
\begin{equation}
g(r)= \pm \sqrt{B''-B'/r\over2B}.
\end{equation}
Conversely, the isotropy condition can be written in terms of $B(r)$ as:
\begin{equation}
\label{second_ode_iso}
B'' - B'/r -2g^2 B = 0.
\end{equation}
This is about as simple as you can make things.

There is an improvement in writing $B(r)= \exp(2 \int h(r) \d r)$ so that
\begin{equation}
g(r)^2 = 2 h(r)^2 + h'(r) - h(r)/r
\end{equation}
which is the basis of \cite{Rahman}.

\subsection{The seventh and eighth theorems}
\begin{theorem}
In Isotropic coordinates, if $\{\zeta(r),B(r)\}$ describe a perfect fluid then so does  $\{\zeta(r)^{-1},B(r)\}$. This is the Buchdahl transformation in disguise.

That is, the geometry defined by holding $B_0(r)$ fixed and setting
\begin{equation}
\d s^2 = - \frac{1}{\zeta_0(r)^2} \; \d t^2 +{\zeta_0(r)^2\over\;B_0(r)^2} \{ \d r^2 + r^2 \d\Omega^2 \}
\end{equation}
is also a perfect fluid sphere. Alternatively, the mapping
\begin{equation}
\gth{7}: \left\{ \zeta_0, B_0  \right\} \mapsto \left\{\zeta_0^{-1} , B_0 \right\}
\end{equation}
takes perfect fluid spheres into perfect fluid spheres.
\end{theorem}
\begin{proof}[Proof for Theorem 7:]
Assuming that $\left\{ \zeta_{0}(r)^{-1},B_{0}(r) \right\}$  solves equation (\ref{ode_for_iso_zeta}).

We consider \emph{L.H.S.} of the equation (\ref{ode_for_iso_zeta}),
\begin{equation}
\zeta_0 \to \zeta_1 :
\left({\zeta_0'\over\zeta_0}\right)^2 \to \left(\left(\zeta_0^{-1}\right)'\over \zeta_0^{-1}\right)^2
=\left(\left(-\zeta_0'\; \zeta_0^{-2}\right)\over\zeta_0^{-1}\right)^2
= \left({-\zeta_0'\over\zeta_0}\right)^2 
=\left({\zeta_0'\over\zeta_0}\right)^2
\end{equation}
This shows that $\{\zeta(r),B(r)\}$ describe a perfect fluid then so does  $\{\zeta(r)^{-1},B(r)\}$.

The transformation $\gth{7}$ defined in Theorem 7 is a ``square root of unity'' in the sense that:
\begin{equation}
\gth{7} \circ\gth{7} = I,
\end{equation}

To see this, consider the sequence
\begin{equation}
\{\zeta_0,B_0\} \mapsto \{\zeta_1,B_0\} \mapsto \{\zeta_2,B_0\} \dots
\end{equation}
In this section we consider a two-step composition to check that $\gth{7}$ is a square root of unity. In particular, to be more explicit about this note that at step one
\begin{equation}
\zeta_{1} = \zeta_{0}^{-1}
\end{equation}
while at the second step
\begin{equation}
\zeta_{2} = \zeta_{1}^{-1} = (\zeta_{0}^{-1})^{-1} =\zeta_{0}
\end{equation}
For reasons of convenience and without loss of generality, we can write for $\zeta(r)$, after applying thorem 1 $n$-times, as :
$\zeta(r) = \zeta(r)^{-1}$, for $n$ = odd number and $\zeta(r) = \zeta(r)$, for $n$ =
even number.
\end{proof}
\begin{theorem}
Let $\{\zeta_0(r),B_0(r)\}$ describe a perfect fluid sphere. Define
\begin{equation}
Z_0 =  \left\{\sigma +\epsilon \int {r \d r \over B_0(r)^2} \right\}.
\end{equation}
Then for all $\sigma$ and $\epsilon$, the geometry defined by holding $\zeta_0(r)$ fixed and setting
\begin{equation} \label{Buchdahl}
\d s^2 = - \zeta_0(r)^2 \; \d t^2 +{1\over \zeta_0(r)^2 \;B_0(r)^2 \; Z_0(r)^2} \{ \d r^2 + r^2 \d\Omega^2 \}
\end{equation}
is also a perfect fluid sphere. That is, the mapping
\begin{equation}
\gth{8}(\sigma,\epsilon): \left\{ \zeta_0, B_0  \right\} \mapsto \left\{ \zeta_0, B_0 \; Z_0(B_0) \right\}
\end{equation}
takes perfect fluid spheres into perfect fluid spheres.
\end{theorem}
\begin{proof}[Proof for Theorem 8]
The proof of theorem 8 is based on the technique of ``reduction in order''.
Assuming that $\left\{ \zeta_{0}(r),B_{0}(r) \right\}$  solves equation (\ref{second_ode_iso}), write
\begin{equation}
B_1(r)= B_{0}(r) \; Z_0(r) \, .
\end{equation}
and demand that $\left\{ \zeta_{0}(r),B_{1}(r) \right\}$ also solves equation (\ref{second_ode_iso}).
We find
\begin{equation}
(B_0 \; Z_0)'' - (B_0 \; Z_0)'/r -2g^2 (B_0 \; Z_0)  = 0.
\end{equation}
we can expand the above equation to
\begin{equation}
(B_0''\; Z_0 + 2 B_0'\;Z_0' + B_0 \; Z_0'') - (B_0' \; Z_0)/r - (B_0 \; Z_0')/r -2g^2 (B_0 \; Z_0)  = 0.
\end{equation}
we can also re-group this same equation as
\begin{equation}
\left\{B_0'' - B_0'/r -2g^2 B_0 \right\} Z_0 + 2 B_0' Z_0' + B_0 Z_0'' - (B_0 Z_0') /r = 0.
\end{equation}
A linear homogeneous 2nd order ODE for $B$ now simplifies to
\begin{equation}
\label{ode_iso_zeta}
(2 B_0' - B_0 / r) Z_0' + B_0 Z_0'' = 0 \, ,
\end{equation}
which is an ordinary homogeneous second-order differential equation, depending only on $Z_0'$
and $Z_0''$. (So it can be viewed as a first-order homogeneous order differential equation in $Z'$, which is solvable.)
Separating the unknown variable to one side,
\begin{equation} \label{de_for_iso_zetaprime}
\frac{Z_0''}{Z_0'}=- 2  \frac{B_0'}{B_0} + \frac{1}{r} \, .
\end{equation}
Re-write $Z_0''/ Z_0' = \d\ln(Z_0')/\d r$, and integrate twice over both sides of
equation (\ref{de_for_iso_zetaprime}), to obtain
\begin{equation} \label{eq_for_iso_zeta_1}
Z_0(r) = \left\{\sigma +\epsilon \int {r \; \d r \over B_0(r)^2} \right\}.
\end{equation}
depending on the old solution $\left\{ \zeta_0 (r) , B_0 (r)  \right\}$, and two
arbitrary integration constants $\sigma$ and $\epsilon$.
\end{proof}
\begin{proof}[``Idempotence'' Proof for Theorem 8]
To see that the transformation $\gth{8}$ defined in Theorem 8 is ``idempotent'' we first show
\begin{equation}
\gth{8} \circ\gth{8} \SIM \gth{8},
\end{equation}
and then iterate.
The precise two-step composition rule is
\begin{equation}
\gth{8}(\sigma_2,\epsilon_2)\circ\gth{8}(\sigma_1,\epsilon_1) = 
\gth{8}\left(\sigma_2\sigma_1, \;\epsilon_1\sigma_2+{\epsilon_2\over\sigma_1} \right).
\end{equation}
In this section we consider a two-step composition rule to find idempotence. In particular, the most outstanding features of these steps are:
\begin{itemize}
\item 
To see ``idempotence'', note that for fixed $\zeta_0(r)$ equation (\ref{second_ode_iso}) has a solution space that is exactly two dimensional.
\item
Since the first application of  $\gth{8}$  takes any specific solution and maps it into the full two-dimensional solution space, any subsequent application of  $\gth{8}$ can do no more than move one around inside this two dimensional solution space --- physically this corresponds to a relabelling of parameters describing the perfect fluid metric you are dealing with, not the generation of new solutions.
\end{itemize}
To be more explicit about this note that at step one
\begin{equation}
B_0 \to B_1 = B_0 \; \left\{
\sigma_1 +\epsilon_1 \lint {r \; \d r\over B_0(r)^2}
\right\},
\end{equation}
while at the second step
\begin{equation}
B_1 \to B_2 = B_1 \; \left\{
\sigma_2 +\epsilon_2 \lint {r \; \d r\over B_1(r)^2}
\right\}.
\end{equation}
That is:
\begin{eqnarray}
B_2 &=& B_0 \;  \left\{
\sigma_1 +\epsilon_1 \lint {r \; \d r\over  B_0(r)^2} 
\right\}
\\
&&
\times \left\{
\sigma_2 +\epsilon_2 \lint {r \; \d r\over B_0(r)^2 \;  \;
[\sigma_1 +\epsilon_1 \lint {r \; \d r/ B_0(r)^2}
]^2  }
\right\}.
\nonumber
\end{eqnarray}
But this can be rewritten as
\begin{eqnarray}
B_2 &=& B_0 \; Z_0 \; 
\left\{ \sigma_2 + {\epsilon_2\over\epsilon_1} \lint { \d Z_0\over Z_0^2 } \right\}
\nonumber
\\
&=&
B_0 \; Z_0 \; 
\left\{ \sigma_2 - {\epsilon_2\over\epsilon_1} \int { \d \left({1\over Z_0}\right) } \right\}
\nonumber
\\
&=&
B_0 \; Z_0 \; 
\left\{ \sigma_2 - {\epsilon_2\over\epsilon_1}  \left[{1\over Z_0 }-{1\over \sigma_1}\right] \right\}.
\end{eqnarray}
Therefore
\begin{equation}
B_2
=
B_0 \;  \left\{  - {\epsilon_2\over\epsilon_1}  + 
\left[\sigma_2+{\epsilon_2\over\epsilon_1} \;{1\over \sigma_1}\right] \; Z_0 \right\}.
\end{equation}
That is
\begin{equation}
B_1 =  - {\epsilon_2\over\epsilon_1}  
+ \left[\sigma_2+{\epsilon_2\over\epsilon_1} \;{1\over \sigma_1}\right]  \; Z_0,
\end{equation}
from which the composition law follows:
\begin{equation}
\gth{8}(\sigma_2,\epsilon_2)\circ\gth{8}(\sigma_1,\epsilon_1) = 
\gth{8}\left(\sigma_2\sigma_1, \;\epsilon_1\sigma_2+{\epsilon_2\over\sigma_1} \right)
\end{equation}
\end{proof}
Note that this almost identical to the computation performed in the Schwarzschild coordinates because idempotence of the theorem is a result of ``reduction of order" not something special to general relativity.


\subsection{Two linking generating theorems:}
Having now found the first and second generating theorems, is now useful to find the connection between them.
We start by taking a perfect fluid sphere solution $\left\{ \zeta_0, B_0  \right\}$ and applying theorem 7 onto it which gives us a new perfect fluid sphere $\left\{ \zeta_1, B_0 \right\}$. The new $\zeta_1 = \zeta_0^{-1}$. We now continue with applying theorem 8, again we get a new solution $\{\zeta_0^{-1}, B_1 \}$, where $B_1$ now depends on the $B_0$.
All together we can consider this as a single process by introducing the following corollary: \\
\begin{corollary}
If $\{\zeta_0,B_0\}$ denotes a perfect fluid sphere, then for all $\sigma$ and $\epsilon$, the two-parameters geometry defined by
\begin{equation} \label{Th1_Th2}
\d s^2 = - \frac{1}{\zeta_0(r)^2} \; \d t^2 +{\zeta_0(r)^2\over\;B_0(r)^2\;Z_0(r)^2} \{ \d r^2 + r^2 \d\Omega^2 \}
\end{equation}
is also a perfect fluid sphere, where $Z_0$ is
\begin{equation}
Z_0(r) = \left\{\sigma +\epsilon \int {r \d r \over B_0(r)^2} \right\}.
\end{equation}
That is
\begin{equation}
\gth{8}\circ\gth{7}: \{\zeta_0,B_0\} \mapsto \{\zeta^{-1}_0,B_0\} \mapsto
\{\zeta^{-1}_0, B_0 \; Z_0 (B_0)\}
\end{equation}
\end{corollary}
Instead of starting with theorem 7 we could first apply theorem 8 on $\left\{ \zeta_0, B_0  \right\}$.
This gives us a new perfect fluid sphere $\left\{ \zeta_0, B_1  \right\}$,
where $B_1= B_0 \; Z_0(B_0)$ is given by equation (\ref{eq_for_iso_zeta_1}). We now continue with theorem 7 which leads to $\{ \zeta_1, B_1 \}$, where $\zeta_1 = \zeta^{-1}$. Again, we can consider this as a single process by introducing the following corollary:\\
\begin{corollary}
If $\{\zeta_0,B_0\}$ denotes a perfect fluid sphere, then for all $\sigma$ and $\epsilon$, the two-parameters geometry defined by
\begin{equation} \label{Th2_Th1}
\d s^2 = - \frac{1}{\zeta_0(r)^2} \; \d t^2 +{\zeta_0(r)^2\over\;B_0(r)^2\;Z_0(r)^2} \{ \d r^2 + r^2 \d\Omega^2 \}
\end{equation}
is also a perfect fluid sphere, where $Z_0$ is
\begin{equation}
Z_0(r) = \left\{\sigma +\epsilon \int {r \d r \over B_0(r)^2} \right\}.
\end{equation}
That is
\begin{equation}
\gth{7}\circ\gth{8}: \{\zeta_0,B_0\} \mapsto \{\zeta_0,B_0 \; Z_0(B_0)\} \mapsto
\{\zeta^{-1}_0, B_0 \; Z_0 (B_0)\}
\end{equation}
\end{corollary}
There are general comments about the relationship between both transformation theorems.
\begin{itemize}
\item
These two transformation operations are \emph{independent} of each other.
\item
The transformation $\gth{7}$ defined in Theorem 7 is a ``square root of unity''.
\item
Theorem 8 is in general idempotent. That means when we apply theorem 8 more than once no further solution will be obtained.
\item
Theorem 7 and theorem 8 are in general commute.
\end{itemize}
In addition, these comments can be described in terms of the equations below.
\begin{equation}
\gth{7} \not\SIM \gth{8}; \qquad \gth{7} \circ \gth{7} \SIM I; \qquad  \gth{8} \circ \gth{8} \SIM \gth{8};
 \qquad \gth{7} \circ \gth{8} \SIM \gth{8} \circ \gth{7}.
\end{equation}
\begin{figure}[!ht]
\centering
\includegraphics[scale = 0.6]{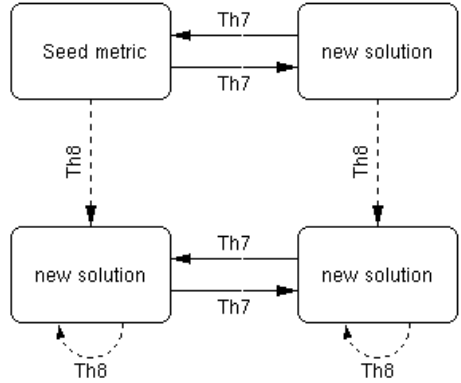}
\caption{\label{Structure graph for theorem 7 and 8}This structure shows that Theorem 7 and Theorem 8 are in general distinct. When we apply Theorem 7 to seed metric, we get a new solution, while applying Theorem 8 to seed metric, we get the other new solution. Furthermore, the diagram shows that Theorem 7 and Theorem 8 in general commute.}
\end{figure}
\def\xxx{
\subsection{Formal properties of theorems seven and eight}
The solution generating theorems we have developed interact in a number of interesting ways and exhibit numerous formal properties that will be useful when classifying generic perfect fluid spheres.
The idempotence of $\gth{7}$ and $\gth{8}$ in this formalism is:
\begin{equation}
\begin{split}
\left( \gth{7} \circ \gth{7} \right) g =& \gth{7}\left(\gth{7}(g) \right) \SIM \gth{7}(g) \, ;\\
\left( \gth{8} \circ \gth{8} \right) g =& \gth{8}\left(\gth{8}(g) \right) \SIM \gth{8}(g) \, .\\
\end{split} 
\end{equation}

Taken together, it is possible to simplify all formulae wherever $\gth{7}$ and $\gth{8}$
appear next to each other more than once. The following examples should demonstrate how this works:
\begin{equation}
\begin{split}
\left( \gth{7} \circ \gth{8} \right) g  \SIM \left( \gth{7} \circ \gth{8} \circ \gth{8} \right) g &\equiv  \left( \gth{8} \circ \gth{7} \circ \gth{8} \right) g \equiv  \left( \gth{8} \circ \gth{8} \circ \gth{7} \right) g \SIM \left( \gth{8} \circ \gth{7} \right) g 
; 
\\
\left( \gth{8} \circ \gth{7} \right) g \SIM \left( \gth{8} \circ \gth{7} \circ \gth{7} \right) g 
&\equiv  \left( \gth{7} \circ \gth{8} \circ \gth{7} \right) g
\equiv  \left( \gth{7} \circ \gth{7} \circ \gth{8} \right) g \SIM \left( \gth{7} \circ \gth{8} \right) g  \, , \\
\end{split}
\end{equation}
}

\section{Gaussian polar coordinates:}
Consider the metric
\begin{equation}
\d s^2 = - \zeta(r)^2 \; \d t^2 +  \d r^2 + R(r)^2 \d\Omega^2 
\end{equation}
and then assume it represents a perfect fluid sphere. That is,
$ G_{\hat\theta\hat\theta} = G_{\hat r\hat r} =G_{\hat\phi\hat\phi} $.
Now $G_{\hat\theta\hat\theta} = G_{\hat\phi\hat\phi}$ is always fulfilled due to spherical symmetry.

We calculate
\begin{equation}
G_{\hat r\hat r} = -\frac{2 \zeta' R R' - \zeta + \zeta (R')^2}{R^2 \zeta}
\end{equation}
and
\begin{equation}
G_{\hat\theta\hat\theta} = -\frac{ \zeta' R' + R'' \zeta + \zeta'' R}{R \zeta}
\end{equation}
We set $G_{\hat r\hat r} = G_{\hat\theta\hat\theta}$, which supplies us with an 2nd order ODE for $\zeta(r)$, which reduces the freedom to choose the two functions to one:
\begin{equation} \label{second_ode_polar}
\zeta'' R^2 - \zeta' R R' + \zeta[1-(R')^2+R'' R]=0 
\end{equation}
We can rewrite this equation as
\begin{equation} \label{second_ode_polar2}
\zeta'' - \zeta' {R' \over R} +\zeta  \left\{ {1-(R')^2+ R''R\over R^2 }\right\} = 0
\end{equation}
\subsection{The ninth theorem}
\begin{theorem}
Suppose $\{ \zeta_0(r), R_0(r) \}$ represents a perfect fluid sphere.
Define
\begin{equation}
\Lambda_0(r) =  \left\{\sigma +\epsilon \int {R_0(r) \d r \over \zeta_0(r)^2} \right\}.
\end{equation}
Then for all $\sigma$ and $\epsilon$, the geometry defined by holding $R_0(r)$ fixed and setting
\begin{equation} 
\d s^2 = - \zeta_0(r)^2 \Lambda_0(r)^2\; \d t^2 +  \d r^2 + R_0(r)^2 \d\Omega^2 
\end{equation}
is also a perfect fluid sphere. That is, the mapping
\begin{equation}
\gth{9}(\sigma,\epsilon): \left\{ \zeta_0, R_0 \right\} \mapsto \left\{\zeta_0 \; \Lambda_0(\zeta_0, R_0) ,R_0 \right\}
\end{equation}
takes perfect fluid spheres into perfect fluid spheres.
\end{theorem}
\begin{proof}[Proof for Theorem 9]
The proof of theorem 9 is again based on the technique of ``reduction in order''.
Assuming that $\left\{ \zeta_{0}(r),R_{0}(r) \right\}$  solves again equation (\ref{second_ode_polar}), write
\begin{equation}
\zeta_1(r) = \zeta_0(r) \; \Lambda_0(r) \, .
\end{equation}
and demand that $\left\{ \zeta_{1}(r),R_{0}(r) \right\}$ also solves equation (\ref{second_ode_polar}).
We find
\begin{equation}
(\zeta_0 \; \Lambda_0)'' - (\zeta_0 \; \Lambda_0)' {R' \over R} + (\zeta_0 \; \Lambda_0) \left\{ {1-(R')^2+ R''R\over R^2 }\right\} = 0
\end{equation}
we can expand the above equation to
\begin{equation}
(\zeta_0'' \; \Lambda_0 + 2 \zeta_0' \; \Lambda_0' + \zeta_0 \; \Lambda_0'') - (\zeta_0' \; \Lambda_0 + \zeta_0 \; \Lambda_0') {R' \over R}
+ (\zeta_0 \; \Lambda_0) \left\{ {1-(R')^2+ R''R\over R^2 }\right\} = 0
\end{equation}
we can also re-group this same equation as
\begin{equation}
\left\{\zeta_0'' - \zeta_0' {R' \over R} + \zeta_0 \left\{ {1-(R')^2+ R''R\over R^2 }\right\}\right\} \Lambda_0 
+ 2 \zeta_0' \; \Lambda_0' + \zeta_0 \; \Lambda_0'' - \zeta_0 \; \Lambda_0' {R' \over R} = 0.
\end{equation}
A linear homogeneous 2nd order ODE for $\zeta$ now simplifies to
\begin{equation}
\label{ode_polar_zeta}
(2 \zeta_0' -  \zeta_0 \; {R' \over R}) \Lambda_0' + \zeta_0 \; \Lambda_0'' = 0 \, ,
\end{equation}
which is an ordinary homogeneous second-order differential equation, depending only on $\Lambda_0'$ and $\Lambda_0''$. (So it can be viewed as a first-order homogeneous order differential equation in $\Lambda_0'$, which is solvable.)
Separating the unknown variable to one side,
\begin{equation} \label{de_for_polar_lambdaprime}
\frac{\Lambda_0''}{\Lambda_0'}=- 2  \frac{\zeta_0'}{\zeta_0} + \frac{R_0'}{R_0} \, .
\end{equation}
Re-write $\Lambda_0''/ \Lambda_0' = \d\ln(\Lambda_0')/\d r$, and integrate twice over both sides of
equation (\ref{de_for_polar_lambdaprime}), to obtain
\begin{equation} \label{eq_for_polar_zeta_1}
\Lambda_0(r) = \left\{\sigma +\epsilon \int {R_0(r) \; \d r \over \zeta_0(r)^2} \right\}.
\end{equation}
depending on the old solution $\left\{ \zeta_0 (r) , R_0 (r)  \right\}$, and two
arbitrary integration constants $\sigma$ and $\epsilon$.
\end{proof}
\begin{proof}[``Idempotence'' Proof for Theorem 9]
To see that the transformation $\gth{9}$ defined in Theorem 3 is ``idempotent'' we first show
\begin{equation}
\gth{9} \circ\gth{9} \SIM \gth{9},
\end{equation}
and then iterate.
The precise two-step composition rule is
\begin{equation}
\gth{9}(\sigma_2,\epsilon_2)\circ\gth{9}(\sigma_1,\epsilon_1) = 
\gth{9}\left(\sigma_2\sigma_1, \;\epsilon_1\sigma_2+{\epsilon_2\over\sigma_1} \right).
\end{equation}
In this section we consider a two-step composition rule to find idempotence. In particular, the most outstanding features of these steps are:
\begin{itemize}
\item 
To see ``idempotence'', note that for fixed $R_0(r)$ equation (\ref{second_ode_polar2}) has a solution space that is exactly two dimensional.
\item
Since the first application of  $\gth{9}$  takes any specific solution and maps it into the full two-dimensional solution space, any subsequent application of  $\gth{9}$ can do no more than move one around inside this two dimensional solution space --- physically this corresponds to a relabelling of parameters describing the perfect fluid metric you are dealing with, not the generation of new solutions.
\end{itemize}
To be more explicit about this note that at step one
\begin{equation}
\zeta_0 \to \zeta_1 = \zeta_0 \; \left\{
\sigma_1 +\epsilon_1 \lint {R_0(r) \; \d r\over \zeta_0(r)^2}
\right\},
\end{equation}
while at the second step
\begin{equation}
\zeta_1 \to \zeta_2 = \zeta_1 \; \left\{
\sigma_2 +\epsilon_2 \lint {R_0 \; \d r\over \zeta_1(r)^2}
\right\}.
\end{equation}
That is:
\begin{eqnarray}
\zeta_2 &=& \zeta_0 \;  \left\{
\sigma_1 +\epsilon_1 \lint {R_0(r) \; \d r\over  \zeta_0(r)^2} 
\right\}
\\
&&
\times \left\{
\sigma_2 +\epsilon_2 \lint {R_0(r) \; \d r\over \zeta_0(r)^2 \;  \;
[\sigma_1 +\epsilon_1 \lint {R_0(r) \; \d r/ \zeta_0(r)^2}
]^2  }
\right\}.
\nonumber
\end{eqnarray}
But this can be rewritten as
\begin{eqnarray}
\zeta_2 &=& \zeta_0 \; Z_0 \; 
\left\{ \sigma_2 + {\epsilon_2\over\epsilon_1} \lint { \d Z_0\over Z_0^2 } \right\}
\nonumber
\\
&=&
\zeta_0 \; Z_0 \; 
\left\{ \sigma_2 - {\epsilon_2\over\epsilon_1} \int { \d \left({1\over Z_0}\right) } \right\}
\nonumber
\\
&=&
\zeta_0 \; Z_0 \; 
\left\{ \sigma_2 - {\epsilon_2\over\epsilon_1}  \left[{1\over Z_0 }-{1\over \sigma_1}\right] \right\}.
\end{eqnarray}
Therefore
\begin{equation}
\zeta_2
=
\zeta_0 \;  \left\{  - {\epsilon_2\over\epsilon_1}  + 
\left[\sigma_2+{\epsilon_2\over\epsilon_1} \;{1\over \sigma_1}\right] \; Z_0 \right\}.
\end{equation}
That is
\begin{equation}
\zeta_1 =  - {\epsilon_2\over\epsilon_1}  
+ \left[\sigma_2+{\epsilon_2\over\epsilon_1} \;{1\over \sigma_1}\right]  \; Z_0,
\end{equation}
from which the composition law follows:
\begin{equation}
\gth{9}(\sigma_2,\epsilon_2)\circ\gth{9}(\sigma_1,\epsilon_1) = 
\gth{9}\left(\sigma_2\sigma_1, \;\epsilon_1\sigma_2+{\epsilon_2\over\sigma_1} \right)
\end{equation}
\end{proof}
Again, note simularities to what happens in Schwarzschild coordinates.
\begin{figure}[!ht]
\centering
\includegraphics[scale = 0.6]{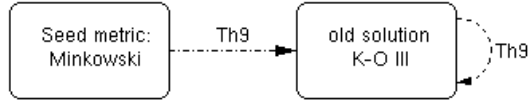}
\caption{\label{Structure graph for theorem 9}This structure shows that Theorem 9 is idempotent. When we apply Theorem 9 to Minkowski as a seed metric, we get the K-O III as a new solution.}
\end{figure}
%
%
%
\bigskip
\section{Buchdahl coordinates:}
Consider the metric
\begin{equation}
\d s^2 = - \zeta(r)^{-1} \; \d t^2 + \zeta(r)\d r^2 + \zeta(r) R(r)^2 \d\Omega^2 
\end{equation}
and then assume it represents a perfect fluid sphere. That is,
$ G_{\hat\theta\hat\theta} = G_{\hat r\hat r} =G_{\hat\phi\hat\phi} $.
Note $G_{\hat\theta\hat\theta} = G_{\hat\phi\hat\phi}$ is always fulfilled due to spherical symmetry.
We calculate
\begin{equation}
G_{\hat r\hat r} = \frac{1}{4} \; \left(\frac{-4 \; \zeta^2 \; (R')^2 + 4 \; \zeta^2 + (\zeta')^2 \; R^2}{\zeta^3 \; R^2}\right)
\end{equation}
and
\begin{equation}
G_{\hat\theta\hat\theta} = - \frac{1}{4} \; \left(\frac{4 \; \zeta^2 \; R'' + (\zeta')^2 \; R}{\zeta^3 \; R}\right)
\end{equation}
We set $G_{\hat r\hat r} = G_{\hat\theta\hat\theta}$, which supplies us with an 2nd order ODE for $\zeta(r)$, which reduces the freedom to choose the two functions to one:
\begin{equation} \label{ode_for_Buchdahl}
\left({\zeta'\over\zeta}\right)^2 =-{2[1-(R')^2+R R'']\over R^2}.
\end{equation}
This coordinate system is a sort of cross between isothermal and proper radius.
\begin{theorem}
If $\{\zeta(r),R(r)\}$ describes a perfect fluid then so does  $\{\zeta(r)^{-1},R(r)\}$. This is the Buchdahl transformation in yet another disguise.

The geometry defined by holding $R_0(r)$ fixed and setting
\begin{equation}
\d s^2 = - \zeta_0(r) \; \d t^2 + {1\over\;\zeta_0(r)}  \d r^2 + \frac{R_0(r)^2}{\zeta_0(r)} \d\Omega^2 
\end{equation}
is also a perfect fluid sphere. That is, the mapping
\begin{equation}
\gth{10}: \left\{ \zeta_0, R_0  \right\} \mapsto \left\{\zeta_0^{-1} , R_0 \right\}
\end{equation}
takes perfect fluid spheres into perfect fluid spheres.
\end{theorem}




\bigskip
\section{Synge isothermal coordinates:}
Consider
\begin{equation}
\d s^2 = - \zeta(r)^{-2} \; \{ \d t^2 - \d r^2\} + \{\zeta(r)^{-2} \;R(r)^2 \d\Omega^2 \}
\end{equation}
and then assume it represents a perfect fluid sphere. That is,
$ G_{\hat\theta\hat\theta} = G_{\hat r\hat r} =G_{\hat\phi\hat\phi} $.
While $G_{\hat\theta\hat\theta} = G_{\hat\phi\hat\phi}$ is always fulfilled due to spherical symmetry.
We calculate
\begin{equation}
G_{\hat r\hat r} = - \frac{3  R^2 \; (\zeta')^2 - 4 R \zeta  R' \; \zeta' + (R')^2 \zeta^2 - \zeta^2}{R^2}
\end{equation}
and
\begin{equation}
G_{\hat\theta\hat\theta} =  - \frac{3 R (\zeta')^2 - 2 \zeta R' \zeta' - 2 R \; \zeta \zeta'' + \zeta^2 R''}{R}
\end{equation}
We set $G_{\hat r\hat r} = G_{\hat\theta\hat\theta}$, which supplies us with an 2nd order ODE for $\zeta(r)$, which reduces the freedom to choose the two functions in equation
\begin{equation} \label{second_order_synge}
\zeta''-\zeta' {R'\over R} - \zeta \left\{ {1-(R')^2+R R''\over 2 R^2}  \right\} = 0.
\end{equation}
This is very similar equation to Gaussian polar.
\begin{theorem}
Suppose $\{ \zeta_0(r), R_0(r) \}$ represents a perfect fluid sphere.
Define
\begin{equation}
A_0(r) =  \left\{\sigma +\epsilon \int {R_0(r) \d r \over \zeta_0(r)^2} \right\}.
\end{equation}
Then for all $\sigma$ and $\epsilon$, the geometry defined by holding $R_0(r)$ fixed and setting
\begin{equation} 
\d s^2 = - \frac{1}{\zeta_0(r)^2 \; A_0(r)^2}\; \left\{\d t^2 - \d r^2\right\} + \frac{R_0(r)^2}{\zeta_0(r)^2 \; A_0(r)^2} \d\Omega^2 
\end{equation}
is also a perfect fluid sphere. That is, the mapping
\begin{equation}
\gth{11}(\sigma,\epsilon): \left\{ \zeta_0, R_0 \right\} \mapsto \left\{\zeta_0 \; A_0(\zeta_0, R_0) ,R_0 \right\}
\end{equation}
takes perfect fluid spheres into perfect fluid spheres.
\end{theorem}
\begin{proof}[Proof for Theorem 11]
The proof of theorem 11 is based on the technique of ``reduction in order''.
Assuming that $\left\{ \zeta_{0}(r),R_{0}(r) \right\}$  solves equation (\ref{second_order_synge}), write
\begin{equation}
\zeta_1(r) = \zeta_0(r) \; A_0(r) \, .
\end{equation}
and demand that $\left\{ \zeta_{1}(r),R_{0}(r) \right\}$ also solves equation (\ref{second_order_synge}).
We find
\begin{equation}
(\zeta_0 \; A_0)'' - (\zeta_0 \; A_0)' {R' \over R} + (\zeta_0 \; A_0) \left\{ {1-(R')^2+ R R'' \over 2 R^2 }\right\} = 0
\end{equation}
we can expand the above equation to
\begin{equation}
(\zeta_0'' A_0 + 2 \zeta_0' \; A_0' + \zeta_0 \; A_0'') - (\zeta_0' \; A_0 + \zeta_0 \; A_0') {R' \over R} 
+ (\zeta_0 \; A_0) \left\{ {1-(R')^2+ R R''\over 2 R^2 }\right\} = 0
\end{equation}
we can also re-group this same equation as
\begin{equation}
\left\{\zeta_0'' - \zeta_0' {R' \over R} + \zeta_0 \left\{ {1-(R')^2+ R R''\over 2 R^2 }\right\}\right\} A_0 
+ 2 \; \zeta_0' \; A_0' + \zeta_0 \; A_0'' - \zeta_0 \; A_0' {R' \over R} = 0.
\end{equation}
A linear homogeneous 2nd order ODE for $\zeta$ now simplifies to
\begin{equation}
\label{ode_polar_zeta2}
(2 \zeta_0' -  \zeta_0 \; {R' \over R}) A_0' + \zeta_0 \; A_0'' = 0 \, ,
\end{equation}
which is an ordinary homogeneous second-order differential equation, depending only on $A_0'$ and $A_0''$. (So it 
can be viewed as a first-order homogeneous order differential equation in $A_0'$, which is solvable.)
Separating the unknown variable to one side,
\begin{equation} \label{de_for_polar_aprime}
\frac{A_0''}{A_0'}=- 2  \frac{\zeta_0'}{\zeta_0} + \frac{R_0'}{R_0} \, .
\end{equation}
Re-write $A_0''/ A_0' = \d\ln(A_0')/\d r$, and integrate twice over both sides of
equation (\ref{de_for_polar_aprime}), to obtain
\begin{equation} \label{eq_for_synge_zeta_1}
A_0(r) = \left\{\sigma +\epsilon \int {R_0(r) \; \d r \over \zeta_0(r)^2} \right\}.
\end{equation}
depending on the old solution $\left\{ \zeta_0 (r) , R_0 (r)  \right\}$, and two
arbitrary integration constants $\sigma$ and $\epsilon$.
\end{proof}




\bigskip
\section{Exponential coordinates:}
You can always choose to effectively use $g_{tt}$ itself as one of the coordinates, by picking the functional form of $g_{tt}$ to be any arbitrarily specified function $f(z)$. For instance, choose 
\begin{equation}
g_{tt} = - \exp(-2z)
\end{equation}
then a useful choice is
\begin{equation}
\d s^2 = - \exp(-2z) \, \d t^2  + \exp(+2z) \left\{ {\d z^2\over B(z)} + R(z)^2 \, \d\Omega^2 \right\}
\end{equation}
and then assume it represents a perfect fluid sphere. That is,
$ G_{\hat\theta\hat\theta} = G_{\hat z\hat z} =G_{\hat\phi\hat\phi} $.
Note $G_{\hat\theta\hat\theta} = G_{\hat\phi\hat\phi}$ is always fulfilled due to spherical symmetry.
Calculate
\begin{equation}
G_{\hat z\hat z} =  \frac{(B R^2 - (R')^2 B + 1) \exp(-2z)}{R^2}
\end{equation}
and
\begin{equation}
G_{\hat\theta\hat\theta} =  - \frac{1}{2} \; \frac{(B' R' + 2 R'' B + 2 B R) \exp(-2z)}{R}
\end{equation}
We set $G_{\hat z\hat z} = G_{\hat\theta\hat\theta}$, which supplies us with an 1st order ODE for $B$, which reduces the freedom to choose the two functions in the metric to one for which the isotropy constraint becomes
\begin{equation} \label{ODE_for_exponential}
B' [R R' ] + B [4 R^2 - 2 (R')^2 + 2 R R''] + 2 = 0
\end{equation}

\begin{theorem}
Suppose $\{B_0(z), R_0(z) \}$ represents a perfect fluid sphere.
Define
\begin{equation}
\Upsilon_0(z) = \frac{R_0(z)^2}{(R_0'(z))^2} \; \exp\left\{{- 4 \int{\frac{ R_0(z)}{R_0'(z)} \, \d z } }\right\} \, .
\end{equation}
Then for all $\lambda$, the geometry defined by holding $R_0(z)$ fixed and setting
\begin{equation}
\d s^2 = - \exp(-2z) \d t^2  + \exp(+2z) \left\{ {\d z^2\over B_0(z) + \lambda \Upsilon_0(z)} + R_0(z)^2 \d\Omega^2 \right\}
\end{equation}
is also a perfect fluid sphere. That is, the mapping
\begin{equation}
\gth{12}(\sigma,\epsilon): \left\{ B_0, R_0 \right\} \mapsto \left\{B_0 \; \Upsilon_0( R_0) ,R_0 \right\}
\end{equation}
takes perfect fluid spheres into perfect fluid spheres.
\end{theorem}
\begin{proof}[Proof for Theorem 12]
Assume that $\left\{ B_0(z), R_0(z) \right\}$ is a solution for equation (\ref{ODE_for_exponential}).

We want to find under what conditions does $\left\{ B_1(z), R_0(z)  \right\}$  also satisfy equation (\ref{ODE_for_exponential})?
Without loss of generality, we write
\begin{equation}
B_1(z)=B_{0}(z) + \lambda \; \Upsilon_0(z) \, .
\end{equation}
Equation ({\ref{ODE_for_exponential}}) can now be used
to determine $\Upsilon_0(z)$.
Substitute $B_1(z)$ in Equation ({\ref{ODE_for_exponential}})
\begin{equation}
(B_0 + \lambda \; \Upsilon_0)' [R_0 R_0' ] + (B_0 + \lambda \; \Upsilon_0) [4 R_0^2 - 2 (R_0')^2 + 2 R_0 R_0''] + 2 = 0
\end{equation}
we can also re-group this same equation as
\begin{eqnarray}
&&[R_0 R_0' ] B_0'+[4 R_0^2 - 2 (R_0')^2 + 2 R_0 R_0'']B_0 + 2 
\nonumber
\\
&&+ \left[ R_0 R_0'  \right] \Upsilon_{0}'
+ \left[ 4 R_0^2 - 2 (R_0')^2 + 2 R_0 R_0'' \right] \Upsilon_0 = 0
\end{eqnarray}
That ordinary \emph{inhomogeneous} first-order differential equation in $B$ now
simplifies to
\begin{equation}
\label{ode_th6_exponenital}
 \left[ R_0 R_0'  \right] \Upsilon_{0}'
+ \left[ 4 R_0^2 - 2 (R_0')^2 + 2 R_0 R_0'' \right] \Upsilon_0 = 0   \, ,
\end{equation}
which is an ordinary \emph{homogeneous} first-order differential equation in $\Upsilon_0$.\\
Now we can calculate $\Upsilon_0$ by following,
\begin{equation}
\left[ R_0 R_0'  \right] \Upsilon_{0}' = -
\left[ 4 R_0^2 - 2 (R_0')^2 + 2 R_0 R_0'' \right] \Upsilon_0.
\end{equation}
Rearrage it into
\begin{eqnarray}
\frac{\Upsilon_0'}{\Upsilon_0} &=& \frac{- \left[ 4 R_0^2 - 2 (R_0')^2 + 2 R_0 R_0'' \right]}{\left[ R_0 R_0' \right]}
\nonumber
\\
&=& -4 \frac{R_0}{R_0'} + 2 \frac{R_0'}{R_0} -2 \frac{R_0''}{R_0'}
\end{eqnarray}

A straightforward calculation, including an integration by parts, leads to
\begin{equation}
\Upsilon_0(z) = \frac{R_0^2}{(R_0')^2} \; \exp\left\{{- 4 \int{\frac{ R_0}{R_0'} \, \d z } }\right\} \, .
\end{equation}
\end{proof}




\section{Discussion}
We have developed several transformation theorems that map perfect fluid
spheres to perfect fluid sphere using other coordinates. We derived the 
seventh
and eighth theorems using Isotropic coordinates. The transformation in
Theorem 7 is a ``square root of unity''. Indeed, Theorem 8 is alomost
identical to the calculation performed in the Schwarzschild coordinates. In
addition, the idempotence of this theorem is a result of ``reduction of
order'', which is not special to general relativity.
We are able to classify the connection between both transformation theorems.
The two transformation operations are independent of each other.
Furthermore, both theorems in general commute. Figure \ref{Structure
graph for theorem 7 and 8} shows that after applying Theorem 7 to the seed
metric, we get a new solution while applying Theorem 8 to seed metric, we
get the other new solution. By the idempotence of Theorem 8 $n$ applications
of $\gth{8}$ to the seed metric still results in the same non-seed metric.
Furthermore, the diagram shows that Theorem 7 and Theorem 8 in general
commute.

In Gaussian polar coordinates, we established another transformation 
theorem,
which has many similarities to the results in Schwarzschild coordinates. 
Figure
\ref{Structure graph for theorem 9} shows that Theorem 9 is idempotent. When
we apply Theorem 9 to Minkowski as a seed metric, we get the K-O III as a
new solution.

Furthermore, we have also established the transformation theorem using
Buchdahl coordinates. Theorem 10 is the Buchdahl transformation in yet
another disguise.

In Synge isothermal coordinates, we presented Theorem 11 is a transformation
theorem. However, this transformation is almost identical to the computation
performed in the Schwarzschild coordinates.

The last solution generating theorem we presented using exponential
coordinates. Indeed, this Theorem 12 also map perfect fluid spheres to
perfect fluid spheres.

\chapter{TOV equation - related theorems}
\section{Introduction}
\emph{The Tolman-Oppenheimer-Volkoff (TOV) equations} describe the hydrostatic equilibrium of perfect fluid stars without rotation in general relativity. We can solve the \emph{TOV} equations by constructing the equation of state for the entire region of nonzero density starting from the higher density at center to the surface density.
In this chapter, we can rephrase Theorems 1, 2, 3 and 4 directly in terms of the TOV equation. In addition, we also can rephrase all these theorems, which apply in Schwarzschild coordinates directly in terms of the pressure profile and density profile.
\bigskip

\section{TOV equation}
\subsection{Interior equation}
In general relativity \cite{Wald}, continuous matter distributions and fields again are described by a stress-energy tensor $T_{ab}$. The stress tensor of a perfect fluid is given by
\begin{equation}
\label{T_{ab}}
T^{ab} = \rho \; u^a u^b + p \; (g^{ab} + u^a u^b)
\end{equation}
where
$\rho(r)$ is density of mass-energy distribution in rest-frame of fluid,
$u$ is 4\emph{-velocity} of fluid, and $p$ is isotropic pressure in rest-frame of fluids.

We know that
\begin{equation}
G_{ab} = 8 \pi G_{N} \, T_{ab}
\end{equation}
We set $G_{N} = 1$ for convenience, and note that:
\begin{equation}
\nabla_{a} G^{ab} = 0
\end{equation}
We know that as contracted Bianchi identity.

The equation of motion for the matter is
\begin{equation}
\label{na_ble}
\nabla_{b} T^{ab} = 0
\end{equation}
We define
\begin{equation}
u^a = {\txt \d x^a \over \txt \d \tau}
\end{equation}
This is the 4-velocity of a fluid element. We also define
\begin{equation}
A^a = {\txt \d ^2 x^a \over \txt \d \tau ^2} = { \txt \d u^a \over \txt \d \tau}
\end{equation}
This is the 4-accerelation.
\\
In the rest frame of the fluid element $u^a = (c,\overrightarrow{0}) \;  \; \text{and}, \; A^a = (0, \overrightarrow{a}).$
\\
It follows that $u_a A^a = 0$.

It is an invariant quantity. In other words, the 4-acceleration of a particle is always orthogonal to its 4-velocity. Writing
\begin{equation}
{\d \over \d s} = (u \,\nabla)
\end{equation}
We see
\begin{equation}
0 = {\d \over \d s}(-1) = {\d \over \d s} (g_{ab} \, u^a u^b) = 2 \, u_{a} \, {\d \over \d s} u^{a} = 2 \, u_{a} \, (u \nabla) u^{a} = 2 \, u_{a} A^{a} = 0
\end{equation}
Now consider $\nabla_{b} T^{ab}$; we can write this as
\begin{eqnarray}
\label{na_ble2}
\nabla_{b} T^{ab} &=&\nabla_{b}\left[\rho u^a u^b + p (g^{ab} + u^a u^b)\right] 
\nonumber
\\
&=&(\nabla_{b} \rho) u^a u^b + (\nabla_{b} p) (g^{ab} + u^a u^b)
\nonumber
\\
&&+ (\rho + p) (\nabla_{b} u^a) u^b + (\rho + p) u^a (\nabla_{b} u^b) 
\nonumber
\\
&=& \left[(\nabla_{b} \rho) u^b + (\rho + p) (\nabla \cdot u)\right] u^a
\nonumber
\\
&& +\left[(\rho + p) A^{a} + \nabla_b p (g^{ab} + u^a u^b)\right]
\nonumber
\\
&=& 0
\end{eqnarray}
Writing out equation (\ref{na_ble}) in terms of $\rho$, $p$, and $u^{b}$, and projecting the resulting equation parallel and perpendicular to $u_{a}$, we can spilt up the equation into two equations:
\begin{eqnarray}
\left[(\nabla_{b} \rho) u^b + (\rho + p) (\nabla \cdot u)\right] u^a &=& 0,
\\
\left[(\rho + p) A^{a} + \nabla_b p (g^{ab} + u^a u^b)\right] &=& 0.
\end{eqnarray}
\begin{figure}[!ht]
\centering
\includegraphics[scale = 0.65]{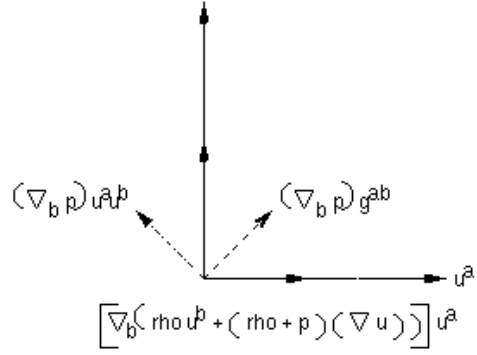}
\label{structure2}
\caption{\label{orthogonal_diagram}The diagram shows that when we write out equation (\ref{na_ble}) in terms of $\rho, p$, and $u^{b}$, and projecting the resulting equation parallel and perpendicular to $u_{a}$, we can spilt up the equation in to two equations.}
\end{figure}
with $u^a u_a =-1$, and $A^a u_a = 0,$ because they are orthogonal to each other.

Now use $(\nabla_{b} p) \, g^{ab} u_a = (\nabla_{b} p) u^b$, and $(\nabla_{b} p) \, u^a u^b u_a = - (\nabla_{b} p) u^b$. This means $(\nabla_{b} p) \, g^{ab} u_a$ and $(\nabla_{b} p) \, u^a u^b u_a $ have the same
magnitude but opposite direction.
We conclude
\begin{equation}
\nabla \cdot (\rho u) + p \, \nabla \cdot u = 0
\end{equation}
and
\begin{equation}
A^a = - {(g^{ab} + u^a u^b) \, \nabla_{b} p \over \rho + p}.
\end{equation}

\section{A perfect fluid sphere in Schwarzschild coordinates}
In the Schwarzschild cooridinates \cite{Wald}, the metric of an arbitrary static, spherically symmetric spacetime can be put into the simple form
\begin{equation}
\d s^2 = -f(r) \; \d t^2 + h(r) \; \d r^2 + r^2(\d \theta^2 + \sin^2 \theta \; \d \phi^2)
\end{equation}
when $G_N = 1$, then $G_{ab} = 8 \pi T_{ab}$.
\\
The Einstein equations yield,
\begin{eqnarray}
\label{T_00}
8 \pi \rho &=& G_{00} = R_{00} + {1 \over 2} \; (R_0{}^0 + R_1{}^1 + R_2{}^2 + R_3{}^3)
\nonumber
\\
&=& (r h^2)^{-1} \; h' + r^{-2} (1 - h^{-1}),
\end{eqnarray}
\begin{eqnarray}
\label{T_11}
8 \pi p &=& G_{11} = R_{11} - {1 \over 2} \; (R_0{}^0 + R_1{}^1 + R_2{}^2 + R_3{}^3)
\nonumber
\\ 
&=& (r f h)^{-1} f' - r^{-2} (1 - h^{-1}),
\end{eqnarray}
\begin{eqnarray}
8 \pi p &=& G_{22} = {1 \over 2} (f h)^{-1/2} {\d \over \d r} [(f h)^{-1/2} f']
\nonumber
\\
&& + {1 \over 2 } (rfh)^{-1} f' - {1 \over 2} (rh^2)^{-1} h'.
\end{eqnarray}
We assume $h$ is
\begin{equation}
\label{h(r)}
h(r) = \left[ 1 - {2 m(r) \over r} \right]^{-1},
\end{equation}
and if we write
\begin{equation}
\label{f(r)}
f = \exp(2 \phi),
\end{equation}
the equation (\ref{T_11}) becomes
\begin{equation}
\label{dphi}
{\d \phi \over \d r} = {[m(r) + 4 \; \pi p(r) \; r^3] \over r^2 \; [1 - 2 m(r)/r]}.
\end{equation}
\begin{proof}[Proof for hydrostatic equilibrium ]
We substitute equations (\ref{h(r)}) and (\ref{f(r)}) into equation (\ref{T_11})
we get,
\begin{eqnarray}
8 \pi p &=& (rfh)^{-1} - r^{-2} (1 - h^{-1})
\nonumber
\\
&=& {f' \over f} \; {1 \over r h}  - {(1 - h^{-1}) \over r^2}
\nonumber
\\
&=& 2 \; {\d \phi \over \d r} \; {1 - {\txt 2 m(r) \over \txt r} \over \txt r}  - {{\txt 2 m(r) \over \txt r} \over r^2}
\nonumber
\\
&=& 2 \; {\d \phi \over \d r} \; {r -2 m(r) \over r^2} - {2 m(r) \over r^3}
\nonumber
\\
2 \; {\d \phi \over \d r} \; {r -2 m(r) \over r^2} &=& 8 \pi p + {2 m(r) \over r^3}
\nonumber
\\
{\d \phi \over \d r} &=& {4 \pi p + {\txt m(r) \over \txt r^3} \over {\txt r -2 m(r) \over \txt r^2}}
\nonumber
\\
&=& {[m(r) + 4 \; \pi p \; r^3] \over r^2 \; [1 - 2 m(r)/r]}
\end{eqnarray}
as previously illustrated.
\end{proof}
To see what happens to the pressure and density it will be useful to consider the TOV equation.
\begin{equation} \label{TOV_pressure_density}
\frac{\d p}{\d r}
= \frac{-[\rho(r) + p(r)] \; [m(r) + 4 \pi p(r) \; r^3]}{r^2 \; [1 - 2 m(r) / r]}
\end{equation}
\begin{proof}
Consider the stress energy of a perfect fluid is in the equation (\ref{T_{ab}})
and it satisfies the equation (\ref{na_ble})
which yield,
\begin{equation}
(p + \rho) u^a \nabla_a u_b + (g_{ab} + u_a u_b) \nabla^a p = 0
\end{equation}
Now consider,
\begin{equation}
\label{e_1}
(e_1)_b \nabla_a T^{ab} = 0,
\end{equation}
where, $(e_{1})_{b} = h^{1/2} (\d r)_b = h^{-1/2} (\partial/ \partial r)_{b}$
\\
We can reduce the equation (\ref{e_1}) into the equation (\ref{tov_eqn}) because $(e_{1})_{b}$ is pointing in the $r$-direction, which is perpendicular to $u^a$.
\begin{eqnarray}
\label{tov_eqn}
(e_1)_b \; [ (p + \rho) \; u^a \nabla_a u^b + (\delta_a^b + u_a u^b) \; \nabla^a p ] &=& 0
\\
h^{-1/2} \; (0,1,0,0)_b [(p + \rho) \; u^a \nabla_a u^b + (\delta_a^b + u_a u^b) \; \nabla^a p] &=& 0
\nonumber
\\
h^{-1/2} (p + \rho) \; (u^a \nabla_a u^b)_{b} + h^{-1/2} \; {\d p \over \d r} &=& 0
\nonumber
\end{eqnarray}
Now we consider,
\begin{eqnarray}
\label{4_velocity}
(u^a \nabla_a u^b) &=& u^a \left[ \partial_a u^b + \Gamma^b_{c a} u^c\right] = u^a \; \Gamma^b_{c a} u^c
\nonumber
\\
&=&{1\over f} \; \Gamma^b_{tt} = {1 \over f} \; {1 \over2} g^{bc} \left[ - g_{tt,e} + g_{te,t} + g_{et,t} \right]
\nonumber
\\
&=& {1 \over f} \;g^{be} \; {1 \over 2} (f)_{,e} = {1 \over 2} \; {1 \over f} \; \nabla^b f
\nonumber
\\
&=& \nabla^b \phi = {\d \phi \over \d r}
\end{eqnarray}
Now we substitute equation (\ref{dphi}) into equation (\ref{tov_eqn}) we get,
\begin{eqnarray}
\label{tov_hydrostatic_equilibrium}
h^{-1/2} (p + \rho) \; {\d \phi \over \d r} + h^{-1/2} \; {\d p \over \d r} &=& 0
\nonumber
\\
{\d p \over \d r} &=& - (p + \rho) \; {\d \phi \over \d r}
\nonumber
\\
{\d p \over \d r} &=& - (p + \rho) \; {[m(r) + 4 \pi p(r) \; r^3] \over r^2 [1 - 2 m(r)/r]}
\end{eqnarray}
Equation (\ref{tov_hydrostatic_equilibrium}) is known as the \emph{Tolman-Oppenheimer-Volkoff equation} of hydrostatic equilibrium.
\end{proof}
This equation applies only for a perfect fluid $(p_r = p_t)$ in Schwarzschild coordinates. There are generalizations appropriate to other coordinate systems. In the non-relativistic Newtonian approximation this simplifies to $(p \ll \rho, \; 2 m(r)/r \ll 1)$:
\begin{equation} \label{non-relativistic}
\frac{\d p}{\d r} = -\frac{[\rho(r) \; m(r)]}{r^2}
\end{equation}
which is something that is very easy to derive in \emph{Newton's theory of gravity}.
\\
If we assume $\rho(r)$ and hence $m(r)$ are given, then the TOV is a nonlinear first-order ODE for $p(r)$ called a Riccatti equation. Given $p(0)$ a unique solution $p(r)$ necessarily exists, and can be used to reconstruct the function $\zeta(r)$ via:
\begin{equation} \label{function_zeta}
\zeta(r) = \exp \left\{ \int{\frac{[m(r) + 4 \; \pi p(r) \; r^3]}{r^2 \; [1 - 2 m(r)/r]} \; \d r}\right\} 
\end{equation}
\begin{proof}
Firstly, start with some static spherically symmetric geometry in Schwarzschild (curvature) coordinates
\begin{equation} \label{line_element_1}
\d s^2 = - \zeta(r)^2 \; \d t^2 + {\d r^2\over B(r)} + r^2 \;\d\Omega^2
\end{equation}
and then assume it represents a perfect fluid sphere.

When 
\begin{eqnarray}
8 \pi p &=& G_{rr} =  {2 B \zeta' r - \zeta + B \zeta \over r^2 \zeta}
\nonumber
\\
8 \pi p &=& {2 B \over r} \; {\zeta' \over \zeta} - {1 \over r^2} + {B \over r^2}
\nonumber
\\
{2 B \over r} \; {\zeta' \over \zeta} &=&  8 \pi p  + {1 \over r^2} - {B \over r^2}
\nonumber
\\
{\zeta' \over \zeta} &=&  {4 \pi p r \over B} + {(1-B) \over 2 B r}
\end{eqnarray}
when $B = 1- 2 m(r)/r$, then we get
\begin{eqnarray}
{\zeta' \over \zeta} &=&  {4 \pi p r \over (1- 2 m(r)/r)} + {1 - (1- 2 m(r)/r) \over 2 \; (1- 2 m(r)/r) r}
\nonumber
\\
{\zeta' \over \zeta} &=& {m(r) + 4 \pi p \; r^3  \over r^2 \; (1- 2 m(r)/r)}
\nonumber
\\
{\d \ln(\zeta) \over \d r} &=& {m(r) + 4 \pi p \; r^3  \over r^2 \; (1- 2 m(r)/r)}
\nonumber
\\
\ln(\zeta) &=& \int \left({ m(r) + 4 \pi p \; r^3  \over r^2 \; (1- 2 m(r)/r)}\right) \; \d r
\nonumber
\\
\exp (\ln(\zeta)) &=& \exp \left\{\int { [m(r) + 4 \pi p \; r^3 ] \over r^2 \; [1- 2 m(r)/r]}\; \d r \right\}
\nonumber
\\
\zeta &=& \exp \left\{\int {[m(r) + 4 \pi p \; r^3 ] \over r^2 \; [1- 2 m(r)/r]} \; \d r \right\}
\end{eqnarray}
as required.
\end{proof}
In this way, starting from $\rho(r)$ you can construct first the function $m(r) = 4 \pi \int \rho(r) r^2 \; \d r$, then (via the TOV) the pressure profile $p(r)$, and finally via the above the gravitational redshift $\zeta(r)$.

Suppose in contrast that the pressure profile $p(r)$ is given, then the TOV equation can be rearranged into a nonlinear first order PDE for $m(r)$:
\begin{equation} \label{first_order_PDE}
\frac{1}{4 \pi r^2} \; \frac{\d m(r)}{\d r} = - p(r) - \frac{r^2 \; [1 - 2 m(r)/r]}{[m(r) + 4 \pi \; p(r) \; r^3]} \frac{\d p}{\d r}
\end{equation}
\begin{proof}
We substitute equations (\ref{h(r)}) and (\ref{f(r)}) into equation (\ref{T_00}) to get,
\begin{equation}
\label{dm(r)}
{ \d m(r) \over \d r} = 4 \pi \rho(r) r^2,
\end{equation}
and rearrange in term of $\rho(r)$ so,
\begin{equation}
\label{rho(r)}
\rho(r) = {1 \over 4 \pi r^2} \; {\d m(r) \over \d r}
\end{equation}
rearrange the equation (\ref{TOV_pressure_density}) into,
\begin{equation}
\label{rho(r)_term}
\rho(r) = - p(r) - \frac{r^2 \; [1 - 2 m(r)/r]}{[m(r) + 4 \pi \; p(r) \; r^3]} \frac{\d p}{\d r}
\end{equation}
now substitute equation (\ref{rho(r)}) into equation (\ref{rho(r)_term}), we get
\begin{equation}
\frac{1}{4 \pi r^2} \; \frac{\d m(r)}{\d r} = - p(r) - \frac{r^2 \; [1 - 2 m(r)/r]}{[m(r) + 4 \pi \; p(r) \; r^3]} \frac{\d p}{\d r}
\end{equation}
\end{proof}
This is an Abel equation (2nd type, class A). There is no simple general solution. However given $p(r)$ and the initial data $m(0) = 0$ this can in principle be solved to determine $m(r)$. Once this is done $\zeta(r)$ is again reconstructed via
\begin{equation} \label{zeta_in_term_of_exp}
\zeta(r) = \exp \left\{\int{\frac{[m(r) + 4 \pi p(r) \; r^3]}{r^2 [1 - 2 m(r)/r]} \; \d r}\right\}
\end{equation}
\section{Solution generating theorems}
\subsection{Four  theorems}
The first four theorems we presented in chapter 3 were first explicitly published in \cite{Petarpa}. We now rephrase the theorems in slightly different formalism, and demonstrate an independent way of proving them. Indeed, we can re-write these theorems directly in terms of pressure and density profiles $p(r)$ and $\rho(r)$.

Firstly, start with some static spherically symmetric geometry in Schwarzschild (curvature) coordinates
\begin{equation} \label{line_element_2}
\d s^2 = - \zeta_0(r)^2 \; \d t^2 + {\d r^2\over 1 - {\txt 2 m_0(r)\over \txt r } } + r^2 \; \d \Omega^2
\end{equation}
and then assume it represents a perfect fluid sphere. Our original four theorems can be written in the form 
\setcounter{theorem}{0}
\begin{theorem}
Suppose $\{ \zeta_0(r), m_0(r) \}$ represents a perfect fluid sphere.
Define
\begin{equation}
m_1(r) = \left({ \zeta_0(r)\over  \zeta_0(r) + r  \;\zeta'_0(r)}\right)^2 \; r^3\; 
\exp\left\{ 2 \int {\zeta'_0(r)\over  \zeta_0(r)} \; { \zeta_0(r)- r\; \zeta'_0(r)\over  \zeta_0(r) + r  \;\zeta'_0(r)} \; \d r\right\}.
\end{equation}
Then for all $\lambda$, the geometry defined by holding $\zeta_0(r)$ fixed and setting
\begin{equation}
\d s^2 = - \zeta_0(r)^2 \; \d t^2 + {\d r^2\over 1 - {\txt 2 [m_0(r) + \lambda \; m_1(r)] \over \txt r}}
+ r^2 \d\Omega^2
\end{equation}
is also a perfect fluid sphere.
\end{theorem}
\begin{theorem}
Let $\{\zeta_0(r),m_0(r)\}$ describe a perfect fluid sphere. For all $\sigma$ and $\epsilon$, the geometry defined by holding $m_0(r)$ fixed and setting
\begin{equation}
\d s^2 = - \zeta_0(r)^2 \; \left\{\sigma +\epsilon \lint {r \; \d r\over  \zeta_0(r)^2\; \sqrt{1 - {\txt 2 m_0(r)\over \txt r}} }\right\}^2
\; \d t^2 + {\d r^2\over {1 - { \txt 2 m_0(r)\over \txt r}}} + r^2 \d\Omega^2
\end{equation}
is also a perfect fluid sphere.
\end{theorem}

\begin{theorem}
Let $\{\zeta_0(r),m_0(r)\}$ describe a perfect fluid sphere. For for all $\sigma$,  $\epsilon$, and $\lambda$, the three-parameters geometry defined by

\begin{eqnarray}
\d s^2 &=& - \zeta_0(r)^2 \; \left\{\sigma +\epsilon \lint {r \; \d r\over  \zeta_0(r)^2\; \sqrt{1 - { \txt 2[ m_0(r) + \lambda \; m_1(r)]\over\txt r}} }\right\}^2 \d t^2 
\nonumber
\\
&+& {\d r^2\over 1 - {{\txt 2[ m_0(r) + \lambda \; m_1(r)]\over\txt r}} } + r^2 \; \d \Omega^2
\end{eqnarray}
is also a perfect fluid sphere, where $m_1(r)$ is defined as above.
\end{theorem}
\begin{theorem}
Let $\{\zeta_0(r),m_0(r)\}$ describe a perfect fluid sphere. For all $\sigma$,  $\epsilon$, and $\lambda$, the three-parameters geometry defined by
\begin{equation}
\d s^2 = - \zeta_0(r)^2 
\left\{
\sigma+\epsilon \lint {r \; \d r\over  \zeta_0(r)^2\; \sqrt{1-{\txt2m_0(r)\over\txt r}}} 
\right\}^2
\; \d t^2 + {\d r^2\over1-{ {\txt 2[m_0(r)+\lambda\; \tilde{m}_1(r)]\over\txt r}} }
+ r^2 \d\Omega^2
\end{equation}
is also a perfect fluid sphere, where $\tilde{m}_1(r)$ is defined as
\begin{equation}
\tilde{m}_1(r)=
\left({ \zeta(r)\over  \zeta(r) + r  \;\zeta'(r)}\right)^2 \; r^3 \; 
\exp\left\{ 2 \int {\zeta'(r)\over  \zeta(r)} \; 
  { \zeta(r)- r\; \zeta'(r)\over  \zeta(r) + r  \;\zeta'(r)} \; \d r\right\} \, ,
\end{equation}
depending on $\zeta = \zeta_0 \zeta_1$, where
\begin{equation}
\zeta_1(r) = \left\{
\sigma+\epsilon \lint {r \; \d r\over  \zeta_0(r)^2\; \sqrt{1-{\txt2m_0(r)\over\txt r}}} 
\right\} \, .
\end{equation}
\end{theorem}
\subsection{Four theorems in terms of pressure and density}
We can re-write these theorems in terms of pressure and density profiles $p(r)$ and $\rho(r)$.

Consider Theorem 1. By definition $\zeta$ is unaffected under the transformation. But we have a formula for $\zeta(r)$ in terms of $p(r)$ and $m(r)$. Let $m_2(r) = m_0(r) + \lambda m_1(r)$ and $p_2(r)$ be the mass profile and pressure profile after applying theorem 1. Then 
\begin{equation}
{[m_0(r) + 4 \pi p_0(r) \; r^3]\over r^2 \; [1 - 2 m_0(r)/r]} = {[m_2(r) + 4 \pi p_2(r) \; r^3] \over r^2 \; [1 - 2 m_2(r)/r]}
\end{equation}
which we can rearrange to yield
\begin{equation}
m_2(r) + 4 \pi p_2(r) \; r^3 = [m_0(r) + 4 \pi p_0(r) \; r^3] \; {[1 -2m_2(r)/r]\over[1 -2 m_0(r)/r]}
\end{equation}
Furthermore, we can rearrange it into
\begin{equation}
\label{p2(r)}
4 \pi p_2(r) \; r^3 = - m_2(r) + [m_0(r) + 4 \pi p_0(r) \; r^3] \; {[1 -2m_2(r)/r]\over[1 -2 m_0(r)/r]}
\end{equation}
Now we substitute $m_2(r) = m_0(r) + \lambda m_1(r)$ into equation (\ref{p2(r)}),
\begin{equation}
4 \pi p_2(r) \; r^3 = -  m_0(r) - \lambda m_1(r) + [m_0(r) + 4 \pi p_0(r) \; r^3] \; {[1 -2(m_0(r) + \lambda m_1(r))/r]\over[1 -2 m_0(r)/r]}
\end{equation}
After a bit of algebra:
\begin{eqnarray}
4 \pi p_2(r) \; r^3  &=&  {-\lambda m_1(r) + 4 \pi p_0(r) - 8 \pi p_0(r) \; r^2 \;  m_0(r) - 8 \lambda \pi p_0(r) \; r^2 \; m_1(r) \over [ 1 - 2 m_0(r)/r]}
\nonumber
\\
&=&  {-\lambda m_1(r) (1 - 8 \pi p_0(r) \; r^2) \over [1 - 2 m_0(r)/r]} + {4 \pi p_0(r) \; r^3 [1 - 2 m_0(r)/r] \over [1 - 2 m_0(r)/r]}
\nonumber
\\
&=& 4 \, \pi p_0(r) \, r^3 - {\lambda m_1(r) (1 + 8 \pi p_0(r) \; r^2) \over [1 - 2 m_0(r)/r]}
\end{eqnarray}
rearrange it in terms of $p_2(r)$,
\begin{equation}
p_2(r) = p_0(r) - \lambda m_1(r) \left[{1 \over 4 \pi r^3}\right] \; \left[{1 + 8 \pi p_0(r) \; r^2 \over 1 -2 m_0(r)/r}\right]
\end{equation}
This allows us to rewrite Theorem 1 as follows:
\setcounter{theorem}{12}
\begin{theorem}[Theorem 1B:]
Suppose we are given a pair of functions $p_0(r)$ and $\rho_0(r)$ that satisfy the TOV.

We use these  functions to construct $m_0(r)$, $\zeta_0(r)$, and $m_1(r)$ as follows:
\begin{equation}
m_0(r) = 4 \pi \int_0^r \rho_0(\bar{r}) \; \bar{r}^2 \; \d \bar{r}
\end{equation}
\begin{equation}
\zeta_0(r) = \exp \left\{\int {[m_0(r) + 4 \pi p_0(r) \; r^3] \over r^2 \; [1 - 2 m_0(r)/r]} dr\ \right\}
\end{equation}
\begin{equation}
m_1(r) = \left({\zeta_0(r) \over \zeta_0(r) + r \; \zeta_0'(r)}\right)^2 \; r^3 \exp \left\{2 \int {\zeta_0'(r) \over \zeta_0(r)} \; {\zeta_0(r) -r \; \zeta_0'(r) \over \zeta_0(r) + r \; \zeta_0'(r)} \; \d r\right\}
\end{equation}
Then for all $\lambda$, the pair of functions
\begin{equation}
p_2(r) = p_0(r) - \lambda m_1(r) \left[{1 \over 4 \pi r^3}\right] \; \left[{ 1 + 8 \pi p_0(r) r^2 \over 1 - 2 m_0(r)/r}\right]
\end{equation}
\begin{equation}
\rho_2 = \rho_0(r) + {\lambda \over 4 \pi r^2} \; {\d m_1(r) \over \d r}
\end{equation}
also satisfies the TOV and so defines a perfect fluid sphere.
\end{theorem}
Similar logic can be applied to theorem 2. Note that in that case you are considering the transformation
\begin{equation}
m_0(r) \to m_2(r) = m_0(r)
\end{equation}
\begin{equation}
\zeta_0(r) \to \zeta_2(r) = \zeta_0(r) \left\{ \sigma + \epsilon \lint {r \; \d r \over \zeta_0(r)^2 \sqrt{1 - {\txt 2 m_0(r) \over \txt r}}} \right\}
\end{equation}
So that $m_0(r)$ remains fixed.
Now from the $G_{\hat{r}\hat{r}}$ component of the Einstein tensor: 
consider the equation below
\begin{equation}
{\zeta' \over \zeta} = {m(r) + 4 \pi p(r) \; r^3  \over r^2 \; (1- 2 m(r)/r)}
\end{equation}
and rearrange this equation in term of $p(r)$
\begin{eqnarray}
{\zeta' \over \zeta} \; r^2 \; (1- 2 m(r)/r) &=& m(r) + 4 \pi p (r)\; r^3
\nonumber
\\
4 \pi p (r)\; r^3 &=& - m(r) + r^2 \; (1- 2 m(r)/r) \; {\zeta' \over \zeta}
\nonumber
\\
p(r) &=& - { m(r) \over 4 \pi r^3} + {1 \over 4 \pi r} \; (1- 2 m(r)/r) \; {\zeta' \over \zeta}
\end{eqnarray}
when
\begin{equation}
{m(r) \over 4 \pi r^3} = {m(r)/3 \over (4 \pi /3) \; r^3} = {1 \over 3} \; \bar\rho(r)
\end{equation}
Now we have $p(r)$ equal to
\begin{equation}
p(r) = - {1 \over 3} \; \bar\rho(r) + {1 \over 4 \pi r} \; (1- 2 m(r)/r) \; {\zeta'(r) \over \zeta(r)}
\end{equation}
Then in particular
\begin{equation}
p_2(r) = - {1 \over 3} \; \bar\rho_0(r) + {1 \over 4 \pi r} \; (1- 2 m_0(r)/r) \; {\zeta'_2(r) \over \zeta_2(r)}
\end{equation}
so that
\begin{equation}
p_2(r) = - {1 \over 3} \; \bar\rho_0(r) + {1 \over 4 \pi r} \; \left(\txt 1 - {\txt 2 m_0(r) \over \txt r}\right) \; {\left[\zeta_0(r) \left\{ \sigma + \epsilon \lint {\txt r \; \d r \over \txt \zeta_0(r)^2 \sqrt{\txt 1 - {\txt 2 m_0(r) \over \txt r}}}\right\}\right]' \over  \zeta_0(r) \left\{ \sigma + \epsilon \lint {\txt r \; \d r \over \txt \zeta_0(r)^2 \sqrt{\txt 1 - {\txt 2 m_0(r) \over \txt r}}} \right\}}
\end{equation}
That is
\begin{eqnarray}
p_2(r) &=& - {1 \over 3} \; \bar\rho_0(r) + {1 \over 4 \pi r} \; \left(1 - {2m_0(r) \over r}\right) \; {\zeta'_0(r) \over \zeta_0(r)} 
\nonumber
\\
&+& {1 \over 4 \pi r} \; \left(\txt 1 - {\txt 2 m_0(r) \over \txt r}\right) \; {\left\{ \txt \epsilon {\txt r \over \txt \zeta_0(r)^2 \sqrt{\txt 1 - {\txt 2 m_0(r) \over \txt  r}}}\right\} \over \left\{ \txt \sigma + \txt \epsilon \lint {\txt r \; \d r \over \txt \zeta_0(r)^2 \sqrt{\txt 1 - {\txt 2 m_0(r) \over \txt r}}} \right\}}
\end{eqnarray}
we know $p(r) = - { \txt 1 \over \txt 3} \; \bar\rho(r) + { \txt 1 \over \txt {4 \pi r}} \; (\txt 1- 2 \txt m(r)/r) \; {\txt \zeta'_0(r) \over \txt \zeta_0(r)}$ , so that
\begin{equation}
p_2(r) = p_0(r) + {\epsilon \over 4 \pi } \; \sqrt{\txt 1 - {\txt 2 m_0(r) \over \txt r}} \; \zeta_0(r)^{-2} \; \left\{ \sigma + \epsilon \lint {r \; \d r \over \zeta_0(r)^2 \sqrt{\txt 1 - {\txt 2 m_0(r) \over \txt r}}} \right\}^{-1}
\end{equation}
This leads us to:
\begin{theorem}[Theorem 2B:]
Suppose we are given a pair of functions $p_0(r)$ and $\rho_0(r)$ that satisfy the TOV.

Use these functions to construct $m_0(r)$ and $\zeta_0(r)$ as follows:
\begin{equation}
m_0(r) = 4 \pi \int_0^r \rho_0(\bar{r}) \; \bar{r}^2 \; \d \bar{r}
\end{equation}
\begin{equation}
\zeta_0(r) = \exp \left\{ \int{[m_0(r) + 4 \pi p_0(r) \; r^3] \over r^2 \; [1 - 2 m_0(r)/r]} \; \d r \right\}
\end{equation}
Then for all $\sigma$ and $\epsilon$, the pair of functions
\begin{equation}
\rho_2(r) = \rho_0(r)
\end{equation}
\begin{equation}
p_2(r) = p_0(r) + {\epsilon \over 4 \pi } \; \sqrt{1 - {2 m_0(r) \over r}} \; \zeta_0(r)^{-2} \; \left\{ \sigma + \epsilon \lint {r \; \d r \over \zeta_0(r)^2 \sqrt{\txt 1 - {\txt 2 m_0(r) \over \txt r}}} \right\}^{-1}
\end{equation}
also satisfies the TOV and so defines a perfect fluid sphere.
\end{theorem}
Having now found these theorems 1B and 2B, is now useful to define a new theorem by composing them.
All together we can consider this as a single process by introducing the following theorem:
\begin{theorem}[Theorem 3B:]
Suppose we are given a pair of functions $p_0(r)$ and $\rho_0(r)$ that satisfy the TOV.

Use these functions to construct $m_0(r)$, $\zeta_0(r)$, and $m_1(r)$ as follows:
\begin{equation}
m_0(r) = 4 \pi \int_0^r \rho_0(\bar{r}) \; \bar{r}^2 \; \d \bar{r}
\end{equation}
\begin{equation}
\zeta_0(r) = \exp \left\{ \int{[m_0(r) + 4 \pi p_0(r) \; r^3] \over r^2 \; [1 - 2 m_0(r)/r]} \; \d r \right\}
\end{equation}
\begin{equation}
m_1(r) = \left({\zeta_0(r) \over \zeta_0(r) + r \; \zeta_0'(r)}\right)^2 \; r^3 \exp \left\{2 \int {\zeta_0'(r) \over \zeta_0(r)} \; {\zeta_0(r) -r \; \zeta_0'(r) \over \zeta_0(r) + r \; \zeta_0'(r)} \; \d r\right\}
\end{equation}
Then for all $\sigma$, $\epsilon$, and $\lambda$, the pair of functions
\begin{eqnarray}
p_2(r) &=& p_0(r) - \lambda m_1(r) \left[{1 \over 4 \pi r^3}\right] \; \left[{1 + 8 \pi p_0(r) \; r^2 \over 1 -2 m_0(r)/r}\right] 
\nonumber
\\
&& + {\epsilon \over 4 \pi } \; \sqrt{1 - {2 m_2(r) \over r}} \; \zeta_0(r)^{-2} \; \left\{ \sigma + \epsilon \lint {r \; \d r \over \zeta_0(r)^2 \sqrt{\txt 1 - {\txt 2 m_2(r) \over \txt r}}} \right\}^{-1}
\end{eqnarray}
\begin{equation}
\rho_2 = \rho_0(r) + {\lambda \over 4 \pi r^2} \; {\d m_1(r) \over \d r}
\end{equation}
where $ m_2(r) = m_0(r) + \lambda m_1(r)$, also satisfy the TOV and so define a perfect fluid sphere.
\end{theorem}
\begin{theorem}[Theorem 4B:]
Suppose we are given a pair of functions $p_0(r)$ and $\rho_0(r)$ that satisfy the TOV.

Use these functions to construct $m_0(r)$, $\zeta_0(r)$, and $m_1(r)$ as follows:
\begin{equation}
m_0(r) = 4 \pi \int_0^r \rho_0(\bar{r}) \; \bar{r}^2 \; \d \bar{r}
\end{equation}
\begin{equation}
\zeta_0(r) = \exp \left\{ \int{[m_0(r) + 4 \pi p_0(r) \; r^3] \over r^2 \; [1 - 2 m_0(r)/r]} \; \d r \right\}
\end{equation}
\begin{equation}
\tilde{m}_1(r)=
\left({ \zeta(r)\over  \zeta(r) + r  \;\zeta'(r)}\right)^2 \; r^3 \; 
\exp\left\{ 2 \int {\zeta'(r)\over  \zeta(r)} \; 
{ \zeta(r)- r\; \zeta'(r)\over  \zeta(r) + r  \;\zeta'(r)} \; \d r\right\} \, ,
\end{equation}
depending on $\zeta = \zeta_0 \zeta_1$, where
\begin{equation}
\zeta_1(r) = \left\{
\sigma+\epsilon \lint {r \; \d r\over  \zeta_0(r)^2\; \sqrt{1-{\txt2m_0(r)\over\txt r}}} 
\right\} \, .
\end{equation}
Then the pair of functions
\begin{equation}
p_2(r) = p_0(r) - \lambda m_1(r) \left[{1 \over 4 \pi r^3}\right] \; \left[{1 + 8 \pi p_0(r) r^2 \over 1 -2 m_0(r)/r}\right] 
\end{equation}
\begin{equation}
\rho_2 = \rho_0(r) + {\lambda \over 4 \pi r^2} \; {\d \tilde{m}_1(r) \over \d r}
\end{equation}
also satisfy the TOV and so define a perfect fluid sphere.
\end{theorem}
\begin{table}[!ht]
\centerline{Some seed geometries and their descendants in term of pressure and density via Theorem 1B}
\bigskip
\hskip-1.5cm
{\footnotesize
\begin{tabular}{|  l  |  l  |  l  |  l  | l |}
\hline
Seed & $p_0(r)$ & $p_{\text{Th1}}(r)$ & $\rho_0(r)$ & $\rho_{\text{Th1}}(r)$\\
\hline
Minkowski & $0$ & $-{\lambda \over 4 \pi}$ & $0$ & ${3 \lambda \over 4  \pi}$ \\
exterior Schwarzschild & $0$ & ${3500 \, \lambda \, (r -a)^2 \over \pi \, r^2}$ & $0$ & $- {3500 \, \lambda \, (r-a) (3 r - 5 a) \over \pi r^2}$ \\
de Sitter& $-{3 \over 8 \pi R^2}$  & $p_0(r) - {\lambda (R^2 - 3 r^2) \over 4 \pi (-R^2 + 2 r^2)}$ & ${3 \over 8 \pi R^2}$ & $\rho_0(r) + {\lambda \over 4 \, \pi} {( 7 R^2 r^2 - 3 R^4 - 6 r^4) \over (-R^2 + 2 r^2)^2}$ \\
Tolman V ($A=0$) & $-{(n+1)^2 \over 8 r^2 \pi (n^2 -2)}$ & $p_0(r) - {\lambda r^{\left(-{(2 n + n^2) \over (n + 2)}\right)} (2 n + 3) \over 4 \, \pi (n + 2)^2 } $ & ${-1 + n^2 \over 8  r^2 \pi (-2 + n^2)}$ & $\rho_0(r) +  {\lambda r^{\left(-{(2 n + n^2) \over (n+2)}\right)} (3 + 2 n) (2 - n) \over 4 \, \pi (n+2)^3}$\\
S1 & $ {1 \over 2 \pi r^2}$ & $p_0(r) - { 5 \lambda \over 36 \, \pi r^{(4/3)}}$ & $0$ & ${ 5 \lambda \over 108 \, \pi  r^{(4/3)}}$ \\
M--W III & $-{(a-6 r) \, (a + 2 r) \over 56 \, \pi \, r^2 a^2 }$ &  $p_0(r) + \lambda {\left(r \over 5\right)}^{({1 \over 3})} \, {(r -a) (2 a  + 3r) \over 9 \, (a + r) (-4 r + 3 a)^{({4 \over 3})}}$ & ${3 \, (a^2 + 4 \, r^2) \over 56 \, \pi r^2  a^2}$ & $\rho_0(r) + {\lambda \over 9 \, \pi} \, {\left(r \over 5\right)} ^{({1 \over 3})} \, {(12 r^2 - 21 a r + 10 a^2) \over (-4 r + 3 a)^{({7 \over 3})}} $ \\
Heint IIa ($C=0$) & $-{9 a (-1 + a r^2) \over 16 \pi (1+a r^2)^2}$ & $p_0(r) - {\lambda (7 a r^2 +1) \over 12 \sqrt{3} \, \pi  (1 + a r^2)^2 \, \sqrt{-4 a r^2 -1}}$ &
${3 a \, (3 + a r^2) \over 16 \pi (1 + a r^2)^2}$ & $\rho_0(r) - {\lambda (1 + 3 a r^2) \over 4 \sqrt{3} \pi (1+ a r^2)^2 \, (-4 a r^2 -1)^{({3 \over 2})}}$\\
\hline
\hline
\end{tabular}
} 
\caption{Seed solutions and their generalizations derived via theorems 1B in terms of pressure and density profile.}
\bigskip
\label{T_1B}
\end{table}
\begin{table}[!ht]
\centerline{Some seed geometries and their descendants in term of pressure and density via Theorem 2B}
\bigskip
\hskip.5cm
\begin{tabular}{|  l  |  l  |  l  |}
\hline
Seed & $p_{\text{Th2}}(r)$ & $\rho_{\text{Th2}}(r)$\\
\hline
Minkowski & ${\epsilon \over 4 \, \pi \left(\sigma + {\epsilon r^2 \over 2}\right)}$   & $0$ \\
exterior Schwarzschild & $ {1 \over 2 \, \pi (r^2 +  5 r a - 30 \, a^2  +  {2 \, \sigma \, \sqrt{r (r - 2 a)} \over \epsilon} + 15 \sqrt{r (r - 2 a)} \, a^2 \ln( r - a + \sqrt{r(r -2 a)}))}$ & $0$\\
de Sitter& $p_0(r) + {\epsilon  \over 4 \pi \sqrt{R^2 - r^2} \left({\sigma \over R} + {\epsilon \, R^2 \over \sqrt{-R^2 + r^2}}\right)}$ & $\rho_0(r)$ \\
Tolman V ($A=0$) & $p_0(r) + {{\epsilon \over 2 -n^2} \over 2 \, \pi r^2 \left(2 B^2 n r^{(1+n)} \sqrt{{1 \over 2-n^2}} -\epsilon\right)} $& $\rho_0(r)$\\
S1 & $p_0(r) + {\epsilon \over {4 \, \pi r^4 \left(\sigma - {\epsilon \over 2 r^2}\right)}}$ & $0$ \\
M--W III & $p_0(r) + {\epsilon \over {2 \sqrt{7} \, \pi A r a  \sigma (r-a) \over \sqrt{a^2 -r^2}} + 7 \, \pi r a \, \epsilon}$ & $\rho_0(r)$ \\
Heint IIa (C=0) & $p_0(r) + { \epsilon \over 8 \, \pi {(1+a r^2)^{(7/2)} \over \sqrt{4 - 2 a r^2}} \left({\sigma + {2 \epsilon \, (5 + 2 a r^2) \over 27 a \, (1 + a r^2)^{(3/2)} \, \sqrt{4 - 2 a r^2}}}\right)}$ &
$ \rho_0(r)$\\
\hline
\hline
\end{tabular}
\caption{Seed solutions and their generalizations derived via theorems 2B in terms of pressure and density profile.}
\bigskip
\label{T_2B}
\end{table}
\begin{table}[!ht]
\centerline{Some seed geometries and their descendants in term of pressure and density via Theorem 3B}
\bigskip
\hskip-2cm
{\footnotesize
\begin{tabular}{|  l  |  l  |  l  |}
\hline
Seed & $p_{\text{Th3}}(r)$ & $\rho_{\text{Th3}}(r)$\\
\hline
\hline
Minkowski & $p_0(r) + {\epsilon \over {4 \, \pi \sigma \over \sqrt{1 - 2  \lambda r^2}} - {2 \, \pi \epsilon \over \lambda}}$ & $\rho_{\text{Th1}}(r)$  \\
\hline
exterior Schwarzschild & [integral] & $\rho_{\text{Th1}}(r)$ \\
\hline
de Sitter& $p_0(r) - {\lambda (R^2 - 3 \, r^2) \over 4 \, \pi (-R^2 + 2 \, r^2)}$  & $\rho_{\text{Th1}}(r)$ \\
&  $+{\epsilon \, \sqrt{1 - {r^2 \over R^2} + {2 \lambda r^2 (-R^2 + r^2)\over (- R^2 + 2r^2)}} \over 4 \, \pi (R^2 -r^2)^2 \left({\sigma \over R^2} + \epsilon \lint {r R^2 \over (R^2 -r^2) \, \sqrt{1 - {r^2 \over R^2} + {2 \lambda r^2 (-R^2 + r^2)\over (- R^2 + 2r^2)}}} \, \d r\right)}$ & \\
\hline
Tolman V ($A=0$) & $p_0(r) - {\lambda r^{{(1+n) (-2n) \over (2+n)}} (3 + 2 n) \over 4 \, \pi (2+n)^2}$
& $\rho_{\text{Th1}}(r)$\\
& $- { \epsilon \sqrt{{1 \over (2 - n^2)} + {2 \lambda r^{{2(2- 2n^2) \over (2 + n)}} \over (2 + n)^2}} \over 4 \, \pi B^2 r^{(1+n)^2}\left( \sigma + \epsilon \lint {r^{(-1- 2 n)} \over B^2 \sqrt{{1 \over (2 - n^2)} + {2 \lambda r^{{2(2- 2n^2) \over (2 + n)}} \over (2 + n)^2}}} \, \d r \right)}$ & \\
\hline
S1 & $p_0(r) + {\lambda \over 12 \, r^{(4/3)} \, \pi} +$ & $\rho_{\text{Th1}}(r)$ \\
& $ {\epsilon \over \left(2 \pi r^4 \left( {6 \sigma \over \sqrt{9 - 2 \, \lambda r^{(2/3)}}}  + \epsilon \left(-{ 1\over r^2} + {5 \lambda \over 3}\left(-{1 \over 6 r^{(4/3)}} + {\lambda \over 9} \left(-{1 \over 2 r^{(2/3)}} - {\lambda \over 3 \sqrt{9 - 2 \, \lambda r^{(2/3)}}} \text{arctanh} \left({3 \over \sqrt{9 - 2 \, \lambda r^{(2/3)}}} \right) \right) \right)\right)\right)\right)}$ & \\
\hline
M--W III & $p_0(r) - {\lambda \over 9}  \left({r \over 5} \right)^{(1/3)} {(2 a - 3 r) \over (-4 r + 3 a)^{(4/3)} }$ & $\rho_{\text{Th1}}(r)$ \\
& $ + {\epsilon \sqrt{{(a^2 -r^2) \over 7 a^2} + {2 \, \lambda r^{(7/3)} \, (r - a) 5^{(2/3)} \over 45 \, (-4 r + 3 a)^{(4/3)}}} \over 2 \pi A r (r-a) \left( \sigma + \epsilon \lint{1 \over A (r-a) \sqrt{{(a^2 -r^2) \over 7 a^2} + {2 \, \lambda r^{(7/3)} \, (r - a) 5^{(2/3)} \over 45 \, (-4 r + 3 a)^{(4/3)}}}} \, \d r \right)}$ & \\
\hline
Heint IIa (C=0) & $p_0(r) - {\lambda (7 a r^2 + 1) \over 12 \, \pi \sqrt{-4 a r^2 -1}}$ & $\rho_{\text{Th1}}(r)$\\
& $-{\epsilon \sqrt{{1 \over (1 + a r^2)} \, \left(( 2 - a r^2) - {4 \, \lambda r^2 \over 3 \sqrt{-12 a r^2 -3}}\right)} \over 4 \, \pi (1+a r^2)^3 \left(\sigma + \epsilon \lint{{r \over (1+a r^2)\sqrt{ \left(( 2 - a r^2) - {4 \, \lambda r^2 \over 3 \sqrt{-12 a r^2 -3}}\right)}} \, \d r} \right)}$ & \\
\hline
\hline
\end{tabular}
} 
\caption{Seed solutions and their generalizations derived via theorems 3B in terms of pressure and density profile. The notation ``[integral]'' denotes a metric so complicated that explicitly writing out the relevant integral is so tedious that it does not seem worthwhile.}
\bigskip
\label{T_3B}
\end{table}

\begin{table}[!ht]
\centerline{Some seed geometries and their descendants in term of pressure and density via Theorem 4B}
\bigskip
\hskip4cm
\begin{tabular}{|  l  |  l  |  l  |}
\hline
Seed & $p_{\text{Th4}}(r)$ & $\rho_{\text{Th4}} (r)$\\
\hline
\hline
Minkowski & $p_{\text{Th1}}(r)$ & ${\lambda (5 \, \epsilon \, r^2 + 6 \, \sigma) \over 4 \, \pi (2 \, \sigma + 3 \, \epsilon \, r^2)^{(5/3)}}$ \\
\hline
exterior Schwarzschild & $p_{\text{Th1}}(r)$ & [integral]\\
\hline
de Sitter& $p_{\text{Th1}}(r)$  & [integral] \\
\hline
Tolman V ($A=0$) & $p_{\text{Th1}}(r)$
& [integral]\\
&  & \\
\hline
S1 & $p_{\text{Th1}}(r)$ & ${\lambda (10 \, \sigma r^2 - 3 \,\epsilon) \over 4 \, \pi (6 \, \sigma r^2 - \epsilon)^{(5/3)}}$ \\
&  & \\
\hline
M--W III & $p_{\text{Th1}}(r)$ & [integral] \\
&  & \\
\hline
Heint IIa (C=0) & $p_{\text{Th1}}(r)$ & [integral]\\
&  & \\
\hline
\hline
\end{tabular}
\caption{Seed solutions and their generalizations derived via theorems 4B in terms of pressure and density profile. The notation ``[integral]'' denotes a metric so complicated that explicitly writing out the relevant integral is so tedious that it does not seem worthwhile.}
\bigskip
\label{T_4B}
\end{table}
\subsection{Discussion}
In this chapter we have rephrased the theorems 1, 2, 3, and 4 of chapter 3 directly in terms of the TOV equation, and consequently in terms of the pressure profile and density profile. We have done this by rephrasing theorem 1 in a slightly different formalism, and developing an independent proof that is easy to rewrite in terms of the pressure and density profiles --- $\rho(r)$ and $p(r)$ --- resulting in the theorem we call theorem 1B. Similar logic, involving the transformation $\gth{2}$, can be applied to theorem 2 --- now leading to theorem 2B. With these new theorems 1B and 2B in hand, we can compose them in various ways to obtain theorems 3B and 4B.

We present several tables that exhibit seed solutions and the geometries generated from them by combining all the various theorems. Table (\ref{T_1B}) shows some seed geometries and their descendants as obtained via theorem 1B, and similarly table (\ref{T_2B}) shows some seed geometries and their descendants obtained via theorem 2B. Finally table (\ref{T_3B}) and (\ref{T_4B}) deal with theorems 3B and 4B. We emphasise that theorem 2B does not affect the density profile, and only the pressure profile changes.
\chapter{Conclusions}
As the range of problems tackled is so diverse, it is hard to think of a 
concise way of summarising all of them. Indeed, to a large extent much of the work has already has summarised them at the end of each section. Instead of 
explaining the problems again and how far through completion they are, I would like to leave with list of things that I believe are useful to understand the overall concept of the thesis.

This thesis has been written with the goal of making it accessible to people 
with a
basic background in general relativity, especially in coordinate conditions,
perfect fluid spheres in general relativity.

\section{The main analysis: Structure of the thesis}
\begin{itemize}
\item
The thesis summarized the basic concept of special relativity to make
general relativity easier to understand. General relativity is the relativistic theory of gravitation. It is required from any new physical theory to satisfy the so called principle of correspondence.
\item
Coordinate conditions are important for the long time scale simulations of relativistic systems. We demonstrated the importance of coordinate conditions in general relativity and proposed a method of looking for nice solutions of the Einstein equations. In addition, a number of examples are given to illustrate the method.
\item
We introduced ``algorithmic" techniques that permit one
to generate large classess of perfect fluid spheres in a purely mechanical
way. Furthermore, we extended these algorithmic ideas, by proving several
solution-generating theorems of varying levels of complexity. We have
developed several transformation theorems that map perfect fluid spheres to
perfect fluid spheres using Schwarzschild coordinates. Furthermore, we
used these transformations as a basis for classifying different types of
perfect fluid sphere solutions.
\item
In chapter 4, we have developed several transformation theorems that map
perfect fluid spheres to perfect fluid sphere using other coordinates.
\item
We rephased Theorems 1, 2, 3 and 4 in chapter 3 directly in terms of the 
TOV equation.
\end{itemize}
\section{Further interesting issues}
I would like to leave with the list of things that I believe are interesting
to continue to analyze. In addition, all the lists below are further issues
that might be interesting to looking at.
\begin{itemize}
\item
For a Riemannian geometry in (2+1) dimensions, we can calculate the Ricci
Scalar, the Ricci tensor and the Riemann tensor. However, all results are 
too difficult to analyze in detail. This is one of the physically interesting cases, and this is a case that I would like to examine in more detail.
\item
Similar for a Riemannian geometry in (3+1) dimensions. This is also one of 
the
physical interesting case, and this is the case that will be most difficult
to analyze in detail. This is a challenging case that I would like to work
through in more detail.
\item
We analysed a generic spacetime using both ADM and reverse-ADM 
(Kaluza--Klein)
decompositions. However, we found that for the general ADM case coordinate
techniques are too complicated to contemplate (page 514 of \cite{Misner}).
The inverse ADM decomposition can calculated the Christoffel symbols from
equation (\ref{Christ_reverse}). This is so interesting to find the 
techniques
for contemplation.
\item
The other things that would be interesting is consider other coordinate
systems:   Generalize the theorems which we derived for Schwarzschild
coordinates or coordinates in Chapter 4 to a number of other coordinate
systems. This may allow us to develop extra relations between the known
solutions.
\item
For Theorem 8 in Isotropic coordinate, it would be interesting to see what
would happen when we feed this theorem a Minkowski space seed. In addition,
this is interesting to find simple ways of generating interesting solutions.
\item
For Theorem 9 in Gaussian polar coordinate, Theorem 10 in Buchdahl
coordinates, Theorem 11 in Synge isothermal coordinates, and Theorem 12 in
Exponential coordinates, it would be interesting to see what would happen
when we feed this theorem a Minkowski space seed. In addition, this is
intersting to find the way that we can do to $R(r)$ to make its ODE look
nicer.
\end{itemize}

All the above suggestions would be interesting and feasible, although some
of them would be more tedious to extend to standard results in
general relativity. However, this thesis would provide a platform for
better understanding coordinate conditions and the perfect fluid sphere in
general relativity. This thesis developed a way of looking for nice new
solutions of the Einstein equations.
Furthermore, the goal of this thesis was to classify the different types of
perfect fluid sphere solutions. This thesis has hopefully served as a brief
introduction the ``Algorithms" ideas by showing how to prove several
solution-generating theorems of varying levels of complexity.
Finally, we illustrated the formalism to develop extra relations between the
known solutions for Schwarzschild coordinates and other coordinates. The 
generating theorems in perfect fluid spheres lead to additional useful ``exact solutions".


\addcontentsline{toc}{chapter}{Bibliography}
\bibliography{references}

\begin{thebibliography}{10}

\bibitem{Adler}
R.~Adler, M.~Bazin, and M.~Schiffer.
\newblock {\em Introduction to general relativity}.
\newblock McGraw-Hill Book company\begin{scriptsize}\end{scriptsize}, 1965.

\bibitem{Wikipedia}
Various authors.
\newblock Special relativity.
\newblock 2005.
\newblock [http://en.wikipedia.org/wiki/Special relativity].

\bibitem{Hojman-et-al}
S.~Berger, R.~Hojman, and J.~Santamarina.
\newblock General exact solutions of {Einstein} equations for static perfect
  fluids with spherical symmetry.
\newblock {\em J.\ Math.\ Phys.}, 28:2649, 1987.

\bibitem{Bondi1}
H.~Bondi.
\newblock Spherically symmetrical models in general relativity.
\newblock {\em Mon. Not. Roy. Astron. Soc.}, 107:410, 1947.

\bibitem{Bondi2}
H.~Bondi.
\newblock Massive spheres in general relativity.
\newblock {\em Mon.~Not.~Roy.~Astron.~Soc.}, 282:303--317, 1964.

\bibitem{Petarpa}
P.~Boonserm, M.~Visser, and S.~Weinfurtner.
\newblock Generating perfect fluid spheres in general relativity.
\newblock {\em Phys. \ Rev.}, D71(124037), 2005.
\newblock [arXiv:gr-qc/0503007].

\bibitem{Buchdahl1}
H.~A. Buchdahl.
\newblock General relativistic fluid spheres.
\newblock {\em Phys.\ Rev.}, 116:1027--1034, 1959.

\bibitem{Buchdahl2}
H.~A. Buchdahl.
\newblock General relativistic fluid spheres {II}: General inequalities for
  regular spheres.
\newblock {\em Ap. J.}, 146:275--281, 1966.

\bibitem{Delgaty}
M.~S.~R. Delgaty and K.~Lake.
\newblock Physical acceptability of isolated, static, spherically symmetric,
  perfect fluid solutions of einstein's equations.
\newblock {\em Comput.\ Phys.\ Commun.}, 115:395, 1998.
\newblock [arXiv:gr-qc/9809013].

\bibitem{Dirac}
P.~A.~M. Dirac.
\newblock {\em General theory of relativity}.
\newblock Princeton University Press, 2 edition, 1975.

\bibitem{Skea}
M.~R. Finch and J.~E.~F. Skea.
\newblock A review of the relativistic static fluid sphere.
\newblock 1998, unpublished.
\newblock [http://www.dft.if.uerj.br/usuarios/JimSkea/papers/pfrev.ps].

\bibitem{Fock}
V.~Fock.
\newblock {\em The Theory of Space, Time, and Gravitation}.
\newblock Pergamon, New York, 1964.

\bibitem{Lake}
K.~Lake.
\newblock All static spherically symmetric perfect fluid solutions of
  {Einstein}'s equations.
\newblock {\em Phys.\ Rev.}, D67:104015, 2003.
\newblock [arXiv:gr-qc/0209104].

\bibitem{Landau}
L.~D. Landau and E.~M. Lifshitz.
\newblock {\em The classical theory of fields}, volume~2.
\newblock Butterworth-Heinemann, Oxford, 4 edition, 1975.

\bibitem{Martin0}
D.~Martin and M.~Visser.
\newblock Bounds on the interior geometry and pressure profile of static fluid
  spheres.
\newblock {\em Class.\ Quant.\ Grav.}, 20:3699, 2003.
\newblock [arXiv:gr-qc/0306038].

\bibitem{Martin}
D.~Martin and M.~Visser.
\newblock Algorithmic construction of static perfect fluid spheres.
\newblock {\em Phys.\ Rev.}, D69:104028, 2004.
\newblock [arXiv:gr-qc/0306109].

\bibitem{Misner}
C.~W. Misner, K.~S. Thorne, and J.~A. Wheeler.
\newblock {\em Gravitation}.
\newblock (W. H. Freeman and Company, San
  Francisco\begin{scriptsize}\end{scriptsize}), 1973.

\bibitem{Rahman}
S.~Rahman and M.~Visser.
\newblock Spacetime geometry of static fluid spheres.
\newblock {\em Class.\ Quant.\ Grav.}, 19:935, 2002.
\newblock [arXiv:gr-qc/0103065].

\bibitem{D'Inverno}
R.D'Inverno.
\newblock {\em Introducing Einstein's Relativity}.
\newblock (Clarendon Press, Oxford\begin{scriptsize}\end{scriptsize}), 2002.

\bibitem{exact}
H.~Stephani, D.~Kramer, M.~MacCallum, C.~Hoenselaers, and E.~Herlt.
\newblock {\em Exact solutions of Einstein's field equations}.
\newblock (Cambridge University Press), 2003.

\bibitem{Synge}
J.~L. Synge.
\newblock {\em Relativity: the General Theory}.
\newblock North-Holland, Amsterdam, 1964.

\bibitem{Matt1}
M.~Visser.
\newblock Math 465: Notes on general relativity.
\newblock {\em Victoria University}, 2005.
\newblock [http://www.mcs.vuw.ac.nz/courses/MATH465/2005T2/Lecture-Notes/].

\bibitem{Carlos}
M.~Visser, C.~Barcelo, and S.~Liberati.
\newblock Analogue models of and for gravity.
\newblock {\em Gen.\ Rel.\ Grav.}, 34:20, 2002.
\newblock [arXiv:gr-qc/0111111].

\bibitem{Wald}
R.~M. Wald.
\newblock {\em General Relativity}.
\newblock The University of Chicago Press, 1984.

\bibitem{Wyman}
M.~Wyman.
\newblock Radially symmetric distributions of matter.
\newblock {\em Phys. Rev.}, D75:1930, 1949.

\end{thebibliography}

\end{document}